\documentclass[letterpaper,twocolumn,10pt]{article}
\usepackage{usenix2019_v3,graphicx,amsmath,amssymb,url}
\usepackage{algorithm}  
\usepackage{algorithmic}
\usepackage{multirow} 
\usepackage{xcolor}
\usepackage{bbm}
\usepackage{graphicx}
\usepackage{enumitem}

\usepackage{tikz}

\usepackage{filecontents}

\newcommand{\R}{\ensuremath{\mathbb{R}}}

\usepackage{array}
\newcolumntype{x}[1]{>{\centering\arraybackslash\hspace{0pt}}p{#1}}

\begin{document}

%don't want date printed
\date{}

%make title bold and 14 pt font (Latex default is non-bold, 16 pt)
\title{\Large \bf Improving Robustness of ML Classifiers against Realizable Evasion Attacks Using Conserved Features}

%for single author (just remove % characters)
\author{
{\rm Liang Tong}\\
Washington University in St. Louis \\
%liangtong@wustl.edu
\and
{\rm Bo Li}\\
UIUC \\
%lxbosky@gmail.com
\and
{\rm Chen Hajaj}\\
Ariel University \\
%chenha@ariel.ac.il
\and
{\rm Chaowei Xiao}\\
University of Michigan \\
%xiaocw@umich.edu
\and
{\rm Ning Zhang}\\
Washington University in St. Louis \\
%zhang.ning@wustl.edu
\and
{\rm Yevgeniy Vorobeychik}\\
Washington University in St. Louis \\
%yvorobeychik@wustl.edu
% copy the following lines to add more authors
% \and
% {\rm Name}\\
%Name Institution
} % end author
\maketitle

% Use the following at camera-ready time to suppress page numbers.
% Comment it out when you first submit the paper for review.
%\thispagestyle{empty}

\subsection*{Abstract}
Machine learning (ML) techniques are increasingly common in security applications, such as malware and intrusion detection.
However, ML models are often
susceptible to \emph{evasion attacks}, 
in which an adversary makes changes to the input (such as
malware) in order to avoid being detected.
%cause erroneous predictions (for example, to
%avoid being detected).
A conventional approach to evaluate ML robustness to such attacks, as
well as to design robust ML, is by considering simplified
\emph{feature-space} models of attacks, where the attacker changes ML
features directly to effect evasion, while minimizing or constraining the magnitude of
this change.
We investigate the effectiveness of this approach to designing robust ML in the face of
attacks that can be realized in actual malware (\emph{realizable
  attacks}).
We demonstrate that in the context of structure-based PDF malware
detection, such techniques appear to have limited effectiveness,
but they are effective with content-based detectors.
In either case, we show that augmenting the feature space models with
\emph{conserved} features (those that cannot be unilaterally modified without compromising malicious functionality) significantly improves performance.
Finally, we show that feature space models enable generalized
robustness when faced with a variety of realizable attacks, as
compared to classifiers which are tuned to be robust to a specific
realizable attack.

\section{Introduction}
\label{introduction}

Machine learning (ML) has come to be widely used in a broad array of settings, including important security applications such as network intrusion, fraud, and malware detection, as well as other high-stakes settings, such as autonomous driving. 
A general approach is to extract a set of \emph{features}, or numerical attributes, of entities in question, collect a training data set of labeled examples (for example, indicating which instances are malicious and which are benign), and learn a model which labels previously unseen instances, presented in terms of their extracted features.
%In an example of PDF malware detection.
%In this case, features may correspond to object paths in a PDF, file size, and counts of \emph{Javascript} objects, while labels indicate whether a file is malicious or benign.
Success of ML in malware detection is particularly striking, with ML-based static detection of malicious entities at times exceeding 99\% accuracy~\cite{ndss2013,oakland2014}.

Nevertheless, ML-based techniques are often susceptible to \emph{adversarial examples}, an important special case of which are \emph{evasion attacks}.
In a prototypical case of an evasion attack, an adversary modifies malware code so that the resulting malware is categorized as benign by ML, but still successfully executes the malicious payload~\cite{Fogla06, asiaccs2013,Grosse17,oakland2014,ndss2016}. 
An even broader class of adversarial examples features attacks that manipulate an object, such as a stop sign, so that a computer vision pipeline misclassifies it as another object (such as a speed limit sign)~\cite{Evtimov18,iclr2015,CCS2016}.

In response, a host of methods emerged for making ML robust to adversarial examples, the most potent of which are those based on game-theoretic approaches, robust optimization (including certified robustness), and adversarial retraining~\cite{kdd2011, iclr2015, li2016,Madry18,Raghunathan18,Wong18,jmlr2009, kdd2012}.
A fundamental ingredient in all of these are \emph{feature-space models of attacks}.
Specifically, the attacker is assumed to directly modify values of features, with either a constraint or a penalty on the aggregate feature change measured in terms of an $l_p$ norm.
%As a typical example, Madry et al.~\cite{Madry18} aim to minimize worst-case error with respect to arbitrary feature perturbations with $l_\infty$ norm at most an exogenously specified bound $\epsilon$.

Such feature-space models of attacks are clearly abstractions of reality.
First, arbitrary modifications of feature values may not be \emph{realizable}.
For example, adding a benign object to a malicious PDF (with no other changes) necessarily increases its size, and so setting the associated feature to 1 (from 0) and simultaneously reducing file size may not be practically feasible.
Second, the key goal for an adversary is to create a target malicious effect, such as to execute a malicious payload.
Limiting feature modifications to be small in some $l_p$ norm clearly need not capture this: one can insert many no-ops (resulting in a large change according to an $l_p$ norm) with no impact on malicious functionality, and conversely, minimal changes (such as removing a Javascript tag) may break malicious functionality.
Nevertheless, an implicit assumption in robust ML approaches is that the feature-space models capture reality sufficiently to yield ML models that are robust even to realizable attacks.
\emph{The goal of our work is to evaluate the validity of this implicit assumption} in the context of PDF malware detection.

%Our first contribution is a general methodological framework for evaluating the validity of mathematical models of ML evasion attacks.
%At the core of the framework is a conceptual model of defense and attack based on a Stackelberg game~\cite{kdd2011}, where we assume to have an attack oracle that can be queried to obtain an attack for a given defense.
%The second ingredient is iterative adversarial retraining that can make use of an arbitrary learning algorithm and automated attack, enabling general applicability of the framework, and a fair comparison in validation (since the defensive approach is the same whether the attack is realizable or in feature space).
%The third feature of the framework is an evaluation measure which quantifies ML robustness by pitting ML against a \emph{realizable attack}, which must avoid being detected \emph{and} preserve malicious functionality as validated using a sandbox~\cite{Cova10, Guarnieri}.

Our first contribution is to evaluate feature-space evasion attack models in the context of PDF malware detection, using EvadeML as a realizable attack~\cite{ndss2016}.
Specifically, we consider four ML-based approaches for PDF malware detection: two based on features that capture PDF file structure (SL2013~\cite{ndss2013} and Hidost~\cite{srndic2016}), and two based on PDF file content (two Mimicus variants of PDFRate~\cite{acsac2012,oakland2014}).
%In all cases, we show that successful defense against a realizable attack is feasible (by retraining with this attack).
%In the case of structure-based detectors, we demonstrate that feature-space models \emph{do not} lead to adequate robustness against realizable attacks.
%In contrast, such models are, indeed, effective in the case of content-based detectors.
In all cases, we show that successful defense against a given realizable attack is feasible (by retraining with this attack). In the case of  structure-based  detectors,  we demonstrate  that adversarial retraining in the feature space does not lead to adequate robustness against realizable attacks. In contrast, adversarial retraining in the feature space is effective in the case of content-based detectors. 
In other words, the nature of the feature space can matter a great deal.

Our second contribution is a method for boosting robustness of feature-space models without compromising their mathematical convenience (crucial for most approaches for robust ML).
The key idea is to identify \emph{conserved features}, that is, features that cannot be unilaterally modified without compromising malicious functionality.
%that is, features the modification of which (essentially) implies compromising malicious functionality.
%While these need not always exist, 
We exhibit such features in our setting, show that they cannot be identified with traditional statistical methods, and develop an algorithm for automatically extracting them.
%Moreover, we show that conserved features cannot be identified using traditional statistical methods (such as sparse regularization).
%, and that using \emph{solely} such features yields robustness to the realizable EvadeML attack.
Finally, we show that by simply constraining that these features remain unmodified in adversarial training, feature-space approaches become effective even for robust structure-based PDF malware detection. 

Our third contribution is to explore the extent to which ML robustness is \emph{generalizable} to multiple \emph{distinct} realizable attacks.
Specifically, we expose both a robust classifier that was retrained by using a realizable attack (EvadeML), and a model hardened using a feature-space attack (accounting for conserved features), to a series of realizable attacks.
Our results reveal a stark difference between the two: ML models hardened using EvadeML are quite fragile; in contrast, ML models hardened using feature-space attacks exhibit uniformly high robustness to the other attacks.
%tends to greatly exceed robustness of a model hardened using EvadeML.
%We show that ML hardened using feature-space attack models are robust to these alternative attacks, whereas ML hardened against a specific realizable attack can fail spectacularly against a different attack.
Remarkably, we demonstrate that ML models hardened using feature-space attacks remain robust \emph{even against realizable attacks that defeat conserved features}.
\section{Machine Learning in Security}

\subsection{Learning and Prediction}

In the (supervised) machine learning literature, it is common to consider the problem abstractly.
We are given a training dataset $\mathsf{D} = \{(x_i,y_i)\}$, where $x_i \in \mathsf{X} \subseteq \R^n$ are numeric feature vectors in some feature space $\mathsf{X}$ and $y_i \in \mathsf{L}$ are labels in a label space $\mathsf{L}$.
Each data point (or example) in $\mathsf{D}$ is assumed to be generated i.i.d.\ according to some unknown distribution $\mathsf{P}$.
We are also given a hypothesis (model) space, $\mathsf{H}$, and our goal is to identify (\emph{learn}) a good model $h \in \mathsf{H}$ in the sense that it yields a small expected error on new examples drawn from $\mathsf{P}$.
In practice, since $\mathsf{P}$ is unknown, one typically aims to find $h \in \mathsf{H}$ which (approximately) minimizes empirical error on training data $\mathsf{D}$.

In security applications---as in others---one is not given numerical features; instead, we start with a collection of entities, such as executables, along with associated labels (we assume henceforth that these are available, as we focus here on supervised learning problems).
We must then \emph{design a collection of feature extractors}, where each feature extractor computes a numerical value of a corresponding feature from an input entity.
For example, we extract a ``size'' feature by computing the size of an executable.
Applying feature extractors to each entity in our dataset, and adding associated object labels, allow us to generate a dataset $\mathsf{D}$ to fit the conventional ML framework.

In this paper we focus on PDF malware detection, where the label space is binary: either a PDF file is benign (which we can code as $-1$), or malicious (which we can code as $+1$).
In addition, several prior efforts presented techniques for defining \emph{feature extractors} (commonly known simply as features) for PDF files~\cite{ndss2013,oakland2014}.
Applying such feature extractors to a PDF file dataset transforms this dataset into one comprised of numerical feature vectors and associated binary labels.
The goal is to predict whether previously unseen PDFs (simulated by holding out a portion of our dataset as \emph{test data}) are correctly labeled as malicious or benign.

\subsection{Evasion Attacks}

In an \emph{evasion attack}, abstractly, one is given a learned model $h(x)$ (e.g., a SVM or neural network) which returns a label $y = h(x)$ (e.g., malicious or benign) for an arbitrary feature vector $x \in \mathsf{X}$ (e.g., extracted from a PDF file).
The attacker additionally starts with an entity $e$ (such as a malicious PDF file), from which we can extract a feature vector $\phi(e)$.
The attacker then transforms $e$ into another entity, $e'$, with an associated feature vector $x' = \phi(e')$ so as to accomplish two goals: first, that $h(x')$ returns an erroneous label (in our running example, labels $e'$ as benign based on its extracted features $\phi(e')$), and second, that $e'$ preserves the functionality of the original entity $e$---which, in our example of PDF malware detection, entails preserving malicious functionality of $e$.
The evasion attack as just described is presumed to transform the \emph{entity itself}, such as the malicious PDF file, albeit accounting for the effect of such transformation on the extracted features $x' =\phi(e')$.
We call attacks of this kind \emph{realizable} evasion attacks.
%In fact, 
The process by which such realizable evasion attacks can be successfully accomplished is quite non-trivial, and typically warrants independent research contributions (e.g.,~\cite{oakland2014,ndss2016}).

In contrast, it is natural to short-circuit the complexity involved, and work directly in the \emph{feature space}, as is conventional in the machine learning literature.
In this case, the attacker is \emph{modeled} as starting with a malicious feature vector $x$ (\emph{not the malicious entity} $e$), and \emph{directly modifying the features} to produce another feature vector $x' \in \mathsf{X}$, so as to yield erroneous predictions, i.e., $y' = h(x')$ (for example, being mislabeled as benign).
Crucially, since we are no longer appealing to original entities, we must abstract away the notion of preserving (malicious) functionality.
This is done through the use of a cost function, $c(x,x')$, whereby the attacker is penalized for greater modifications to the given feature vector $x$, commonly measured using an $l_p$ norm difference between the original malicious instance and the modified feature vector~\cite{pkdd2013,li2016}.
We term these the \emph{feature-space models} of evasion attacks.
Crucially, \emph{essentially all approaches for robust ML, particularly the most successful ones, such as those based on robust optimization, leverage these models}.

\subsection{Evasion Defense}

A large number of approaches have been proposed for defending against evasion attacks or, more broadly, adversarial examples~(e.g., \cite{pkdd2013, Bruckner12,kdd2011, jmlr2009, Papernot16, Papernot18, Raghunathan18,Vorobeychik18book,Wong18}).
While many have been shown inadequate~\cite{Athalye18,Carlini17}, the three generally effective approaches are: (a) game-theoretic reasoning, (b) robust optimization (a special case of (a) where the game is zero-sum), and (c) iterative adversarial retraining.\footnote{Otherwise known as adversarial training, it can be viewed as an approach for obtaining approximate game-theoretic or robust optimization solutions~\cite{Vorobeychik18book,li2016,Madry18}.}
Game-theoretic methods in general, and robust optimization in particular, are not general-purpose, as solving these directly requires special structure, such as a continuous feature space and differentiability~\cite{pkdd2013, Bruckner12,kdd2011}, and often additional structure of the learning model, such as linearity~\cite{jmlr2009} or neural network architecture and activation functions~\cite{Raghunathan18,Wong18}.
Finally, to date all have used the mathematical feature-space attack model at their core.
In contrast, retraining can be performed without making assumptions about the nature of the learning algorithm or the adversarial model~\cite{li2016}.
Since our study below involves realizable attacks (in addition to the mathematical models of attacks), non-linear SVM and, in all cases but one, binary features, iterative retraining is the sole defense that can be applied uniformly (which we require to ensure that our results are directly comparable).
\section{Validating Models of ML Evasion Attacks}
\label{S:framework}

%Our main goal is to evaluate whether robust ML approaches that make use of feature-space models of evasion attacks are, indeed, robust against \emph{real}---realizable---attacks.
We have two major goals: 1) \emph{validation}: to evaluate whether robust ML approaches that make use of feature-space models of evasion attacks are, indeed, robust against \emph{real}---realizable---attacks, and 2) \emph{generalizability}: to study generalizability of evasion defenses.
%To this end, 

%\begin{figure}[h]
%\centering
%\includegraphics[width=0.5\linewidth]{figures/framework}
%\caption{A conceptual model of how an attack (either realizable or using a feature-space model) can be used in improving ML security.  Let defense be parametrized by $\theta$, and an attack \emph{reacting} to the particular defense $\theta$ (e.g., attacker evades the learned ML model $h(x)$).  We wish to choose the best defense $\theta$ against such a reactive attacker, as captured by our attack model.}
%\label{F:framework}
%\end{figure}

We start with a conceptual model of defense and attack
% illustrated in %Figure~\ref{F:framework}.
%We can view this conceptual model 
as a Stackelberg game between ML (``defender''), who first chooses a defense $\theta$ (in our case, the learned model $h(x)$) and the attacker, who finds an optimal attack that \emph{reacts} to the particular defense $\theta$.
An \emph{attack model} captures how the attacker changes behavior in response to the defense $\theta$.
The defender's goal is to choose the best defense $\theta$ against such a reactive attacker, as captured by the attack model.
Indeed, this is a common way to model the adversarial evasion problem in prior literature~\cite{kdd2011,nips2014,Vorobeychik18book}.
This model has two useful features.
First, the attack is treated as an oracle in the sense that it returns an attack for an arbitrary defense $\theta$.
This allows us, in principle, to design a defense against an arbitrary evasion attack, making no distinction between feature-space attack models and realizable attacks.
Second, we can separately consider \emph{defense} against a specific attack (for example, a feature-space attack), and evaluation, which can use another attack (e.g., a realizable attack).

To be more precise, let $O(h;\mathsf{D})$ be an arbitrary attack which returns evasions given a dataset $\mathsf{D}$ and a classifier $h$, and let $u(h;O(h;\mathsf{D}))$ be the measure that the defender wishes to optimize (for example, accuracy on data \emph{after} evasions).
Then defense against the attack $O(h;\mathsf{D})$ amounts to solving the following optimization problem:
\begin{equation}
\label{E:robustML}
\max_{h} u(h;O(h;\mathsf{D})).
\end{equation}

%The crucial feature of this model is that the attack is treated as an oracle in the sense that it returns an attack for an arbitrary defense $\theta$.
%This allows us, in principle, to design a defense against an arbitrary evasion attack, making no distinction between feature-space attack models and realizable attacks.
%This oracle is clearly not universally available for realizable attacks; however, we demonstrate below how to instantiate it in the context of ML model validation.

In practice, we need a means for approximately solving the optimization problem in Equation~\eqref{E:robustML} for an arbitrary attack.
To this end, we make use of \emph{iterative retraining}, an approach previously proposed for hardening classifiers against evasion attacks~\cite{li2016,icml2016}.
%\begin{figure*}[t]
%	\centering
%	\includegraphics[width=0.8\textwidth]{figures/retraining2.pdf}
%	\caption{A general framework for retraining a classifier}
%	\label{retraining}
%\end{figure*}
In particular, we use a variant of iterative retraining with provable guarantees~\cite{li2016}, which is outlined as follows:
\begin{enumerate}
\item Start with the initial classifier.
\item Execute the \emph{evasion attack} for each malicious instance in
  training data to generate a new feature vector.
\item Add all new data points to training data (removing any
  duplicates), and retrain the classifier.
\item Terminate after either a fixed number of iterations, or when no
  new evasions can be added.
\end{enumerate}

Now, we describe our approach to validation and generalizability evaluations.

In \emph{validation}, consider a model of an evasion attack, $\tilde{O}(h;\mathsf{D})$ (e.g., a feature-space attack model), which is a proxy for a ``real'' (realizable) attack, $O(h;\mathsf{D})$; note that each attack evades a given ML model $h$.
We first find the defense against $\tilde{O}$ using the retraining procedure above; let the resulting robust classifier be $\tilde{h}$.
Next, we \emph{evaluate} $\tilde{h}$ by running the target realizable attack $O(\tilde{h};\mathsf{D})$.
Finally, we create a \emph{baseline} $h^*$, which is a robust classifier against a target realizable attack $O$. 
We then evaluate how well $\tilde{h}$ performs, compared to $h^*$, against the target attack.
For example, if we find that $\tilde{h}$ is ineffective against the target attack, we say that $\tilde{O}$ is a poor attack proxy, whereas if it remains robust, we view $\tilde{O}$ as a good proxy for the target attack $O$.
We focus on validation in Sections~\ref{S:validation} and~\ref{CR}.

In evaluating \emph{generalizability}, the approach is slightly different.
Again, we consider a proxy attack $\tilde{O}$ (which may now be either a feature-space model, or some particular realizable attack), and find a defense $\tilde{h}$ against this attack.
For evaluation, we consider a \emph{collection} of target attacks $\{O_i)\}$, and run each of these attacks against $\tilde{h}$.
We say that our proxy attack is generalizable if $\tilde{h}$ remains robust to all, or most of the attacks $i$; otherwise, it fails to generalize.
We consider generalizability in Section~\ref{sec:alternative}.

%Putting everything together, we propose the following framework for validating the effectiveness of ML evasion models.
%Choose the ML algorithm that we wish to make robust.
%Next, consider a model of an evasion attack, $\tilde{O}(h)$ (e.g., a feature-space attack model), which is a proxy for a ``real'' (realizable) attack, $O(h)$; note that each attack evades a given ML model $h$.
%Let $u(h;O)$ be a measure of robustness of an ML model $h$ \emph{against the realizable attack}.
%Now,
%\begin{enumerate}
%\item Perform iterative retraining, using the model, $\tilde{O}(h)$; let $\tilde{h}$ be the resulting ``hardened'' ML model;
%\item Perform iterative retraining, using the realizable attack, $O(h)$; let $h^*$ be the ML model hardened against this attack;
%\item The effectiveness of $\tilde{O}(h)$ \emph{relative to} $O(h)$ (for which it is a proxy) is $\max\{u(h^*;O) - u(\tilde{h};O),0\}$.
%\end{enumerate}
%Note that we generally expect that ML hardened using the realizable attack will be more robust against this attack than ML hardened using some other proxy (e.g., feature-space) attack.
%However, this is not always the case, and we do not require it; we simply use $u(h^*;O)$ as a fair baseline, and claim the model to be effective as long as it's nearly as good as this baseline, and certainly when it's better.

%In the sequel, we use our framework to evaluate robustness of conventional feature-space approaches for hardening ML when they are confronted with a realizable attack.
\section{Experimental Methodology}
\label{S:exp}

%Our main goal is to evaluate the efficacy of PDF malware classifier evasion
%models.
%In particular, we aim to compare the elegant and commonly used
%feature-space models, which allow an attacker to modify features
%directly, with attacks that actually modify PDF files and are validated to have preserved %malicious functionality.

We use malicious PDF detection as a case study to investigate robustness of ML hardened using feature-space models of evasion attacks.
We now describe our experimental methodology.
We start with some background on PDF structure, and proceed to describe the specific ML-based detectors, evasion attacks (both realizable, and feature-space), datasets, and evaluation metrics used in our experiments.

\subsection{PDF Document Structure}

%\begin{figure*}[t]
%\centering
%\includegraphics[width=0.6\textwidth]{figures/pdf-structure2.pdf}
%\caption{Various representations of the PDF structure: file structure (left), physical layout (middle), and logical structure (right)}
%\label{pdf-structure}
%\end{figure*}

The Portable Document Format (PDF) is an open standard format used to present content and layout on different platforms. A PDF file structure consists of four parts: \emph{header}, \emph{body}, \emph{cross-reference table} (CRT), and \emph{trailer}.
The header contains information such as the magic number and format version.  The body is the most important element of a PDF file, which comprises multiple PDF objects that constitute the content of the file. 
These objects can be one of the eight basic types: Boolean, Numeric, String, Null, Name, Array, Dictionary, and Stream. They can be referenced from other objects via indirect references. There are other types of objects, such as JavaScript which contains executable JavaScript code. The CRT indexes objects in the body, while the trailer points to the CRT. 

%The syntax of the body of a PDF file is shown in the middle of Figure \ref{pdf-structure}. In this example, the PDF body contains three indirect objects referenced by others. 
%The first is a \verb|Catalog| object, which contains two additional entries: \verb|OpenAction| and \verb|Pages|. 
%These entries are dictionaries. 
%The \verb|OpenAction| entry has two internal entries: \verb|S| and \verb|JS|, which are JavaScript codes to be executed.
%The \verb|Pages| entry refers to the second object with the type \verb|Pages|. 
%The second object contains an entry "Kids" which refers to the third object, which is a \verb|Page| object and refers back to the second object. 

The relations between objects with cross-references can be described as a directed graph that presents their logical structure by using edges representing reference relations and nodes representing different objects.% as shown (right-hand-side of Figure \ref{pdf-structure}).  
As an object can be referred to by its child node, the resulting logical structure is a directed cyclic graph. 
%For example in Figure \ref{pdf-structure}, the second and third objects refer to each other and constitute a directed cycle. 
To eliminate the redundant references, the logical structure can be reduced to a structural tree with the breadth-first search procedure. 

\subsection{Target Classifiers}

\begin{table}[]
\centering
\scalebox{0.8}{
\begin{tabular}{|c|c|c|}
\hline
\textbf{Classifier} & \textbf{Feature type} & \textbf{Number of features} \\ \hline\hline
SL2013     & Binary       & 6,087              \\ \hline
Hidost     & Binary       & 961              \\ \hline
PDFRate-R  & Real-valued  & 135                \\ \hline
PDFRate-B  & Binary  & 135                \\ \hline
\end{tabular}
}
\caption{Target classifiers.}
\label{tab:target_classifier}
\end{table}

Several PDF malware classifiers have been proposed \cite{Cova10,acsac2012, ndss2013, srndic2016}. 
For our study, we selected SL2013~\cite{ndss2013}, Hidost~\cite{srndic2016} and two variants of PDFRate~\cite{acsac2012} (termed PDFRate-R and PDFRate-B respectively), displayed in Table~\ref{tab:target_classifier}. 
SL2013 and its revised version, Hidost, are \emph{structure-based} PDF classifiers, which use the logical structure of a PDF document to construct and extract features used in detecting malicious PDFs.
PDFRate, on the other hand, is a \emph{content-based} classifier, which constructs features based on \emph{medadata} and \emph{content} information in the PDF file to distinguish benign and malicious instances.
Evasion attacks on both SL2013 and PDFRate classifiers, particularly of the realizable kind, have been developed in recent literature \cite{ndss2013, oakland2014, ndss2016, srndic2016}, providing a natural evaluation framework for our purposes.
 
\subsubsection{Structure-Based Classifiers} 

\noindent{\bf SL2013:} SL2013 is a well-documented and open-source machine learning system using Support Vector Machines (SVM) with a radial basis function (RBF) kernel, and was shown to have state-of-the-art performance~\cite{ndss2013}.
It employs structural properties of PDF files to discriminate between malicious and benign PDFs. 
Specifically, SL2013 uses the presence of particular \emph{structural paths} as binary features to present PDF files in feature space.  A structural path of an object is a sequence of edges in the reduced (tree) \emph{logical structure}, starting from the catalog dictionary and ending at this object. Therefore, the structural path reveals the shortest reference path to an object. 
SL2013 uses 6,087 most common structural paths among 658,763 PDF files as a uniform set for classification.

%SL2013 uses a uniform set of structural paths to classify a PDF file. To get these paths, it first obtains a total of 658,763 benign and malicious PDF files (around 595 GB), then selects only the structural paths that occur in at least 1,000 PDF files. 
%This process reduced the number of features from over 9 million to 6,087. Trained using 5,000 malicious and 5,000 benign PDF files, SL2013 was shown to have $99.8\%$ accuracy on test data and AUC $>99.9\%$~\cite{ndss2013}.
%\vspace{-0.2in}
\noindent{\bf Hidost:} Hidost is an updated version of SL2013. 
It inherits all the characteristics of SL2013 and employs \emph{structual path consolidation} (SPC), a technique to consolidate features which have the same or similar semantic meaning in a PDF.
As the semantically equivalent structural paths are merged, Hidost reduces polymorphic paths and still preserves the semantics of logical structure, so as to improve evasion-robustness of SL2013~\cite{srndic2016}.

%Hidost used a dataset of 407,037 benign and 32,567 malicious PDF files collected over 14 weeks for training, and obtained $99.5$\% accuracy on test data, with AUC $>99.5$\%.
In our work, we employ the 961 features identified in the latest version of Hidost. 

\subsubsection{PDFRate: A Content-Based Classifier}

%\paragraph{PDFRate}
The original PDFRate classifier uses a random forest algorithm, and
employs PDF \emph{metadata} and \emph{content} features.
The metadata features include the size of a file, author name, and creation date, while content-based features include position and counts of specific keywords.
All features were manually defined by Smutz and Stavrou \cite{acsac2012}.

PDFRate uses a total of 202 features, but only 135 of these are publicly documented~\cite{GMUreport}. 
Consequently, in our work we employ the Mimicus implementation of PDFRate which was shown to be a close approximation~\cite{oakland2014}. 
Mimicus trained a surrogate SVM classifier with the documented 135 features and the same dataset as PDFRate, using both the SVM and random forest classifiers, both performing comparably.
We use the SVM implementation in our experiments to enable more direct comparisons with the structure-based classifiers that also use SVM.
An important aspect of Mimicus is \emph{feature standardization} on extracted data points performed by subtracting the mean of the feature value and dividing by standard deviation, transforming all features to be real-valued and zero-mean (henceforth, PDFRate-R).
This surrogate was shown to have $\sim99$\% accuracy on the test data~\cite{acsac2012}.
%In addition, we construct a \emph{binarized} variant of PDFRate (henceforth, PDFRate-B), where each feature is transformed into a binary feature (where needed) by using the presence of corresponding raw features.
In addition, we construct a \emph{binarized} variant of PDFRate (henceforth, PDFRate-B), where each feature is transformed into a binary feature by assigning 0 whenever the feature value is 0, and assigning 1 whenever the feature value is non-zero. 

\subsection{Realizable Evasion Attacks}

\subsubsection{EvadeML}

%To evaluate the robustness of a PDF classifier against adversarial evasion attacks, we adopt EvadeML \cite{ndss2016}, an automated method to craft evasion instances of PDF malware in problem space. 
%There is good reason to believe \emph{a priori} that this is the most potent evasion attack against ML-based PDF malware detectors available: nearly all other attacks are different forms of \emph{mimicry}, attempting to add features/objects from benign files.
%EvadeML, in contrast, considers both addition and deletion of PDF objects in constructing a successful evasion.

The primary realizable attack in our study is EvadeML \cite{ndss2016}, which allows insertion, deletion, and swapping of objects, and is consequently a stronger attack than most other realizable attacks in the literature, which typically only allow insertion to ensure that malicious functionality is preserved.
EvadeML assumes that the adversary has black-box access to the classifier and can only get classification scores of PDF files, and was shown to effectively evade both SL2013 and PDFRate~\cite{ndss2016}.
It employs genetic programming (GP) to search the space of possible PDF instances to find ones that evade the classifier while maintaining malicious features. 
First, an initial population is produced by randomly manipulating a malicious PDF repeatedly. 
The manipulation is either a deletion, an insertion, or a swap operation on PDF objects. 
%A deletion operation deletes a target object from the seed malicious PDF file. 
%An insertion operation inserts an object from external benign PDF files (provided exogenously) after the target object.
%A swap operation replaces the entry of the target object with that of another object in the external benign PDFs. 
After the population is initialized, each variant is assessed by the Cuckoo sandbox~\cite{Guarnieri} and the target classifier to evaluate its fitness. 
The sandbox is used to determine if a variant preserves malicious behavior, such as API or network anomalies. 
The target classifier provides a classification score for each variant.
%If the score is above a threshold, then the variant is classified as malicious. 
%Otherwise, it is classified as benign.  
If a variant is classified as benign but displays malicious behavior, or if GP reaches the maximum number of generations, then GP terminates with the variant achieving the best fitness score and the corresponding mutation trace is stored in a pool for future population initialization. 
Otherwise, a subset of the population is selected for the next generation based on their fitness evaluation. 
Afterward, the variants selected are randomly manipulated to generate the next generation of the population.

%EvadeML was used to evade SL2013 in \cite{ndss2016}. The reported results show that it can automatically find evasive variants for all 500 selected malicious test seeds.

We use EvadeML as the primary realizable evasion model for the first part of the paper.
We set the GP parameters in EvadeML as the same as in the experiments by Xu et al.~\cite{ndss2016}. 
The population size in each generation is 48. 
The maximum number of generations is 20. The mutation rate for each PDF object is 0.1. 
The mutation traces that lead to successful evasion and promising variants are stored and applied in our experiments.
The fitness threshold of a classifier is 0. 
We use the same external benign PDF files as Xu et al.~\cite{ndss2016} to provide ingredients for insertion and swap operations.  

%We also evaluate the effectiveness of robustML against two other realizable evasion attacks, Mimicry and MalGAN.
%We describe these next.

\subsubsection{The Mimicry Attack}
Mimicry assumes that an attacker has full knowledge of the features employed by a target classifier. 
The mimicry attack then manipulates a malicious PDF file so that it mimics a particular selected benign PDF as much as possible.
The implementation of Mimicry is simple and independent of any particular classification model. 

Our mimicry attack uses the Mimicus~\cite{oakland2014} implementation, which was shown to successfully evade the PDFRate classifier. 
To improve its evasion effectiveness, Mimicus chooses 30 different target benign PDF files for each attack file.
It then produces one instance in feature space for each target-attack pair by merging the malicious features with the benign ones.
The feature space instance is then transformed into a PDF file using a \emph{content injection approach}. 
The resulting  30 files are evaluated by the target classifier, and only the PDF with the best evasion result is selected, which was submitted to WEPAWET~\cite{Cova10} to verify malicious functionality.
To make Mimicry consistent with our framework, we employ the Cuckoo sandbox~\cite{Guarnieri} in place of WEPAWET (which was in any case discontinued) to validate maliciousness of the resulting PDF file.

In addition to the original version of Mimicry, we implement an enhanced variation, \emph{Mimicry+}, with two modifications.
First, Mimicry+ chooses the 30 most benign PDF files predicted by the target classifier as target files (instead of randomly selecting those, as in Mimicry).
Second, for each attack file, all the resulting 30 files are evaluated by the sandbox and only those verified to have malicious functionality are selected to evade the target classifier.

\subsubsection{MalGAN}

MalGAN~\cite{Hu2017} is a Generative Adversarial Network~\cite{Goodfellow2014} framework to generate malware examples which can evade a black-box malware detector with binary features.  
It assumes that an attacker knows the features, 
%has full knowledge of the feature set of the malware detector, 
but has only black-box access to the detector decisions.
%, and it can repeatedly query the classification results of submitted PDF files. 
MalGAN comprises three main components: a generator which transforms malware to its adversarial version, a black-box detector which returns detection results, and a substitute detector which 
%has no knowledge of the black-box detector but 
is used to fit the black-box detector and train the generator.
The generator and substitute detector are feed-forward neural networks which work together to evade the black-box detector.
The results of \cite{Hu2017} show that MalGAN is able to decrease the \emph{True Positive Rate} on the generated examples from $>90$\% to 0\%.
We note that strictly speaking, MalGAN variants are not implemented as actual PDF files; however, we still treat it as a realizable attack since it only adds features to a malicious file, which can be implemented (at least in structure-based detection) by adding the associated objects into the PDF file.

\subsubsection{Reverse Mimicry}

The \emph{Reverse Mimicry} attack assumes that an attacker has zero knowledge of the target classifier.
The basic idea is to inject malicious payloads into target benign files to minimize the structural difference between the resulting examples and targets.  
Our Reverse Mimicry attack employs the adversarial examples provided by Maiorca et al.~\cite{asiaccs2013} which was shown to successfully evade PDF classifiers based on structural analysis.
Specifically, we use the 500 PDF files produced by injecting a malicious JavaScript code that does not contain references to other objects to target benign PDF files.
We selected the 376 files out of 500 that display malicious behaviors detected by the Cuckoo sandbox.

\subsubsection{The Custom Attack}

We implemented a custom attack which exploits a feature extraction vulnerability in the Mimicus implementation of PDFRate.
Normally, the characters used in the Name objects of a PDF file are limited to a specific set.
Since PDF specification version 1.2, a lexical convention has been added to represent a character with its hexadecimal ANSI-code, e.g., \#xx.
Such a modification enables us to create an arbitrary string in the form of  \#xx\#xx\#xx.
In our implementation, we replaced a set of entries in the attack PDF files with their hexadecimal representations (see Table~\ref{table:custom}). These features were selected with the goal to obfuscate tags crucial to the code execution in PDF, which are frequently used for feature extraction.  With this technique, the scanner would not be able to detect malicious code without dynamically reconstructing the PDF structure. While it is theoretically possible to replace all the ASCII text inside the document, we chose not to do that due to the concern on the expansion of file size.
%We summarize the transformations of these entries in Table~\ref{table:custom}.

\begin{table}[]
\centering
\scalebox{0.8}{

\begin{tabular}{|c|c|}

\hline
\textbf{Entry}       & \textbf{Hexadecimal Representation}                \\ \hline\hline
/Action     & /\#41\#63\#74\#69\#6f\#6e                 \\ \hline
/Filter     & /\#46\#69\#6c\#74\#65\#72                 \\ \hline
/Length     & /\#4c\#65\#6e\#67\#74\#68                 \\ \hline
/JavaScript & /\#4a\#61\#76\#61\#53\#63\#72\#69\#70\#74 \\ \hline
/JS         & /\#4a\#53                                 \\ \hline
/S          & /\#53                                     \\ \hline
/Type       & /\#54\#79\#70\#65                         \\ \hline
\end{tabular}
}
\caption{Transformation of entry names in the custom attack.}
\label{table:custom}
\end{table}  

\subsection{Feature-Space Evasion Model}

In typical realizable attacks, including EvadeML, a consideration is not merely to move to the benign side of the classifier decision boundary, but to appear as benign as possible.
This naturally translates into the following multi-objective optimization in feature space:
\begin{equation}
\begin{aligned}
& \underset{x}{\text{minimize}}
& & Q(x) = f(x) + \lambda c(x_M,x),\\
\end{aligned}
\label{evasion-model}
\vspace{-0.05in}
\end{equation}
where $f(x)$ is the score of a feature vector $x$, with the actual classifier (such as SVM) $g(x) = \mathrm{sgn}(f(x))$, $x_M$ the malicious seed, $x$ an evasion instance, $c(x_M,x)$ the cost of transforming $x_M$ into $x$, and $\lambda$ a parameter which determines the feature transformation cost.
We use $l_2$ norm distance between $x_M$ and $x$ as the cost function: $c(x_M,x) = \sum_i |x_i - x_{M,i}|^2$.
Since in most of our experiments features are binary, the choice of $l_2$ norm (as opposed to another $l_p$ norm) is not critical.
%Below we consider two variations of this cost function: first, using uniform weights, with $\alpha_i = 1$ for all features $i$, and second, using non-uniform weights (which we term \emph{weighted distance (WD)} below).

As the optimization problem in Equation~\eqref{evasion-model} is non-convex and variables are binary in three of the four cases we consider, we use a stochastic local search method designed for combinatorial search domains, \textsl{Coordinate Greedy} (alternatively known as iterative improvement), to compute a local optimum (the binary nature of the features is why we eschew gradient-based approaches)~\cite{Hoos04,li2016}.
In this method, we optimize one randomly chosen coordinate of the feature vector at a time, until a local optimum is reached.
To improve the quality of the resulting solution, we repeat this process from several random starting points.
This approach has been shown to be extremely effective for computing evasion instances in binary domains~\cite{li2016}.

\subsection{Datasets}

The dataset we use is from the \emph{Contagio Archive}.\footnote{Available at the following URL: \url{http://contagiodump.blogspot.com/2013/03/16800-clean-and-11960-malicious-files.html}.}
We use 5,586 malicious and 4,476 benign PDF files for training, and another 5,276 malicious and 4,459 benign files as the non-adversarial test dataset. The training and test datasets also contain 500 seeds selected by Xu et al.~\cite{ndss2016}, with 400 in the training data and 100 in the test dataset. 
These seeds are filtered from 10,980 PDF malware samples and are suitable for evaluation since they are detected with reliable malware signatures by the Cuckoo sandbox~\cite{Guarnieri}. We randomly select 40 seeds from the training data as the retraining seeds and use the 100 seeds in the test data as the test seeds. 

\subsection{Implementation of Iterative Adversarial Retraining}

We made a small modification to the general iterative retraining approach described in Section~\ref{S:framework} when it uses EvadeML as the realizable attack $O(h;\mathsf{D})$.
Specifically, we used only 40 malicious seeds to EvadeML to generate evasions, %to remain consistent with the prior use of EvadeML, 
to reduce running time and make the experiment more consistent with realistic settings where a large proportion of malicious data is not adapting to the classifier.
As shown below, this set of 40 instances was sufficient to generate a model robust to evasions from held out 100 malicious seed PDFs.

We distribute both retraining and adversarial test tasks on two servers (Intel(R) Xeon(R) CPU E5-2695 v4 @ 2.10GHz, 18 cores and 64 GB memory, running Ubuntu 16.04). For retraining using EvadeML as the attack, we assign each server 20 seeds; each seed is processed by EvadeML to produce the adversarial evasion instances. We then add the 40 examples obtained to the training data, retrain the classifier, and then split the seeds between the two servers in the next iteration. In the evaluation phase, we assign each server 50 seeds from the 100 test instances, and each seed is further used to evade the classifier by using EvadeML.

\subsection{Evaluation Metrics}

We evaluate performance in two ways: 1) evaluation of evasion robustness (which is central to our specific inquiry), and 2) traditional evaluation.
To evaluate robustness, we compute the proportion of 100 malicious test seed PDFs for which EvadeML successfully evades the classifier; this is our metric of \emph{evasion robustness}, evaluated with respect to EvadeML.
Thus, evasion robustness of 0\% means that the classifier is successfully evaded in every instance, while evasion robustness of 100\% means that evasion fails every time.
Our traditional evaluation metric uses test data of malicious and benign PDFs, where no evasions are attempted.
On this data, we compute the ROC (receiver operating characteristic) curve and the corresponding AUC (area under the curve).
% \section{Case Study with a Structure-Based PDF Malware Classifier}
\section{Efficacy of Feature-Space Attack Models}
\label{S:validation}
% \label{hidost}

We now undertake our first task: evaluation of the effectiveness of robust ML obtained by using the abstract feature-space models of attack.
%To this end, we obtain a baseline for a uniform comparison
We compare to a baseline classifier obtained by retraining with the most %powerful 
potent attack on our menu, EvadeML (which, in addition to inserting content, as done by other attacks~\cite{oakland2014, Hu2017, asiaccs2013}, also allows the attacker to delete and swap PDF objects).
We can think of our baseline as assuming that the defender knows that EvadeML is employed by the attacker, along with its hyperparameters.
Throught this and next section, we also use EvadeML to evaluate the effectiveness of classifiers hardened using a feature-space model, in comparison with the above baseline.
%As we will see, the success of this baseline illustrates that it is \emph{possible} to effectively harden the classifiers against EvadeML in our setting, and observation of poor performance with the feature-space attack model is not due to a fundamental difficulty of the problem.

\subsection{Structure-Based PDF Malware Classification}

Our first case study uses a state-of-the-art PDF malware classifier which engineers features based on PDF \emph{structure}.
Indeed, we evaluate two versions of this classifier: an earlier version, which we call \emph{SL2013}, and a more recent version,
%reengineered specifically to be more robust to a class of evasion attacks (specifically, \emph{mimicry} attacks),
which we call \emph{Hidost}.
The experiments by Xu et al.~\cite{ndss2016} demonstrate that SL2013 can be successfully evaded.
%that although SL2013 was designed to be resistant to evasion attacks, it can be successfully evaded.
Since Hidost was a recent redesign attempting in part to address its vulnerability to mimicry attacks by significantly reducing the feature space, no data exists on its vulnerability to evasion attacks.
Below we demonstrate that Hidost
%despite a deliberate effort to harden Hidost,
is also vulnerable to evasion attacks (indeed, more so than SL2013).
%Significantly, we demonstrate that retraining with a problem space evasion attack (PSR henceforth) based on EvadeML significantly hardens both classifiers against evasion.
%, \emph{not only against EvadeML itself, but also against two other evasion models}.
%On the other hand, we show that similar hardening through retraining by using a \emph{feature space} evasion model (FSR henceforth) \emph{fails to adequately improve classifier robustness}.

From the perspective of defense, we show that it is possible to harden both SL2013 and Hidost against a powerful realizable EvadeML attack by simply retraining with this attack (\emph{RAR}, for \emph{realizable-attack retraining}, henceforth refers to a model hardened using EvadeML).
This serves as a baseline we use to evaluate the efficacy of a retraining defense with a feature-space attack model (henceforth, FSR for \emph{feature-space retraining}).
We then show that for both SL2013 and Hidost, FSR significantly underperforms RAR.

In our experiments, we empirically set the \textsl{RBF} parameters for training both SL2013 and Hidost to $C=12$ and $\gamma = 0.0025$.

%with nearly $0\%$ accuracy on selected malicious instances in problem space. 

%\subsection{Experiments}

%\subsubsection{Experiment setup}

%\subsection{Experiments}

\subsubsection{SL2013}

\paragraph{Retraining with a Powerful Realizable Attack}

\begin{figure}[t]
\centering
  \begin{tabular}{cc}
    \includegraphics[width=0.22\textwidth]{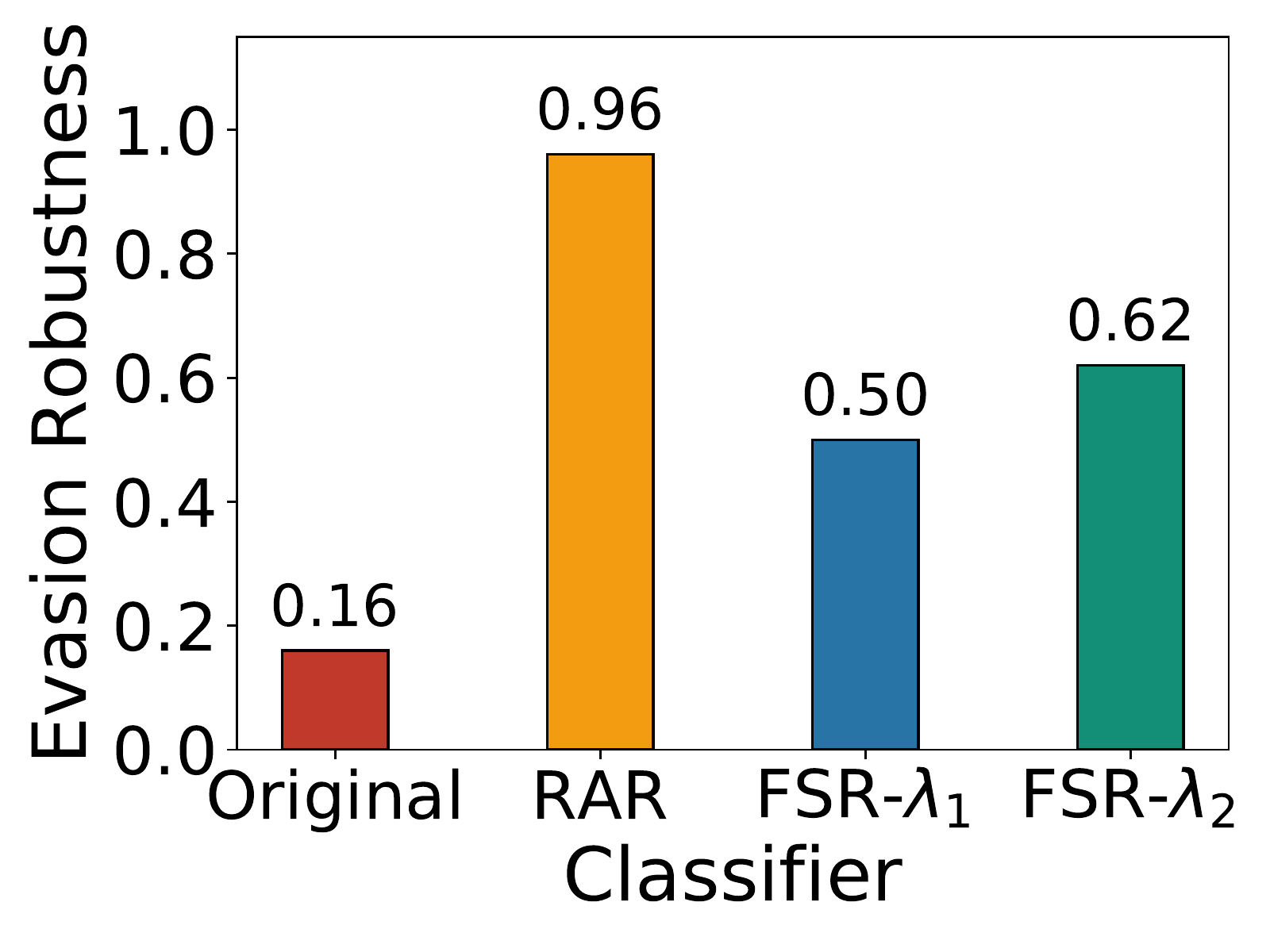} &
     \includegraphics[width=0.22\textwidth]{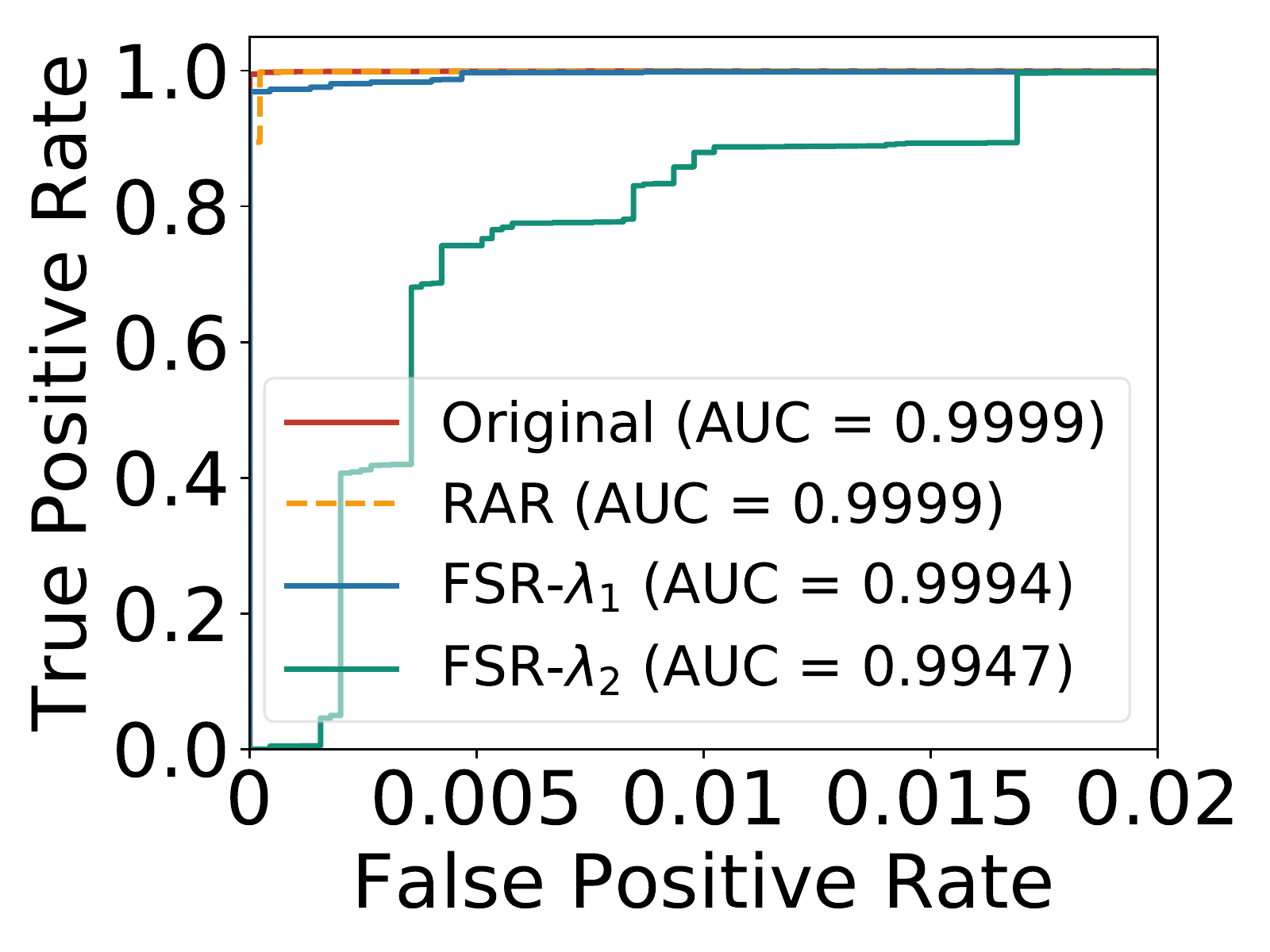}

\end{tabular}
	\caption{Evasion robustness under EvadeML test (left) and performance on non-adversarial data (right) of different classifiers for SL2013.}
	\label{SL2013-sys}
      \end{figure}

First, we replicated the EvadeML attack on the original SL2013; the classifier achieves only a 16\% evasion robustness.\footnote{This result differs from the experiments in \cite{ndss2016} which show a 0\% evasion robustness. 
We found a flaw in the implementation of feature extraction in EvadeML which causes evaluation to be performed using the wrong feature vectors. 
This bug has been fixed in the GitHub version of EvadeML.}
%\footnote{This result is different from the experiments in \cite{ndss2016} which show a 0\% evasion robustness, because found a flaw in the implementation of feature extraction in EvadeML which has been reported to the authors and fixed in our experiments.}
Next, to create a baseline, we conduct experiments in which EvadeML is employed to retrain SL2013.
%Before the target classifier SL2013 was retrained, its robustness was first evaluated by EvadeML. For all the 100 adversarial examples produced by EvadeML, SL2013 could only achieve a 16\% evasion robustness.
%This result provides a baseline with which we compare the robustness after retraining SL2013. 
%We conducted an EvadeML test to evaluate the robustness of SL2013 after each iteration of retraining.
%This experiment of retraining SL2013 with EvadeML took approximately six days to execute.
The process terminated after 10 iterations at which point no evasive variants of the 40 retraining seeds could be generated.
We observe (Figure \ref{SL2013-sys} (left)) that the retrained classifier (RAR) obtained by this approach achieves a 96\% evasion robustness.
Moreover, RAR is essentially as accurate as the baseline SL2013 on non-adversarial data (Figure \ref{SL2013-sys} (right)).
Thus, it is clearly possible to be highly robust to this evasion attack without significantly compromising effectiveness on data not featuring explicit evasion attacks.

\begin{figure}[t]
\centering
  \begin{tabular}{cc}
    \includegraphics[width=0.22\textwidth]{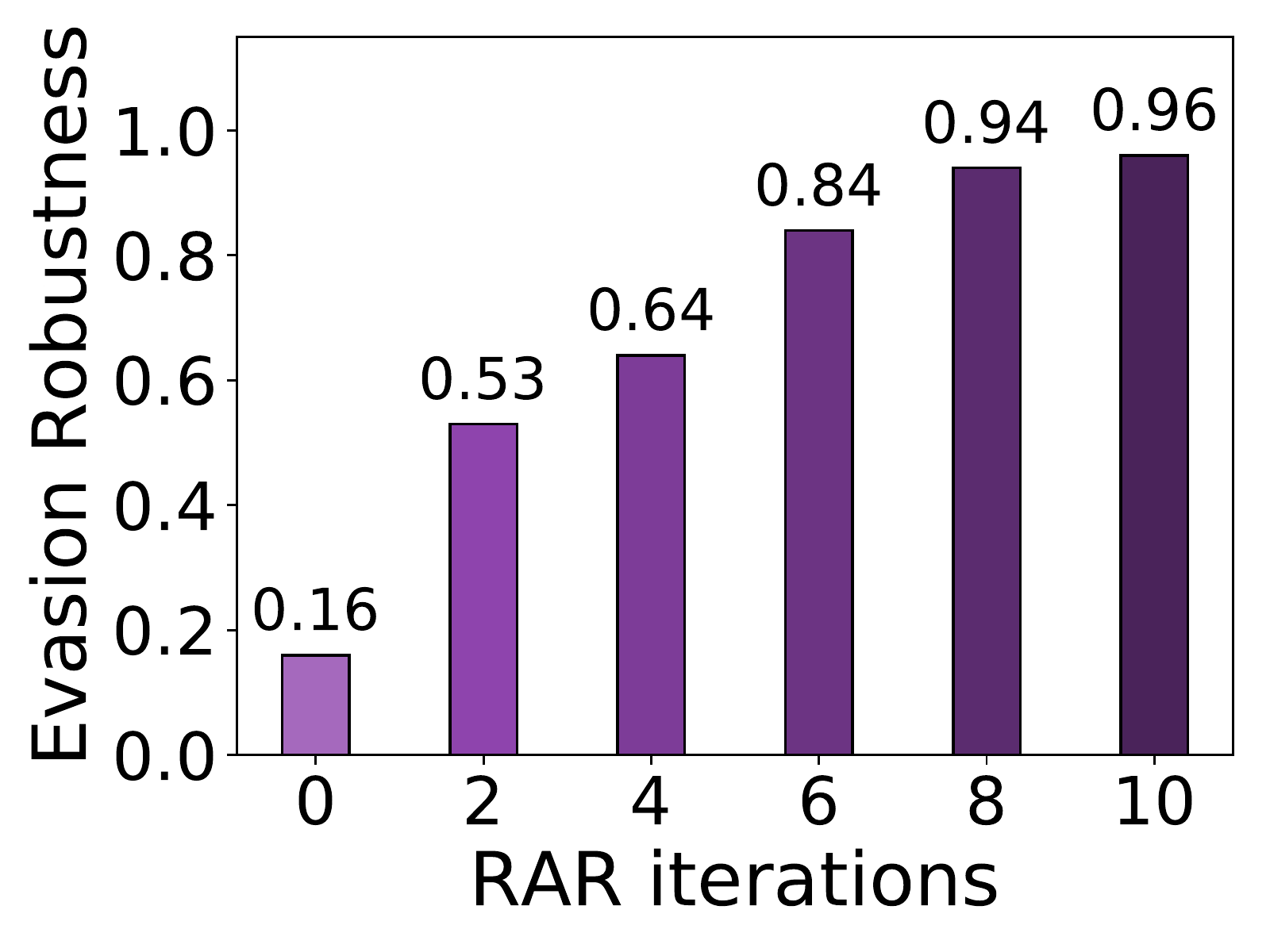} &
     \includegraphics[width=0.22\textwidth]{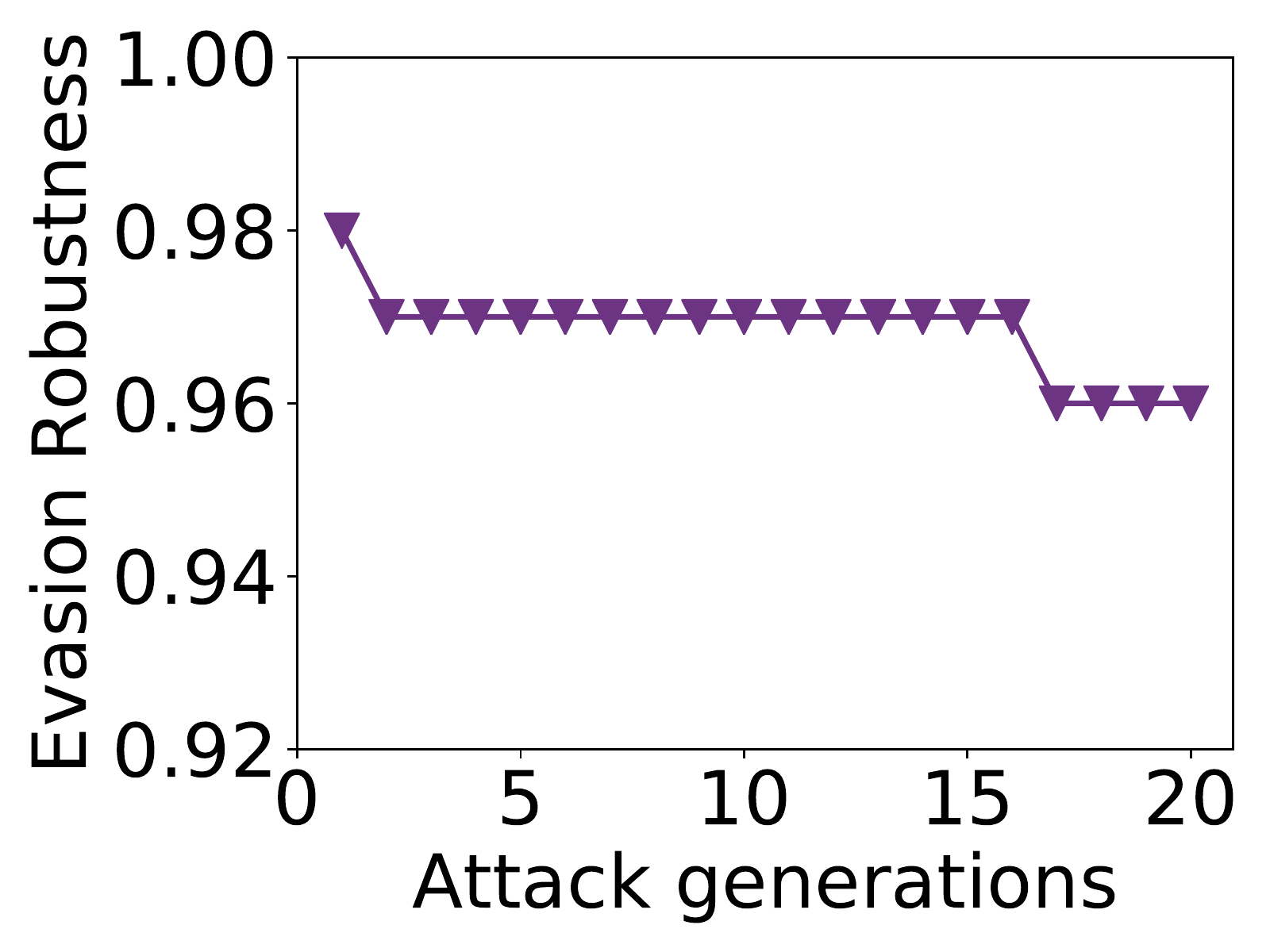}

\end{tabular}
	\caption{Evasion robustness with retraining iterations (left) and generations of the EvadeML attack test (right).}
	\label{SL2013-evademl}
      \end{figure}

Figure~\ref{SL2013-evademl} (left) shows the gradual improvement of evasion robustness over the 10 retraining iterations.
        This plot demonstrates non-trivial effectiveness of EvadeML: the first few iterations are clearly insufficient, as re-running EvadeML creates many new evasions that cannot be correctly detected by the classifier.
        Only after 6 iterations does EvadeML optimization loop begin to show significant signs of failing.
%It indicates that by iteratively retraining SL2013, its robustness against EvadeML is gradually improved. 
        Figure~\ref{SL2013-evademl} (right) shows how increasing the number of generations in EvadeML attacks affects robustness of the RAR classifier.  At this point, we can see that increasing the capability of the attack has minimal impact.
%        presents the evasion robustness of RAR to EvadeML attacks using different generations in genetic programming.
%        It shows that there is little difference after 20 generations of EvadeML attacks.
%        \textcolor{red}{TODO: update the figure!}   
%This provides us with a baseline to evaluate the effectiveness of the feature space evasion model in achieving evasion robustness.
%Figure \ref{SL2013-sys} shows the improvement of evasion robustness as a function of the number of iterations. 
%It indicates that by iteratively retraining SL2013, its robustness against evasions in problem space gradually improves, and only 10 iterations are necessary for the process to converge. 

%\begin{figure}
%	\includegraphics[width=0.48\textwidth]{figures/SL2013-accu.pdf}
%	\caption{Evasion robustness of the baseline and different retraining approaches for SL2013 under EvadeML \cite{ndss2016} test.}
%	\label{SL2013-accu}
%\end{figure}

\paragraph{Feature-Space Retraining}

% We next conduct experiments to evaluate the effectiveness of the retraining approach to boost evasion robustness which uses a feature space evasion model.
Next, we experimentally evaluate the effectiveness of retraining with a feature-space model of evasion attacks in obtaining robust ML in the face of the EvadeML realizable attack.
%As above, we use EvadeML to \emph{evaluate} how evasion robust the resulting classifier is after retraining.
We consider the setting with $\lambda = 0.05$ and $\lambda = 0.005$ in Equation~\ref{evasion-model} (henceforth, FSR-$\lambda_1$ and FSR-$\lambda_2$). 

The robustness results are shown in Figure \ref{SL2013-sys} (left).  
Compared to the SL2013 baseline, feature-space retraining (FSR) boosts evasion robustness from 16\% to 62\%.
%, although it only achieves a 10\% evasion robustness when weighted distance (denoted as WD in Figures \ref{SL2013-sys} (left) and \ref{SL2013-sys} (right)) is employed. 
Crucially, \emph{the robustness of the resulting classifier is far below the classifier achieved by RAR.}
This illustrates that defense that relies on feature-space models of adversarial examples may not in fact lead to robustness when it is faced with a real attack.
%In other words, \emph{the feature space model is not an adequate representation of real evasion attacks.}
% with a problem space oracle, which achieves 96\% evasion robustness.
%The additional cause for concern is that the smaller value of $\lambda$, while resulting in a more robust classifier, induces a significantly slower training process, with training time approaching that for systematic problem space retraining due to the large number of iterations required for convergence.

%\begin{figure}
%	\includegraphics[width=0.48\textwidth]{figures/SL2013-roc.pdf}
%	\caption{ROC curves on non-adversarial test data of the baseline and classifier retrained on SL2013}
%	\label{SL2013-roc}
% \end{figure}

We again consider performance of FSR classifier on non-adversarial test data (Figure \ref{SL2013-sys} (right)).
%Evaluating the quality of FSR,
We can see that robustness boosting again does not much degrade performance, with AUC remaining above 99\%.
However, we do see a substantial degradation as we move from $\lambda = 0.05$ to $0.005$; thus, as we increase adversarial power in the feature-space model, while we do obtain a slightly more robust model, we incur a nontrivial hit in performance on non-adversarial data.

\subsubsection{Hidost}
%We now evaluate the robustness of both the problem space and feature space retraining approach for Hidost, the updated version of SL2013.
%We use the same setup as in our study of SL2013.
% and employ EvadeML to evaluate the classifiers.
We now repeat our experiments above with another structure-based classifier, Hidost.
We set the retraining parameter $\lambda = 0.005$, which appears to strike a reasonable balance between robustness and accuracy on non-adversarial data.
% , as described in Section \ref{S:exp}.
As before, we first evaluated the robustness of the original Hidost \cite{srndic2016} by EvadeML. 
The result shows a 2\% robustness---remarkably, significantly worse than SL2013.

%We first evaluate the evasion robustness of the Hidost baseline classifier and other classifiers retrained with EvadeML and the feature space attack model (The resulting classifiers are termed as PSR and FSR). 

%\begin{figure}
%	\includegraphics[width=0.48\textwidth]{figures/Hidost-sys.pdf}
%	\caption{Evasion robustness of problem space retraining (PSR) on Hidost as a function of iterations.}
%	\label{Hidost-sys}
%\end{figure}

\begin{figure}[t]
\centering
     \begin{tabular}{cc}
	\includegraphics[width=0.22\textwidth]{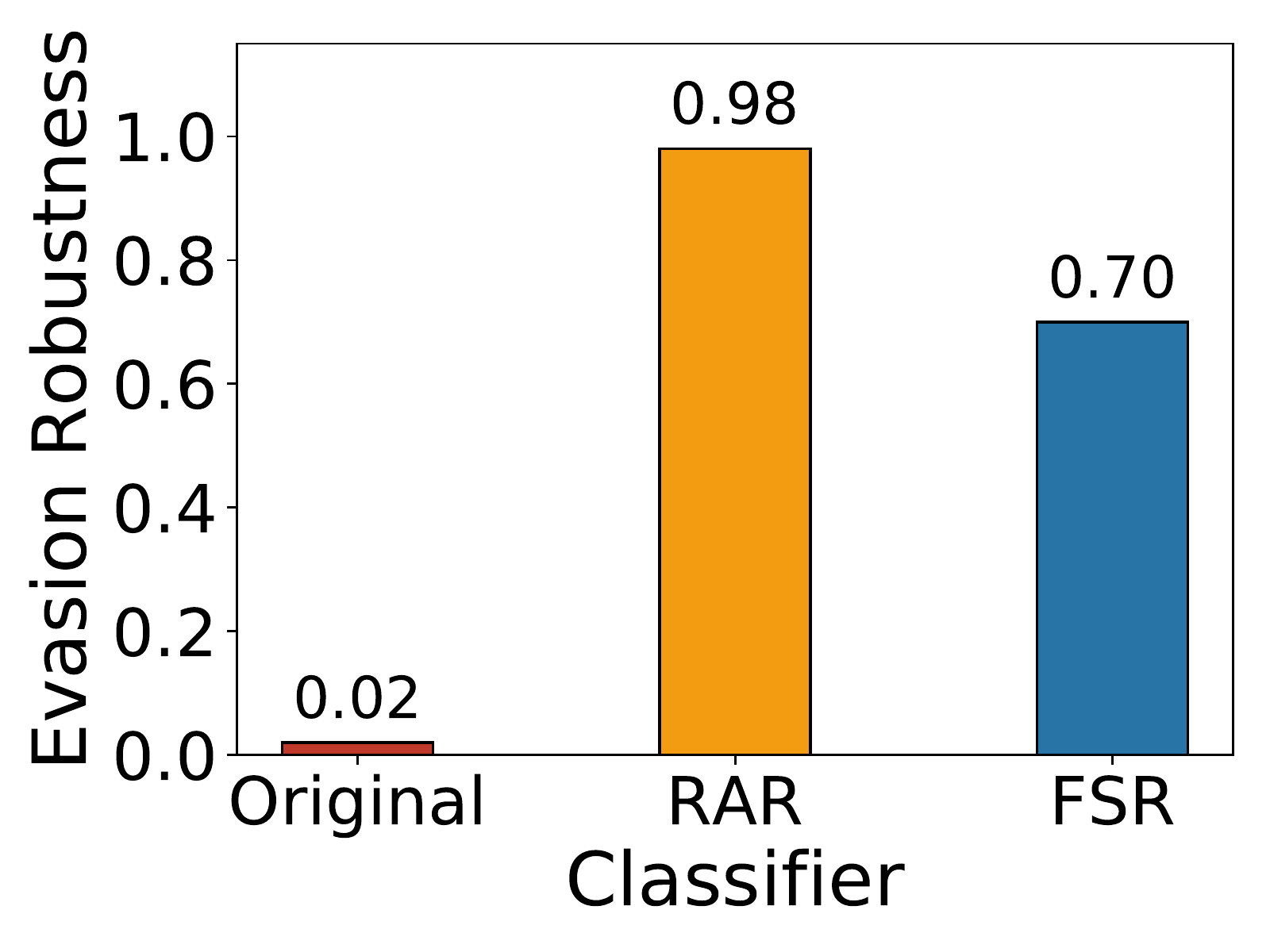} &
\includegraphics[width=0.22\textwidth]{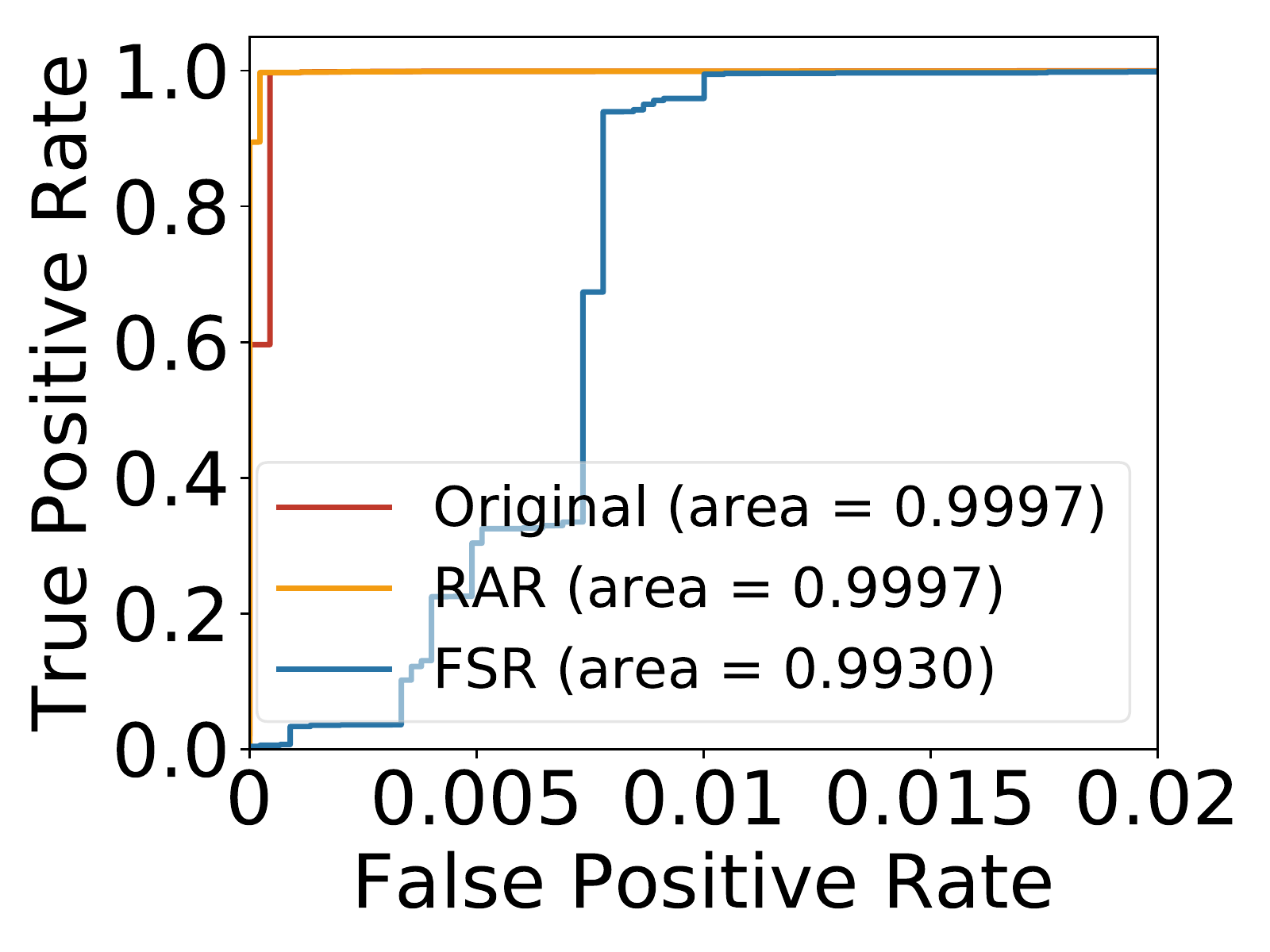}
\end{tabular}
	\caption{Evasion robustness under EvadeML test (left) and performance on non-adversarial data (right) of different classifiers for Hidost.}
	\label{Hidost-accu}
\end{figure}

%\begin{figure}
%	\includegraphics[width=0.48\textwidth]{figures/Hidost-roc.pdf}
%	\caption{ROC curves on non-adversarial test data of the baseline and classifier retrained on Hidost}
%	\label{Hidost-roc}
%\end{figure}

%The results of PSR and FSR approaches applied to Hidost are summarized in Figure \ref{Hidost-sys}, \ref{Hidost-accu} and \ref{Hidost-roc}. 
Evasion robustness of Hidost, as well as improvements achieved by RAR and FSR, are shown in Figure \ref{Hidost-accu} (left), and the results are consistent with our observations for SL2013.
%Figure \ref{Hidost-accu} shows that 
%We can observe that although Hidost is designed to be more robust than SL2013, it is actually even more vulnerable to EvadeML evasion attacks, with only a $2$\% evasion robustness (compared to $15\%$ for SL2013).
%, which is even worse than SL2013.
First, by retraining with the realizable attack, evasion robustness is boosted to $98\%$, a rather dramatic improvement, and clear demonstration that successful defense is possible.
% (that is, EvadeML terminates without successfully finding an evasive variant in 99\% of the test instances).
In contrast, FSR achieves a $70\%$ evasion robustness, a significant boost over the original Hidost, to be sure, but far below the evasion robustness of RAR.

Evaluating these classifiers on non-adversarial test data in terms of ROC curves (Figure \ref{Hidost-accu} (right)), we can observe that
%The results show that 
RAR achieves comparable accuracy ($>99.9$\% AUC) with the original Hidost classifier on non-adversarial data, and provides even better \emph{True Positive Rate (TPR)} when \emph{False Positive Rate (FPR)} is close to zero.
On the other hand, FSR achieves $>99$\% AUC, but yields a significant degradation of TPR when FPR$<0.01$.

\subsection{Content-Based PDF Malware Classification}
%\label{sec:pdfrate}

Our next case study concerns another two PDF malware classifiers which use features based on PDF file content, rather than logical structure.
%Aside from the nature of features, another important difference between SL2013/Hidost and PDFRate is feature types: PDFRate uses features of mixed type, including binary, integer, and real-valued.
%However, the surrogate representations of PDFRate previously used to evaluate its robustness had normalized the features to be real-valued and zero-mean~\cite{oakland2014,ndss2016} (henceforth, PDFRate-R).
%In addition, we construct a \emph{binarized} version of PDFRate (henceforth, PDFRate-B), where each feature is transformed into a binary feature (where needed) by thresholding.
%Thus, our evaluation enables a direct comparison between binary and
%real-valued versions of the same classifier, facilitating
%generalizable insights.
%%%%%The original PDFRate classifier uses a collection of features which have varying domains, including binary features, integer-valued, and real-valued. 
%In order to facilitate the most generalizable study, we transform this feature space into two classes: binarized, where each feature (before being normalized) is then binarized, and real-valued, where the feature normalization step is performed. 
%The binarized version of PDFRate (PDFRate-B henceforth) can now be comparable in concept to Hidost, which also uses binary features, whereas the real-valued version (PDFRate-R henceforth) allows us to explore implications of having a fundamentally different feature space on evasion models and evasion robustness. 
We trained both real-valued and binarized PDFRate (henceforth, PDFRate-R and PDFRate-B) on the same dataset as SL2013 and Hidost, and achieved $>99.9$\% AUC for both classifiers on test data.
In our experiments, we empirically set the SVM \textsl{RBF} parameters for training  to $C=10$ and $\gamma = 0.01$.
%Hence, the binarized and real-valued PDFRate can efficiently classify non-adversarial data as the original PDFRate.
%\subsection{Experiments}
%We now evaluate the robustness of both the problem space and feature
%space retraining of the two versions of PDFRate, with the same settings as in our investigation of Hidost. 
In our evaluation of ML robustness, we again set the feature-space model parameter $\lambda$ to be $0.005$.%

\subsubsection{PDFRate with Real-Valued Features}

We begin with the variant of PDFRate---PDFRate-R---which has been constructed in
previous evaluations and shown comparable in performance to the
original implementation~\cite{oakland2014}.
We again begin by replicating the EvadeML evasion robustness evaluation of
the baseline classifier.
As expected, we find the classifier quite vulnerable, with only 2\%
evasion robustness.

\begin{figure}[t]
\centering
\begin{tabular}{cc}
	\includegraphics[width=0.22\textwidth]{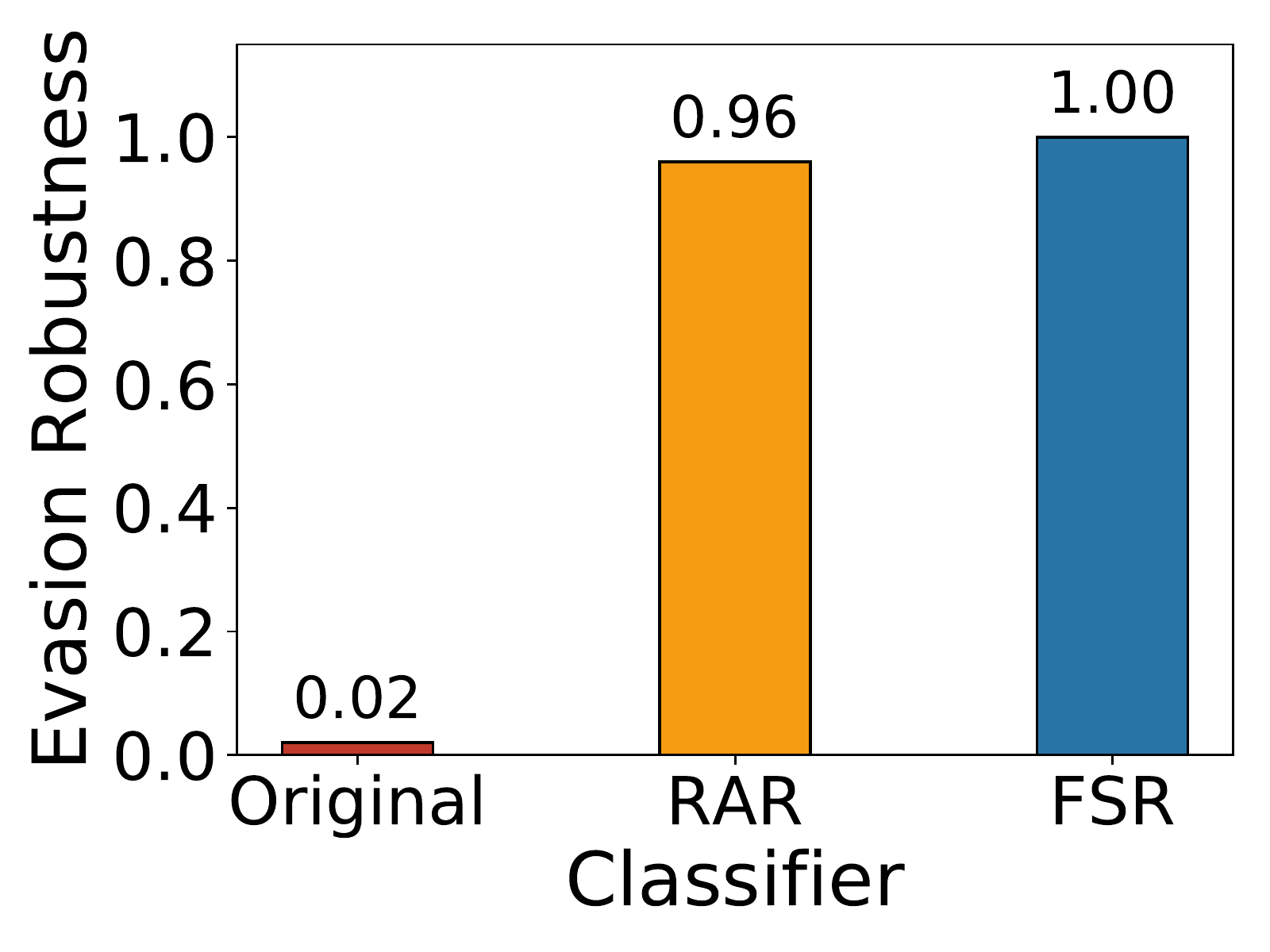} &
\includegraphics[width=0.22\textwidth]{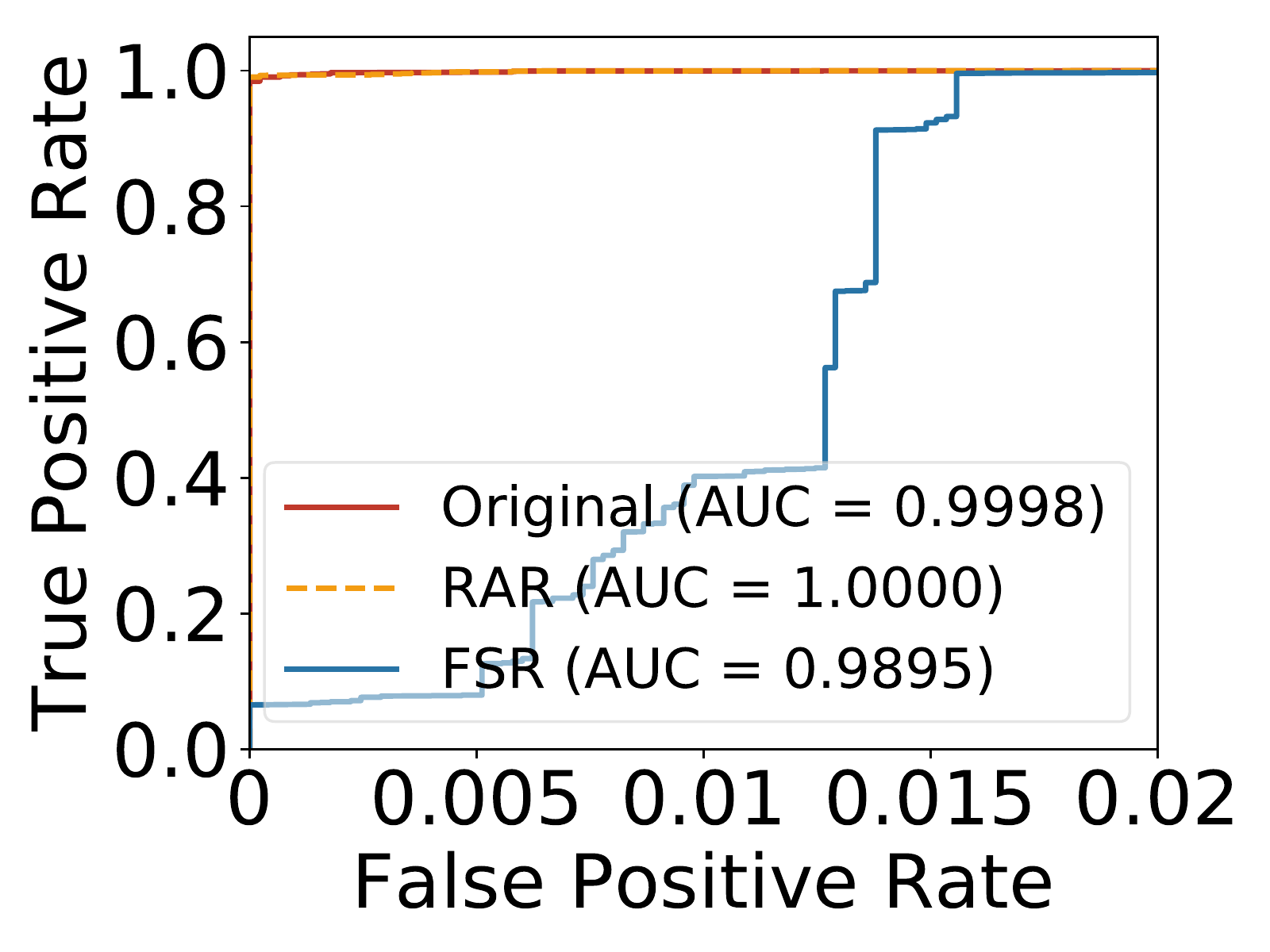}
\end{tabular}
	\caption{Evasion robustness under EvadeML test (left) and performance on non-adversarial data (right) of different classifiers for PDFRate-R.}
	\label{PDFRate-R-accu}
\end{figure}

Next, we retrain PDFRate-R with EvadeML for 10 iterations (RAR
baseline), and perform feature-space retraining using the conventional
feature space model above.
% Our results are quite surprising.
Our results are shown in Figure \ref{PDFRate-R-accu} (left).
%, are rather surprising.
Observe that while RAR indeed achieves a highly robust classifier (96\% robustness), FSR actually performs \emph{even better}, with 100\% robustness.
%both PSR and FSR achieve high
%evasion robustness; indeed, FSR is slightly better at 100\%
%robustness, with PSR at 96\%.

Comparing RAR and FSR performance on non-adversarial data
(Figure~\ref{PDFRate-R-accu} (right)), we observe that the high robustness of FSR does incur a cost:  
%begin to observe some meaningful
%degradation for FSR:
while RAR remains exceptionally effective
($>$99.99\% AUC), FSR achieves AUC slightly lower than 99\%, although
most significantly, the degradation is rather pronounced for low
FPR regions (below $0.015$).
%Thus, there remains a gap between PSR and FSR, but this gap now has
%different nature: both are quite robust to problem space evasion, but
%FSR degrades baseline performance significantly more than PSR.

\subsubsection{PDFRate with Binarized Features}
\label{subsec:PDFRate-B}

\begin{figure}[t]
\centering
\begin{tabular}{cc}
	\includegraphics[width=0.22\textwidth]{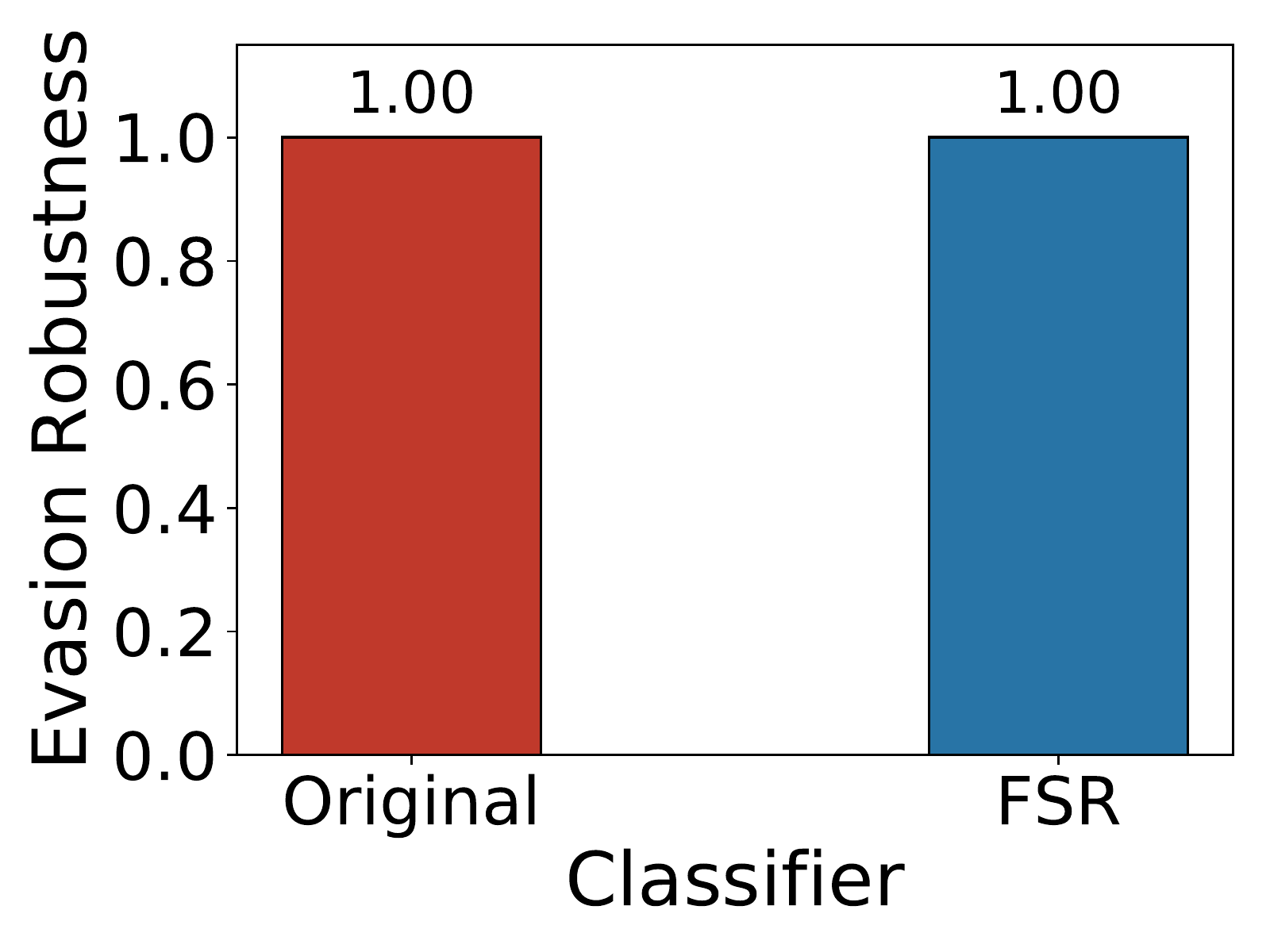} &
\includegraphics[width=0.22\textwidth]{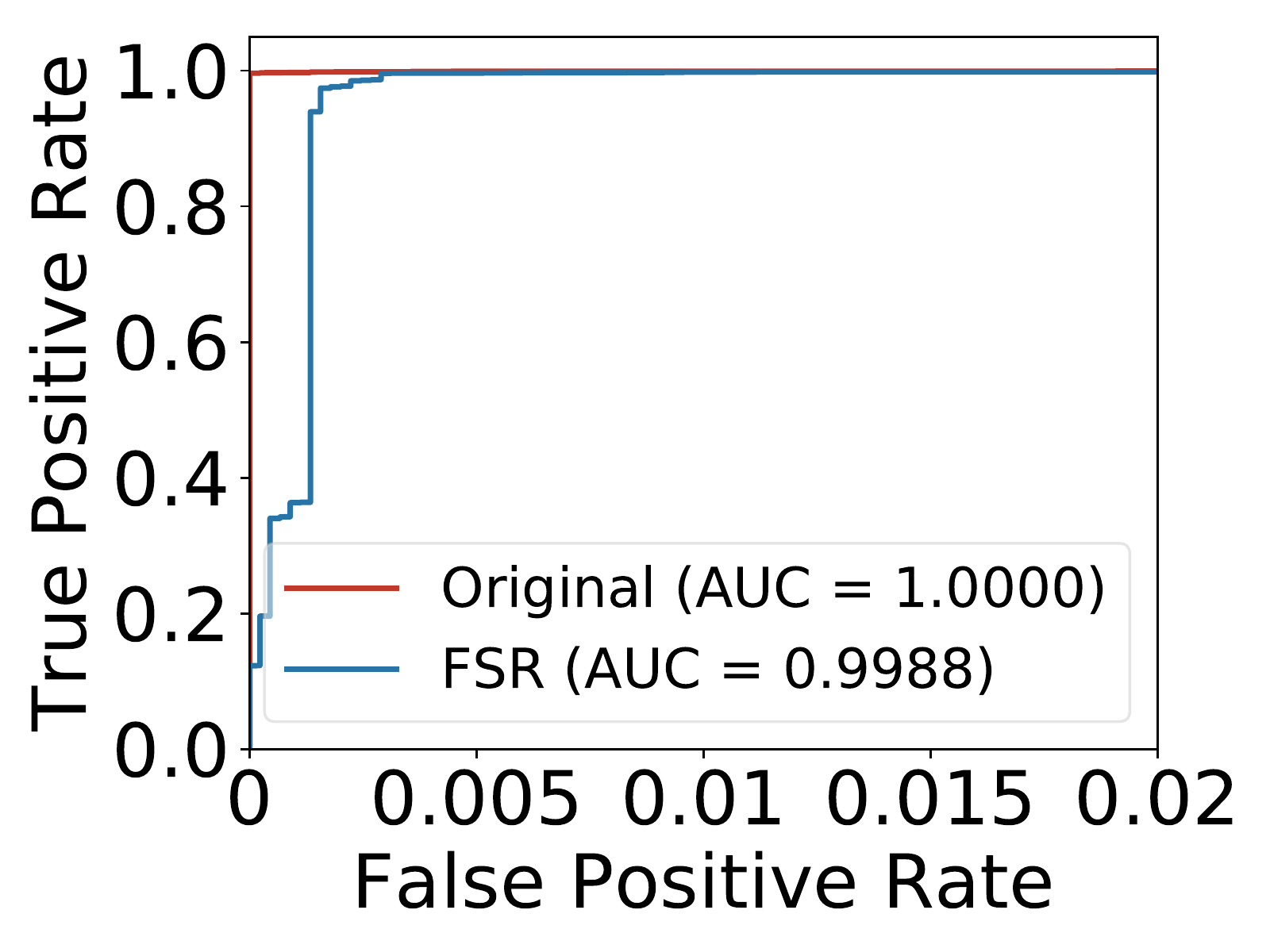}
\end{tabular}
	\caption{Evasion robustness under EvadeML test (left) and performance on non-adversarial data (right) of different classifiers for PDFRate-B.}
	\label{PDFRate-B-accu}
\end{figure}

%\begin{figure}
%\centering
%	\includegraphics[width=0.23\textwidth]{figures3/PDFRate-B-roc.pdf}
%	\caption{Performance on non-adversarial data of different classifiers for PDFRate-B.}
%	\label{PDFRate-B-roc}
%\end{figure}

One of our great surprises is the robustness of the binarized PDFRate: despite the fact that the real-valued PDFRate is
quite vulnerable, \emph{the same classifier using binary features was 100\% robust to EvadeML} (Figure~\ref{PDFRate-B-accu} (left)).
%Consequently, the RAR effectively terminates with no retraining at
%all.
Consequently, this will serve as our robust baseline (equivalently, RAR would terminate with no iterations).
Feature-space retrained PDFRate-B also exhibits 100\% evasion
robustness, although it does require a number of iterations to converge.

Considering now the performance of PDFRate-B and FSR on
non-adversarial test data (Figure \ref{PDFRate-B-accu} (right)), we can make
two interesting observations.
First, the baseline PDFRate-B is remarkably good even on this data; in a sense, it appears to hit the sweet spot of adversarial
robustness and non-adversarial performance.
Second, FSR retrained classifier is competitive in terms of AUC ($\sim
99.9\%$), but is observably worse than the baseline classifier for
very low false positive rates.

\section{Evasion-Robust Classification with Conserved Features}
\label{CR}

%So far, we learned that we can successfully ``robustify'' a classifier against evasions with systematic iterative retraining when attacks are modeled in problem space, but not necessarily when attacks are abstracted in feature space.
%So far, we observed that the widely used and mathematically convenient feature-space evasion models may only poorly represent realistic evasion attacks.
Thus far, we had observed that ML hardened with the standard mathematically convenient feature-space evasion attack model may in some cases not yield satisfactory robustness against real attacks.
The key issue is that feature-space models are entirely disembodied from the domain.
This is crucial to enable us to have mathematical formulations of attacks, but clearly has limitations.
The key question is whether we can devise a simple way of anchoring feature-space attacks in the application domain to allow us to meaningfully and minimally constrain abstract attacks to reflect some of the constraints that real attacks face.
Next, we propose a refinement of the feature-space model that aims to do just that.
% Next we propose a simple idea for bridging the gap between problem and feature space (where one exists): explicitly accounting for a subset of features which are \emph{conserved} during evasions.

Specifically, we introduce the idea of \emph{conserved features}, which we define to be \emph{features, the unilateral modification of which compromises malicious functionality.}
%More precisely, conserved features are those which are invariant under (are unaffected by) evasion attacks.
We develop this idea specifically for \emph{binary features}, as this notion is particularly crisp in such a case (e.g., such features tend to correspond to the existence of particular objects in PDF).
%; when features are real-valued, it is not clear how we can even define conserved features, and in any case we observed no appreciable evasion robustness gap between PSR and FSR in that setting.

Next, we present three major findings.
First, conserved features do exist in all three of our classifiers over the binary feature space, and can be effectively identified (see our algorithm for identifying conserved features in Appendix~\ref{sec:CF}).
%Previous accounts have generally been skeptical of the ability to find, or use, such attack invariants for defense.
%Second, we show that a classifier using \emph{only} the conserved features is completely robust to the EvadeML attack (essentially by construction).
%, and (b) quite effective on test data not involving evasion attacks.
Second, conserved features cannot be recovered using statistical feature reduction (in our case, sparse regularization), and feature reduction methods do not lead to robust classifiers.
The reason is that conservation is connected to the relationship between features and malicious functionality, rather than statistical properties of non-evasion data; for example, features which are strongly correlated with malicious behavior are often a consequence of attacker ``laziness'' (such as whether a PDF file has an author), and are easy for attackers to change.
Third, we demonstrate that the limitations of feature-space robust ML can be substantially alleviated by incorporating conserved features as attack invariants in the feature-space evasion model.
%, and the resulting classifier outperforms the one using only conserved features on non-evasion data.
%Finally, we present a systematic approach for identifying the conserved features \textcolor{red}{in Section \ref{sec:CF}}.
%Subsequently (in Section~\ref{sec:CF}) we present a novel automated approach for identifying conserved features in a binary feature space.

%\subsection{Conserved Features}

To develop intuition about the nature of conserved features, consider SL2013, which employs structural paths as features to discriminate between malicious and benign PDFs. 
On the one hand, the structural paths like \verb|/Type| are unessential to preserve malicious behaviors, and we do not expect them to be conserved.
On the other hand, as the shellcode which triggers malicious functionality is embedded in certain PDF objects, those corresponding structural paths are likely to be conserved in each variant crafted from the same malicious seed (e.g., \verb|/OpenAction/JS|).
In addition, structural paths that facilitate embedded script in PDF files also can be conserved features as removing them can break the script (e.g., \verb|/Names| and \verb|/Pages|).
This further illustrates that conserved features are not necessarily optimal for statistically distinguishing benign and malicious instances (indeed, these may be common to both); rather, they serve to anchor the feature-space attack model in the domain by connecting features to malicious functionality.
%Thus, it is actually possible that some conserved features are commonly found in both malicious and benign files.

%As the shellcode which triggers malicious functionality is embedded in certain PDF objects, those corresponding structural paths are likely to be conserved in each variant crafted from the same malicious seed (e.g., \verb|/OpenActionJS|). 
%On the other hand, the structural paths like \verb|/Type| are unessential to preserve malicious behaviors, and we do not expect them to be conserved.

\subsection{Classifying Using Only Conserved Features}
\label{S:conserved}

We begin by exploring the effectiveness of using \emph{only} conserved features for classification.
%; in Section~\ref{sec:CF} we describe a systematic approach for identifying these.
%For each classifier, we use a uniform conserved feature set and the 40 malicious seeds discussed in Section~\ref{S:exp}. 
We identified 8 conserved features for SL2013 (out of $\sim$6000), 7 for Hidost (out of $\sim$1000), and 4 for PDFRate-B (out of 135); these are detailed in Table~\ref{table:cf} of the appendix, while our algorithm for identifying conserved features is presented in Appendix~\ref{sec:CF}.
%Our algorithms for identifying conserved features are provided in Appendi~\ref{sec:CF}, and the lists of conserved features for all classifiers are in.
%The conserved features and our approach for identifying these are detailed in Appendix \ref{sec:CF}.
%Particularly, we identify a conserved feature set with only 8 features out of 6,087 for SL2013, a conserved feature set with only 7 features out of 961 for Hidost, and a conserved feature set with only 4 features out of 135 for both PDFRate-B. 
%These are detailed in Appendix \ref{appendix:cf}.

%The conserved feature sets raise four questions:
We start by considering four natural questions pertaining to conserved features:
1) are they sufficient to make a classifier robust to evasions, 2) do they effectively discriminate between benign and malicious instances, 3) can they be identified using standard statistical methods (such as sparse regularization), and 4) are they just detecting the presence of JavaScript in PDF?

\begin{figure}[t]
\centering
\begin{tabular}{cc}
\includegraphics[width=0.22\textwidth]{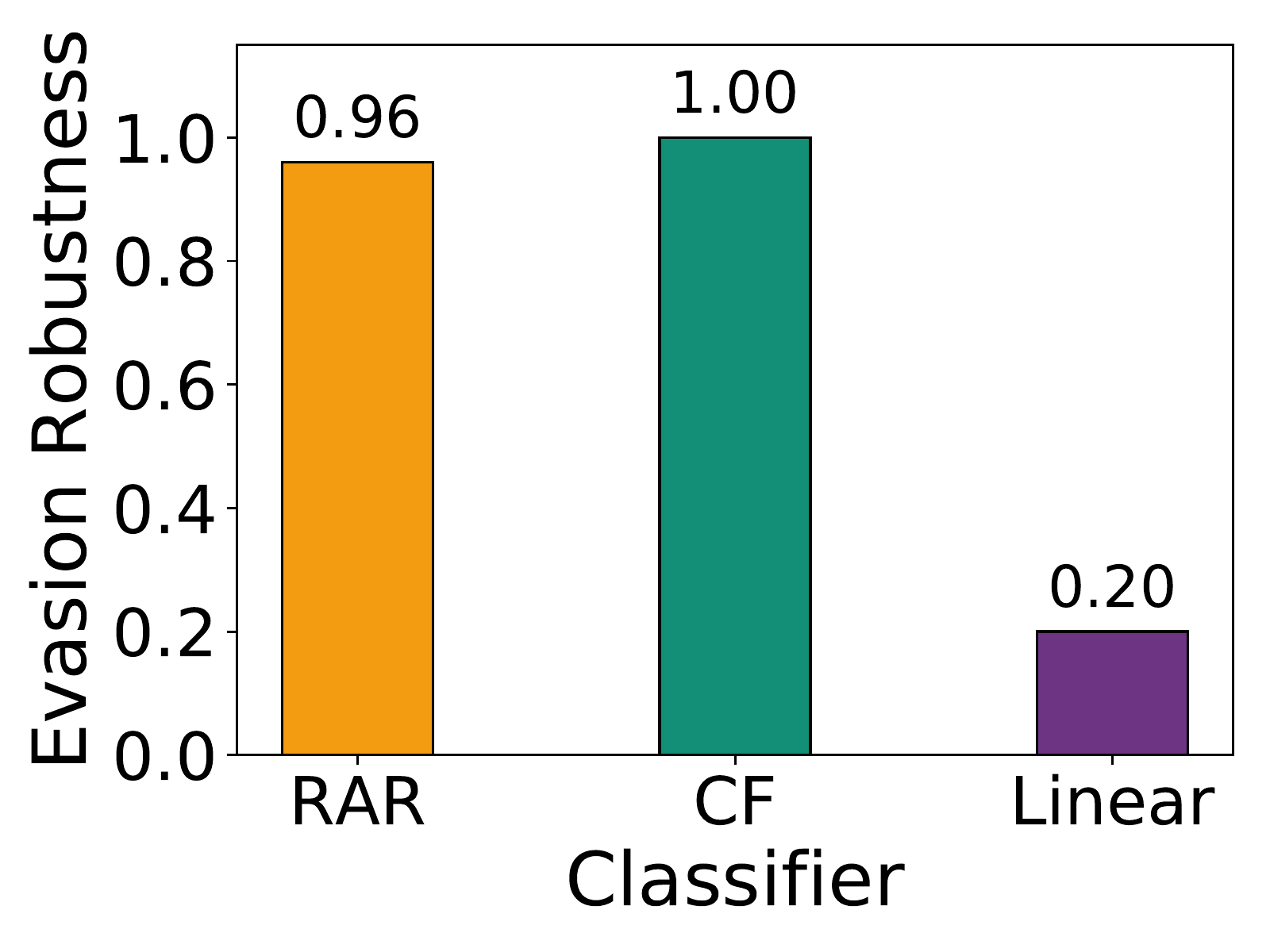} &
\includegraphics[width=0.22\textwidth]{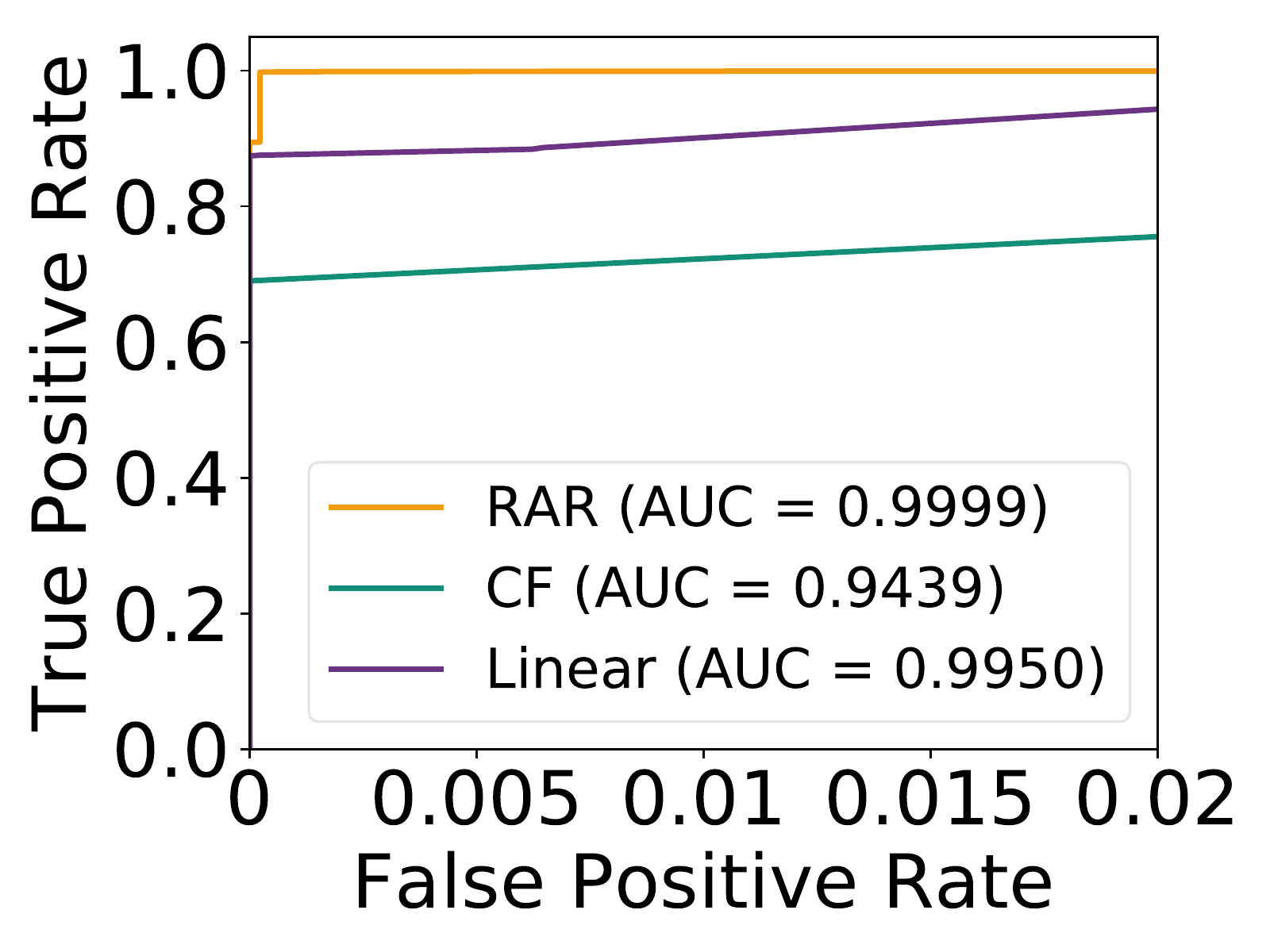}
\end{tabular}
\caption{Classifying with conserved features: comparing evasion robustness (left) and ROC curves (right).}
\label{SL2013-CF-accu}
\end{figure}
We explore these for SL2013.
%To address these questions, 
Specifically, we trained a classifier using \emph{only} the 8 conserved features (\emph{CF} henceforth).
As we can see in Figure~\ref{SL2013-CF-accu} (left), this classifier is 100\% robust to EvadeML attacks, appearing to resolve the first question.
However, we emphasize that conserved features alone need not capture the full spectrum of adversarial behavior and constraints.
Indeed, in Section~\ref{sec:alternative} we show that classifiers based solely on conserved features can also be evaded, particularly if attacks are \emph{specifically designed to evade them}.
Rather, as we show presently, they provide a \emph{sufficient anchoring} in the problem domain for feature-space attack models to succeed.

To address question (2), consider Figure~\ref{SL2013-CF-accu} (right): clearly, if we desire a low false positive rate, using only conserved features for classification yields subpar performance on non-adversarial data.
To address the third question, we learn a linear SVM classifier for SL2013 with $l_1$ regularization (henceforth, Linear) where we empirically adjust the SVM parameter $C$ to perform feature reduction until the number of the features is also 8; we find that only 3 of these are conserved features (see Appendix~\ref{appendix:regularized-features} for a more detailed analysis of the relationship between statistically useful and conserved features).
%(the features obtained by regularization are detailed in Appendix~\ref{appendix:regularized-features}.)
%We use both EvadeML and non-adversarial dataset to evaluate these classifiers.    
%The \emph{Baseline} is, as before, the \textcolor{red}{RAR version of} the SL2013 classifier.
As we can see in Figure~\ref{SL2013-CF-accu} (left), this classifier exhibits poor robustness; thus, statistical methods are insufficient to identify good conserved features.

%As shown in Figure~\ref{SL2013-CF-accu} (left), the classifier using only the 8 conserved features (CF) is 100\% resistant to evasion attacks by EvadeML.
%In contrast, the linear SVM with sparse ($l_1$) regularization yielding only 8 most statistically important features is easily evaded.
%Figure~\ref{SL2013-CF-accu} (right) shows performance of the classifiers on non-evasive test data.
%We observe that linear SVM using only 8 features yields better than 99\% AUC, approaching the performance of the baseline classifier.
%Moreover, even \emph{CF}, which is robust to evasions, achieves AUC just under 95\%!
%Nevertheless, the \emph{CF} classifier can be seen to perform relatively poorly in the region of the ROC curve where the false positive rate is low, which in practice is the most consequential part of it.
%Below we consider whether the feature space retraining procedure can be improved by incorporating conserved features into the evasion model without compromising its mathematical elegance.

%\begin{table}[h]
%\centering
%\begin{tabular}{|c|c|c|}
%\hline
%Classifier & FPR  & FNR  \\ \hline\hline
%JS         & 0.04 & 0.14 \\ \hline
%CF         & 0.04 & 0.11 \\ \hline
%\end{tabular}
%\caption{Performance on non-adversarial data for JS and CF.}
%\label{table:JS}
%\end{table}

  To address the fourth question, we create a classifier using only one boolean feature which identifies the presence of JavaScript in a PDF file (henceforth, we refer to this feature as \emph{JS}).
  We find that this classifier is also robust to EvadeML.
%We evaluate this simple classifier on the same non-adversarial data as CF.
%As displayed in Table~\ref{table:JS},
On non-adversarial data, JS achieves FPR of 0.04 and FNR of 0.14 (in other words, 4\% of the benign files in the non-adversarial dataset use JavaScript, while 14\% of malicious instances use alternative attacks to Javascript).\footnote{We observe similar results for 5,000 benign PDFs obtained by using Google web searches~\cite{oakland2014}, where 3\% of benign files use Javacript.}
To create an apples-to-apples comparison with the CF classifier, we empirically adjust the classification threshold of CF until we get the same FPR with JS.
The resulting CF classifier exhibits FNR of 0.11, considerably better than JS.
Nevertheless, it is clear that using either CF (only conserved features), or only JS, is impractical, since both FNR and FPR of these are quite high.
Moreover, as we show in Section~\ref{sec:alternative}, classifiers based only on conserved features can be defeated by other realizable attacks.
Next, we show that identification of conserved features is nevertheless crucial in creating highly effective feature-space attack models.
%As shown in Table~\ref{table:JS}, using the full batch of conserved features for classification can do better than JS on non-adversarial data, with FNR of 0.11.
%In Section~\ref{sec:alternative}, we demonstrate that using JS classifier for malware detection can be easily defeated by new attacks, whereas the ones hardened with conserved features are not.

\subsection{Feature-Space Model with Conserved Features}

%\subsubsection{Modified Feature-Space Model}

As discussed above, the feature-space evasion model in Equation (\ref{evasion-model}) may 
%can fail in presenting representative evasion attacks in problem space, and does 
not sufficiently boost ML robustness. 
%Since conserved features represent malicious functionality in feature space, 
Since conserved features allow us to minimally tie the abstract feature-space representation to malicious functionality, we offer a natural modification of the model in Equation~\eqref{evasion-model}, imposing the constraint that conserved features cannot be modified by the attacker.
%are preserved in evasive instances.
%they can be applied to evasion models in feature space to generate adversarial examples which are actually malicious. The modified feature space evasion model is then translated into an optimization problem with constraints in feature space:
We formally capture this in the new optimization problem 
%Formally, the new optimization problem is shown 
in Equation~\eqref{evasion-model-cf},
where $\mathsf{S}$ is the set of conserved features:
\begin{equation}
\begin{aligned}
& \underset{x}{\text{minimize}}
& & Q(x) = f(x) + \lambda c(x_M,x),\\
& \text{subject to}
& & x_i = x_{M,i}, \; \forall i \in \mathsf{S}.
\end{aligned}
\label{evasion-model-cf}
\end{equation}
Other than this modification, we use the same \emph{Coordinate Greedy} algorithm with random restarts as before to compute adversarial examples.
%an adversarial evasion in feature space.
We adopt the evasion model in Equation (\ref{evasion-model-cf}) to retrain the target classifier using the retraining procedure from Section~\ref{S:exp}.
We denote the classifier obtained by the retraining procedure using a feature-space model grounded by conserved features by \emph{CFR}.
We also study the effectiveness of our automated procedure for identifying conserved features as compared to using a subset that only considers Javascript features (we can think of these as expert-identified conserved features, as this is what an expert would naturally consider).
To this end, we repeat the procedure above by replacing the conserved feature set $\mathsf{S}$ in Eq.~\ref{evasion-model-cf} with a subset that involves Javascript.
The classifier resulting from such restricted adversarial retraining with ``expert''-identified conserved features is termed \emph{CFR-JS}.

%To compare the performance of CFR with a classifier that only uses a subset of such features pertaining to the existence of Javascript in a PDF, we consider a classifier retrained as above, but with the set $\mathsf{S}$ in \ref{evasion-model-cf} only containing the Javascript-related features.
%\textcolor{red}{
%In addition,  we replace $\mathsf{S}$ in Eq.~\ref{evasion-model-cf} with a subset of conserved features that pertain to JavaScript (detailed in Table~\ref{table:cf} in Appendix~\ref{appendix:cf} ) as part of the full machinery of retraining with conserved features.
%We denote the resulting classifier by \emph{CSF-JS}.
%}

%Since we use uniform conserved features for each malicious seed,  malicious functionality of some PDFs may still be removed.
%To mitigate such degradation, we optionally also generate \emph{mimicry instances} in feature space and add these instances into the training data after the termination of retraining with conserved features.
%We then retrain the classifier again. 
%In our case, we generate such instances by combining feature vectors of malicious and benign PDFs. For each benign file in the training data, we randomly select a malicious PDF, then copy all the features of the malicious PDF to the benign one.
%% Therefore, the resulting feature vector has both conserved and non-conserved features, by which its malicious functionality is preserved, it appears more benign compared to the malicious PDF. 
%The resulting classifier is termed \emph{Mimicry-CFR}. 

%\subsubsection{Experiments}

\subsubsection{SL2013}

We now evaluate the robustness and effectiveness of the feature space retraining approach, which uses conserved features.
We set the parameter $\lambda = 0.005$ as before.
%The iterative retraining process converges after 220 iterations, producing 4,461 adversarial instances.
%Afterward, we produce 4,496 mimicry instances and add them to the training data, and then retrain the classifier again. 
\begin{figure}[t]
\centering
\begin{tabular}{cc}
\includegraphics[width=0.22\textwidth]{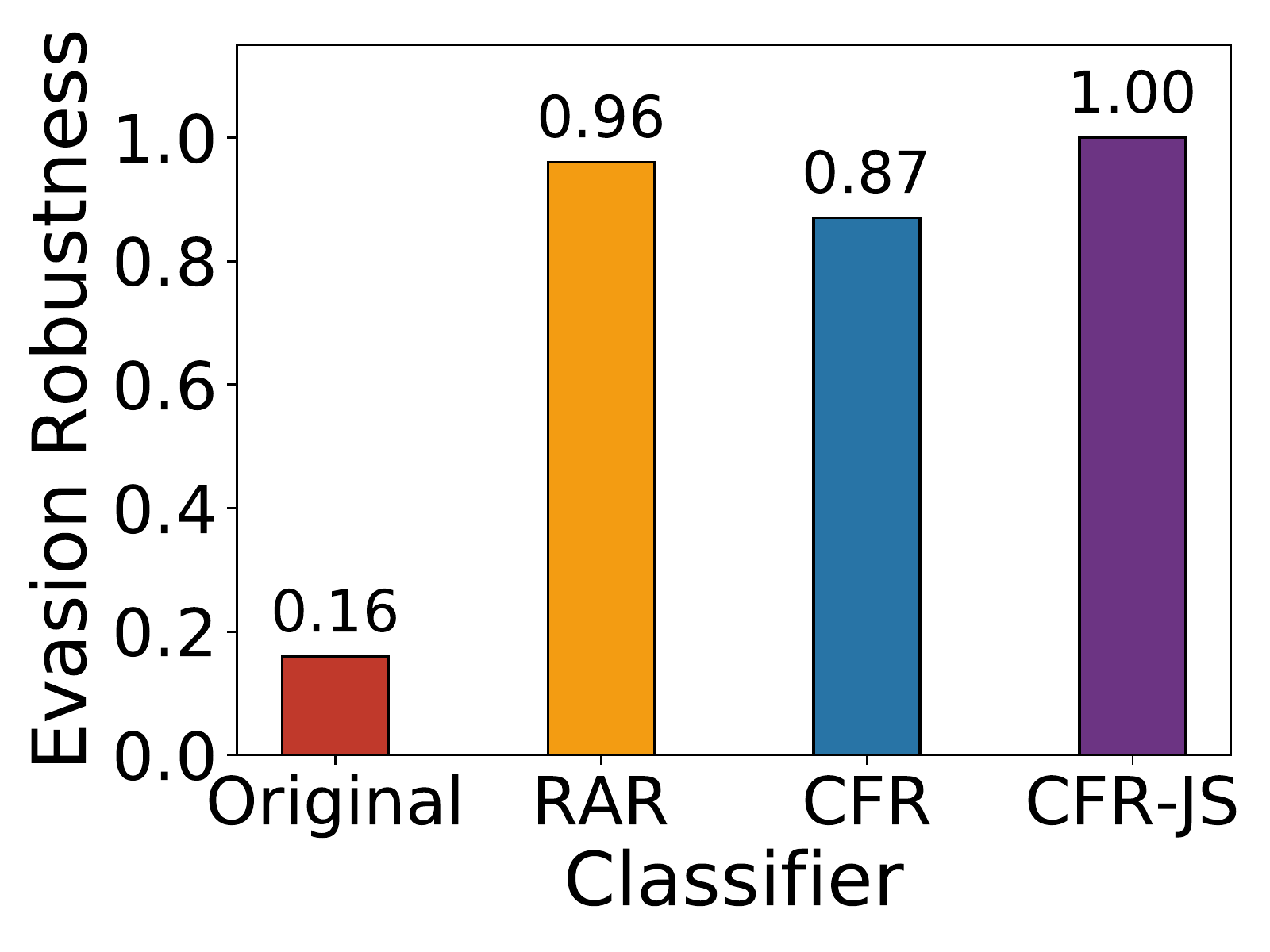} &
\includegraphics[width=0.22\textwidth]{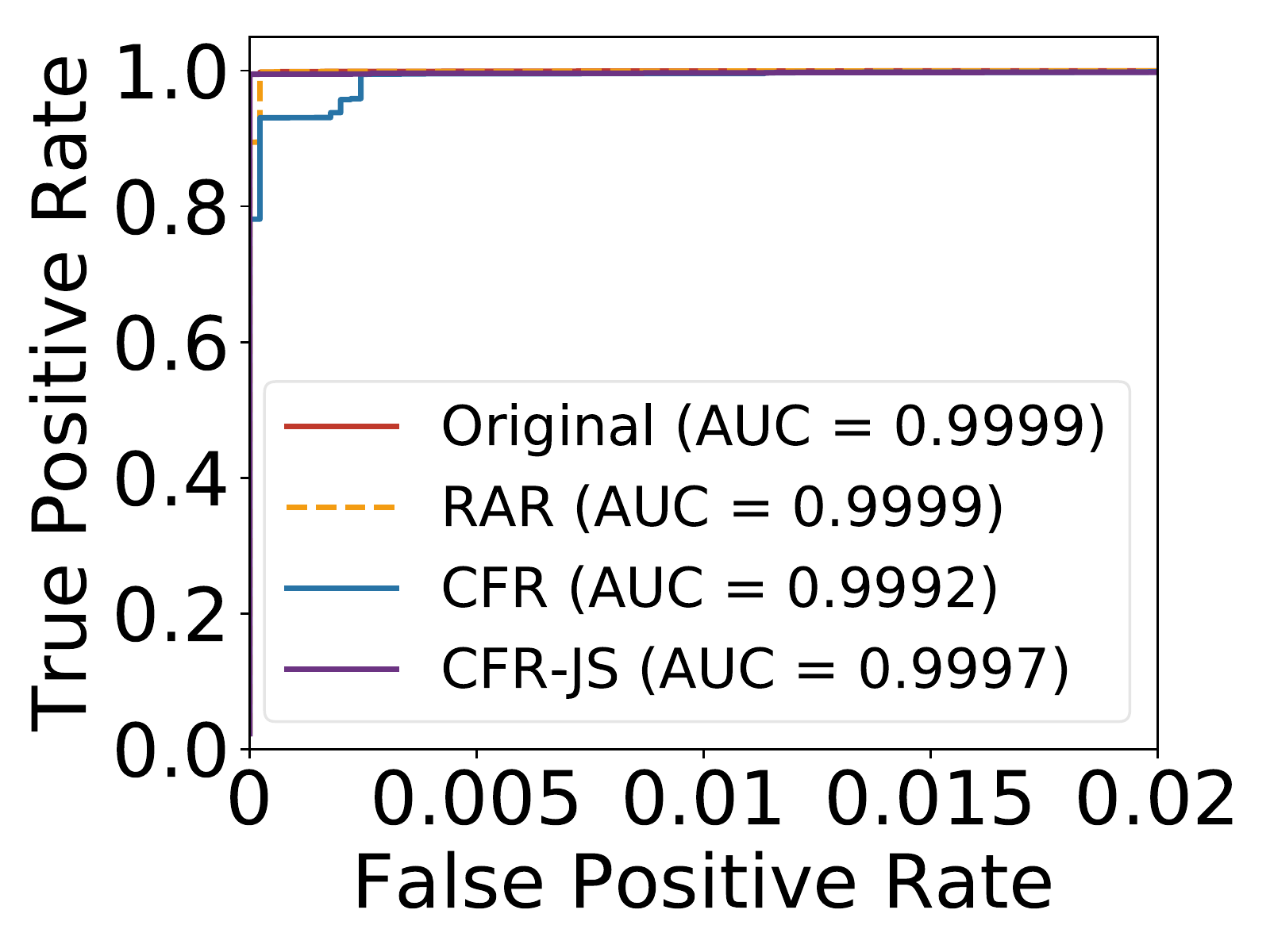}
\end{tabular}
\caption{Evasion robustness (left) and performance on non-adversarial data (right) of different variants of SL2013.}
\label{SL2013-CFR-accu}
\end{figure}
%We first evaluate the evasion robustness of the SL2013 baseline classifier and other classifiers obtained by different retraining. 
The robustness results are presented in Figure \ref{SL2013-CFR-accu} (left).  
Observe that \emph{CFR} now significantly improves robustness of the original classifier, with evasion robustness rising from 16\% to 87\%.
Moreover, \emph{CFR-JS} achieves a 100\% evasion robustness against EvadeML.
%By adding mimicry instances to the training data and retraining \emph{CFR} again, evasion robustness is further improved to 94\%, which is comparable with the approach that retrains with the realizable attack.
% in Section~\ref{S:validation}. 
These results demonstrate that by leveraging the conserved features, the feature-space evasion models are now quite effective as a means to boost evasion robustness of SL2013.

In Figure \ref{SL2013-CFR-accu} (right) we evaluate the quality of these classifiers on non-adversarial test data in terms of ROC curves.
In all cases, be it original, RAR, CFR, and CFR-JS, AUC is $>99.9\%$, although we can see a slight degradation of CFR for extremely low false positive rates compared to the others.
It is noteworthy that CFR performs much better than FSR (robust ML using a standard feature-space approach, recall Figure~\ref{SL2013-sys} (right)).
%We can observe that all variants of \emph{CFR} classifiers are now comparable in terms of performance on non-adversarial instances to both the baseline and RAR classifiers, %as well as the one obtained using problem space evasions, 
%with AUC $\sim99.9\%$.
%\textcolor{red}{Figure~\ref{SL2013-CFR-accu} (right) also indicates that there is a degradation for low FPRs ($<$0.005) that comes with improved robustness. 
%Compared to \emph{FSR} displayed in Figure~\ref{SL2013-sys} (right), such degradation is negligible. }

\subsubsection{Hidost}
\begin{figure}[t]
\centering
\begin{tabular}{cc}
\includegraphics[width=0.22\textwidth]{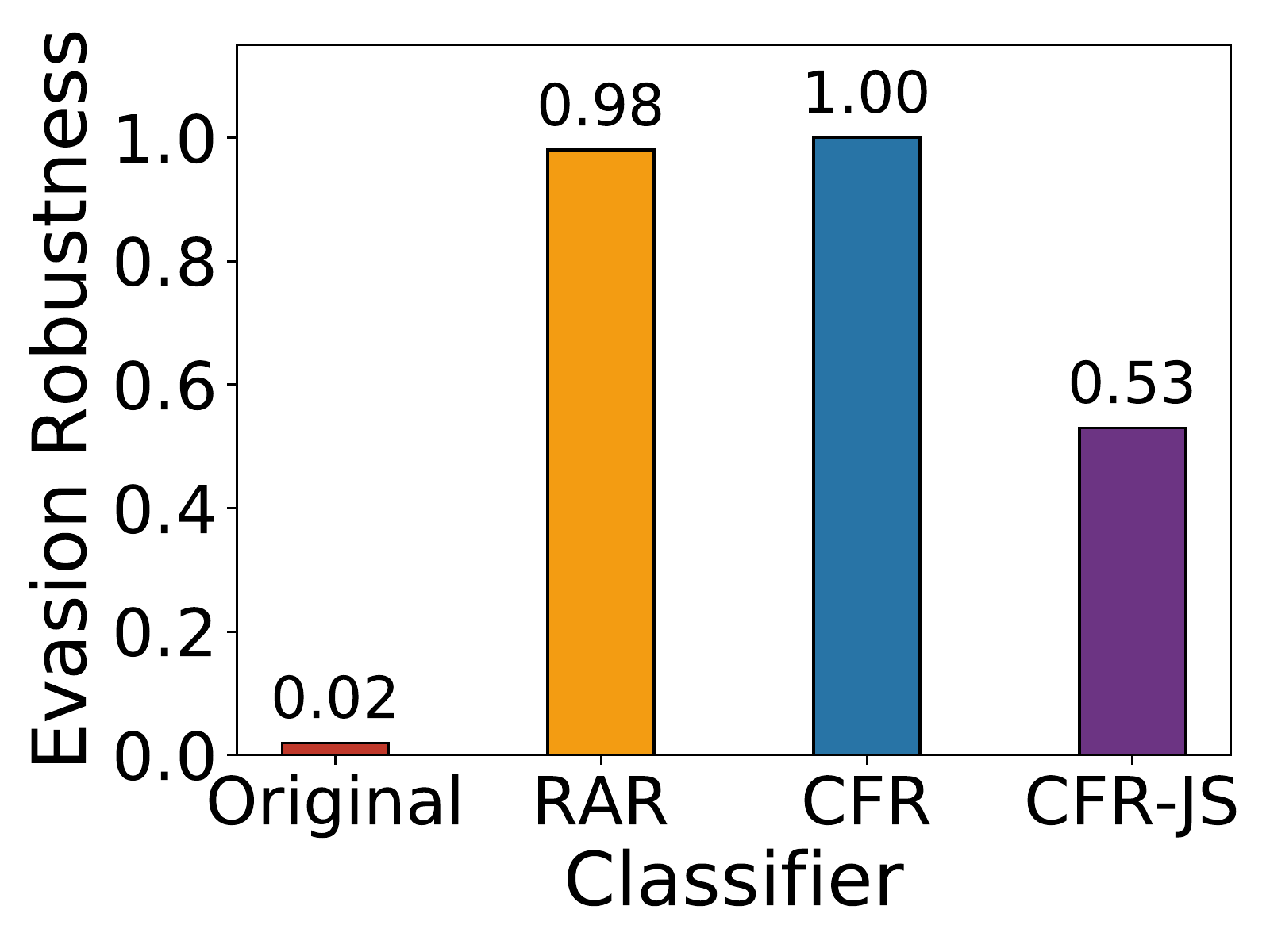}&
\includegraphics[width=0.22\textwidth]{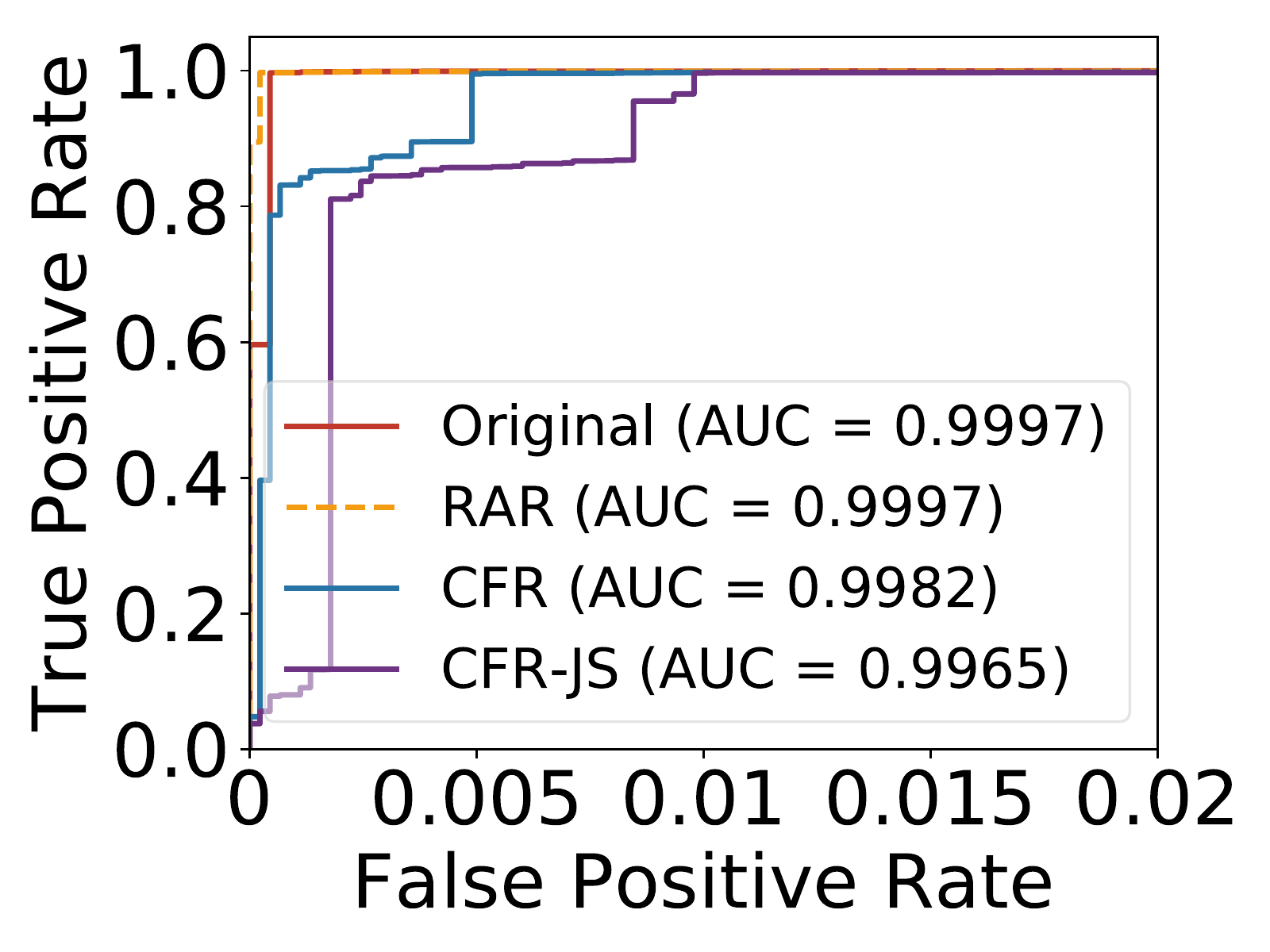}
\end{tabular}
\caption{Evasion robustness (left) and performance on non-adversarial data (right) of different variants of Hidost.}
\label{Hidost-CFR-accu}
\end{figure}

%We next conduct experiments to evaluate the robustness of Hidost using CFR, as compared to the other robust and baseline variants.
%baseline classifier, PSR and CFR.
Next, we evaluate the effectiveness of CFR for Hidost.
The results are shown in Figure \ref{Hidost-CFR-accu} (left) and are largely consistent with SL2013. 
In particular,
% we can observe that by using the modified feature-space model defined in Equation~\eqref{evasion-model-cf},
CFR boosts evasion robustness from 2\% to 100\% (slightly better than RAR), well above conventional FSR (recall Figure ~\ref{Hidost-accu} (left)).
% shown in Figure \ref{Hidost-accu} (left).
In contrast, CFR-JS only boosts robustness to 53\%, showing that our algorithmic approach can in some cases offer a considerable advantage to expert-chosen conserved features.
%, showing a substantial advantage of our algorithmic approach for identifying conserved features over features that an expert would naturally choose.
%\textcolor{red}{In contrast, evaion robustness is only boosted to 53\% with \emph{CFR-JS}, which is far below using the full set of conserved features.}
% shown in Figure~\ref{Hidost-accu}(left).

Evaluating the performance of CFR and CFR-JS on non-adversarial test data in terms of ROC curves in Figure~\ref{Hidost-CFR-accu} (right), we find that the CFR classifier can achieve $\sim99.8$\% AUC.
This is somewhat worse than RAR, particularly for very low false positive rates, but better than CFR-JS---again, in this case using the full batch of conserved features exhibits a significant advantage over solely looking for Javascript.
%\textcolor{red}{Again, we observe a negligible degradation of TPR for low FPRs compared to \emph{FSR} shown in Figure~\ref{Hidost-accu} (right).
%Figure~\ref{Hidost-CFR-accu} (right) also shows that \emph{CFR-JS} yields a significant degradation of TPR when FPR  $<0.002$.
%Hence, feature space models refined with the full collection of conserved features achieves better robustness as well as performance on non-adversarial data compared to those refined with the subset of conserved features for Hidost.  
%}
%, and a significant boost in TPR when FPR$<0.01$ compared to FSR.

%These results show that, by leveraging conserved features in feature space attack models, the performance on both adversarial and non-adversarial data can be significantly improved.

\subsubsection{Binarized PDFRate}
\begin{figure}[t]
\centering
\begin{tabular}{cc}
\includegraphics[width=0.22\textwidth]{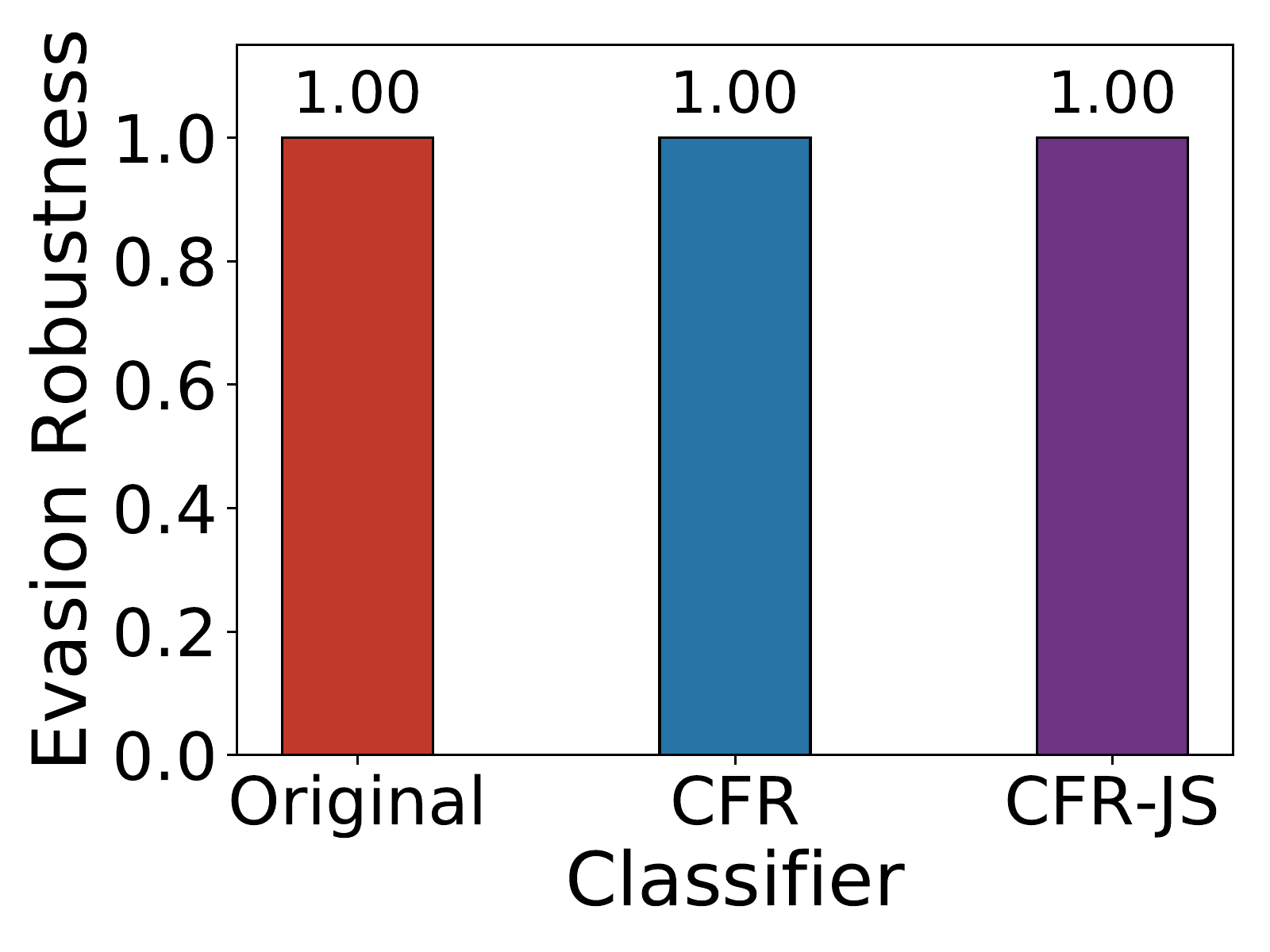}&
\includegraphics[width=0.22\textwidth]{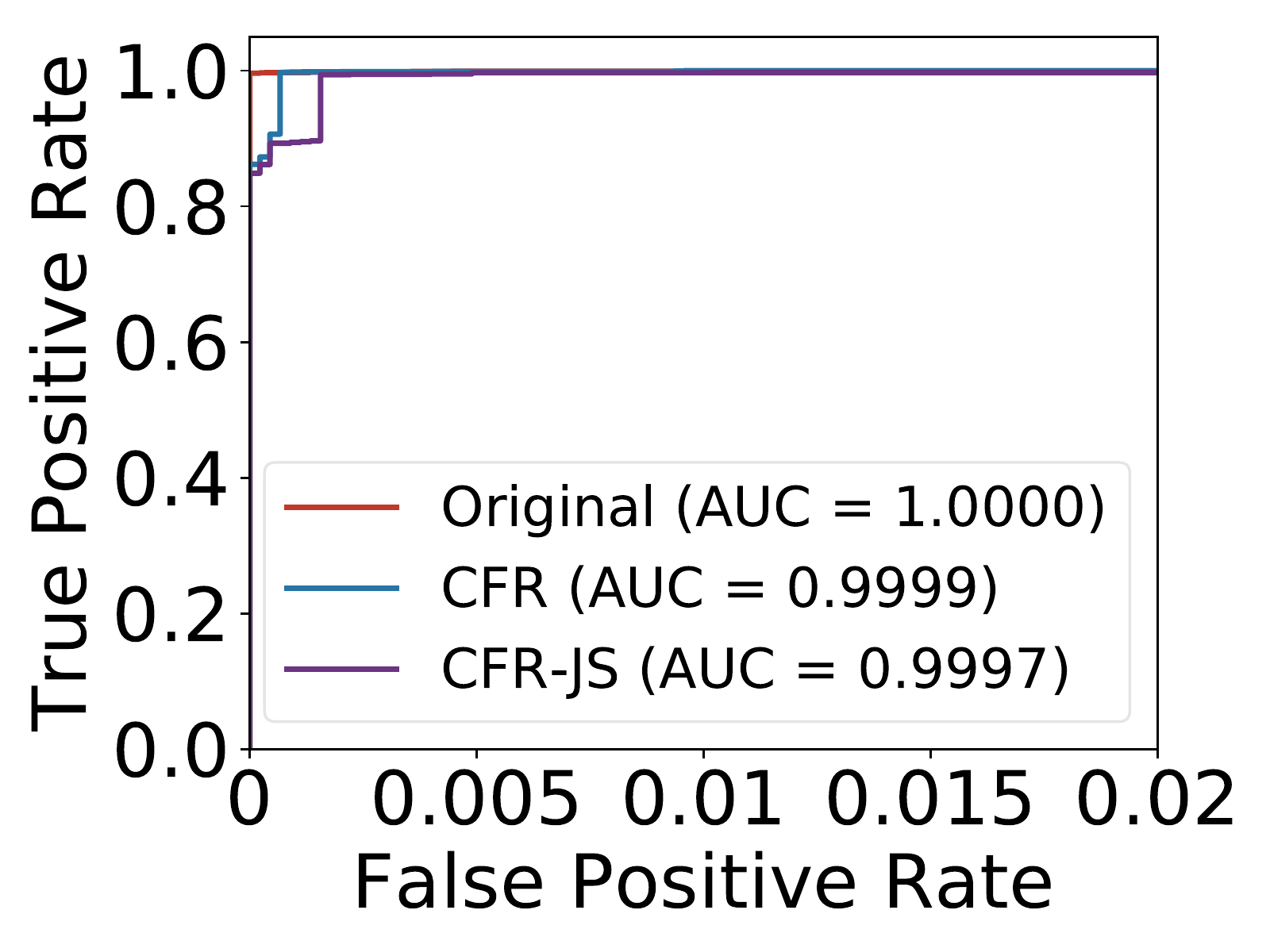}
\end{tabular}
\caption{Evasion robustness (left) and performance on non-adversarial data (right) of different variants of PDFRate-B.}
\label{PDFRate-B-CFR-accu}
\end{figure}

Finally, we evaluate the effectiveness of the CFR variants of PDFRate-B.
%retrained by the feature space attack model in Equation~\eqref{evasion-model-cf}. 
  We observe that both the \emph{CFR} and \emph{CFR-JS} classifiers in the PDFRate-B family achieve 100\% evasion robustness against EvadeML (Figure~\ref{PDFRate-B-CFR-accu} (left)), just as the RAR and FSR counterparts had.
  % investigated in Section~\ref{subsec:PDFRate-B}.
%  \begin{figure}[h]
%\centering
%\includegraphics[width=0.23\textwidth]{figures3/PDFRate-B-CFR-roc.pdf}
%\caption{ROC curves of different variants of PDFRate-B on non-adversarial test data.}
%\label{PDFRate-B-CFR-roc}
%\end{figure}

However, a close look at Figure~\ref{PDFRate-B-CFR-accu} (right) demonstrates that CFR and CFR-JS achieve far better performance on non-adversarial data, with $>$99.9\% AUC, where improvements are particularly significant for small false positive rates compared to FSR (recall Figure~\ref{PDFRate-B-accu} (right)).
Moreover, in this experiment, CFR achieves slightly higher TPR than CFR-JS for low FPR regions (below 0.003).
The main takeaway here is that although the feature-space approach already yields high robustness in this setting, introducing conserved features significantly mitigates its degradation in performance on non-adversarial data.
%that can come with improved robustness.

\section{Additional Realizable Evasion Attacks}
\label{sec:alternative}
So far we used EvadeML as the primary realizable attack in our experiments.
This choice is defensible, as EvadeML explores a significantly larger
attack space than many other evasion methods (e.g., Mimicry~\cite{oakland2014}), allowing deletions and swaps, in addition to insertions.
%is a powerful and generic attack which
%makes few artificial restrictions on the attack space, in contrast to
%mimicry attacks, which tend to focus on adding benign features to
%malicious instances.
Nevertheless, it is natural to wonder whether classifiers robust to
EvadeML remain robust to other classes of evasion attacks.
A particularly intriguing question is how the classifiers hardened
against EvadeML fare in comparison with classifiers hardened against
feature-space models, when faced with different realizable attacks.
%A particularly intriguing question is whether a classifier hardened
%against EvadeML remains effective against other attacks.
%We also wish to explore whether classifiers hardened using feature-space models of evasion attacks are robust to mimicry, whatever their effectiveness was against EvadeML.

To answer these questions, we consider \emph{five} additional realizable attacks:
\emph{Mimicry}~\cite{oakland2014}, which was one of the first realizable attacks on PDF malware detectors, 
\emph{Mimicry+}, an enhanced variant of Mimicry,
\emph{MalGAN}~\cite{Hu2017}, which uses Generative Adversarial Networks (GANs) to create evasion
attacks (but only targets binary classifiers),
\emph{Reverse Mimicry}~\cite{asiaccs2013}, which inserts malicious payloads into target benign files,
and a new custom attack aimed at defeating PDFRate-B
conserved features.
The Mimicry/Mimicy+ attacks are designed specifically for PDFRate, and cannot be usefully applied to SL2013 or Hidost, whereas the Reverse Mimicry attack and our custom attack require \emph{zero knowledge} of target classifiers.
%In addition, MalGAN attack targets only binary classifiers.

\subsection{Mimicry and Mimicry+ Attacks}

We start by considering the Mimicry and Mimicry+ attacks for both real-valued and binarized variants of PDFRate,
with the same 100 malicious seeds employed in
  Section~\ref{S:validation} and~\ref{CR} as attack files.
\begin{figure}[t]
\centering
\begin{tabular}{cc}
  \includegraphics[width=0.22\textwidth]{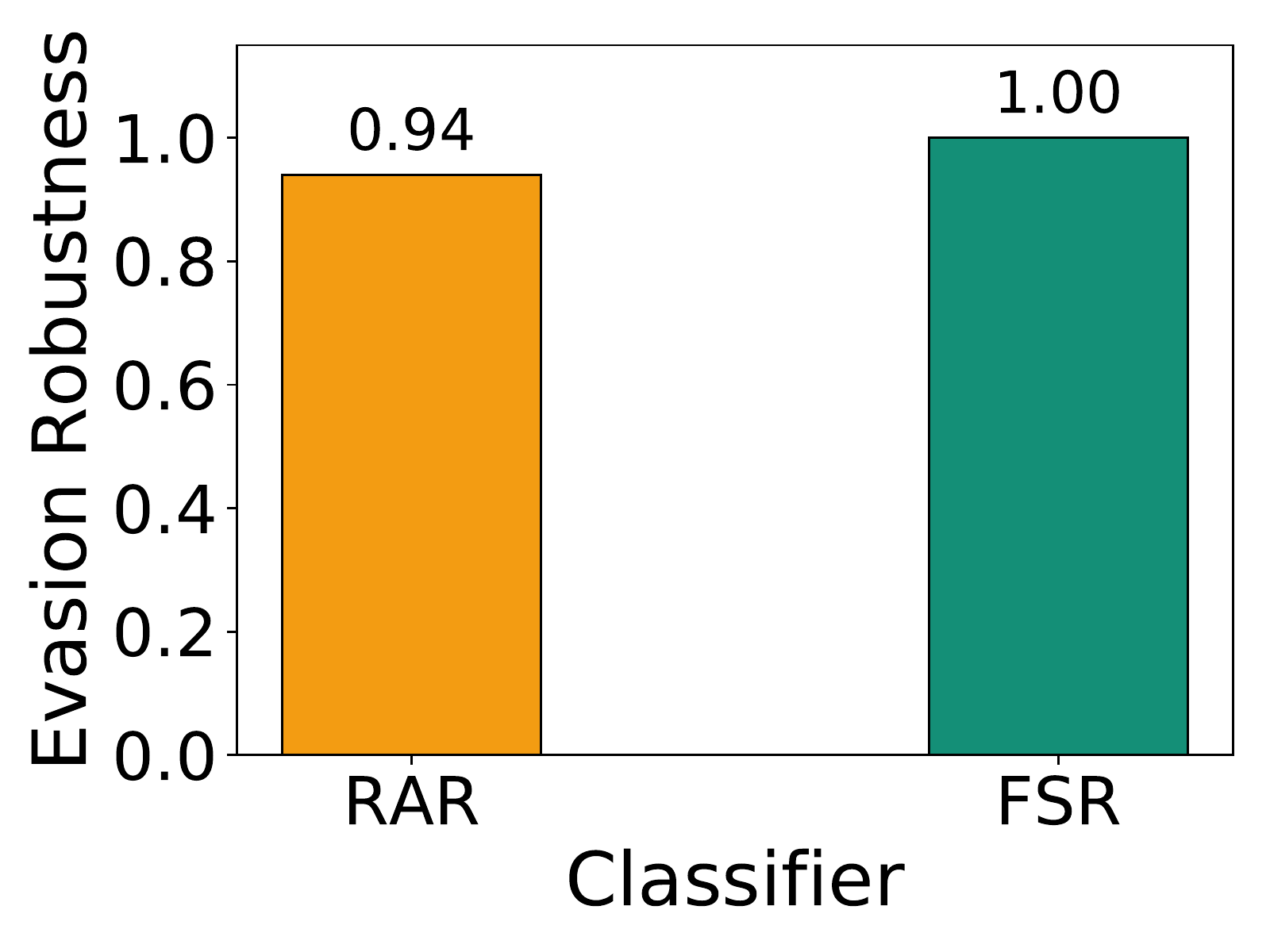} &
 \includegraphics[width=0.22\textwidth]{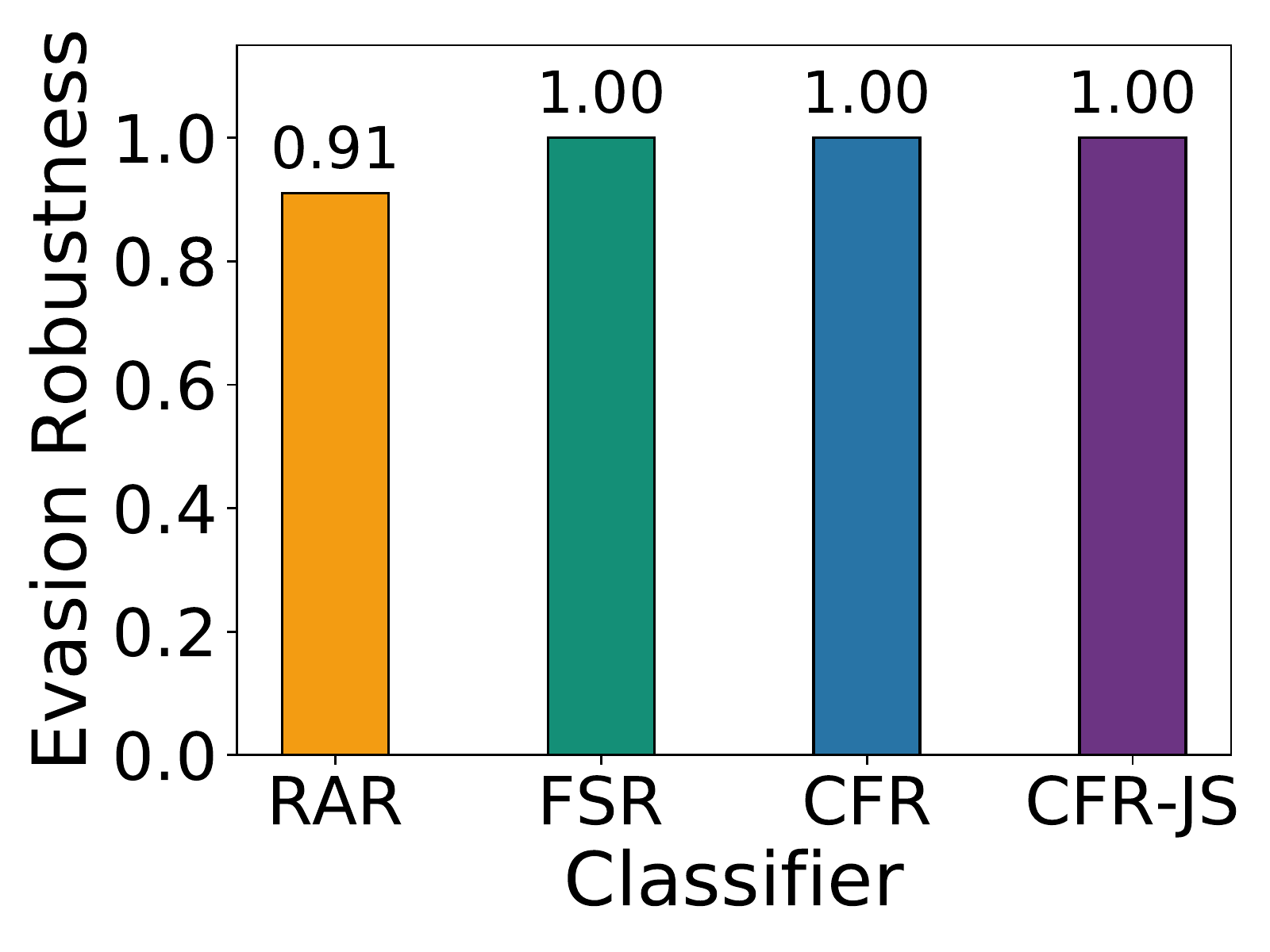}\\
\end{tabular}
	\caption{Robustness to Mimicry attack.  Left: PDFRate-R (note
          that our notion of CFR is not applicable here).  Right: PDFRate-B.}
	\label{F:mimicry}
\end{figure}

\begin{figure}[t]
\centering
\begin{tabular}{cc}
  \includegraphics[width=0.22\textwidth]{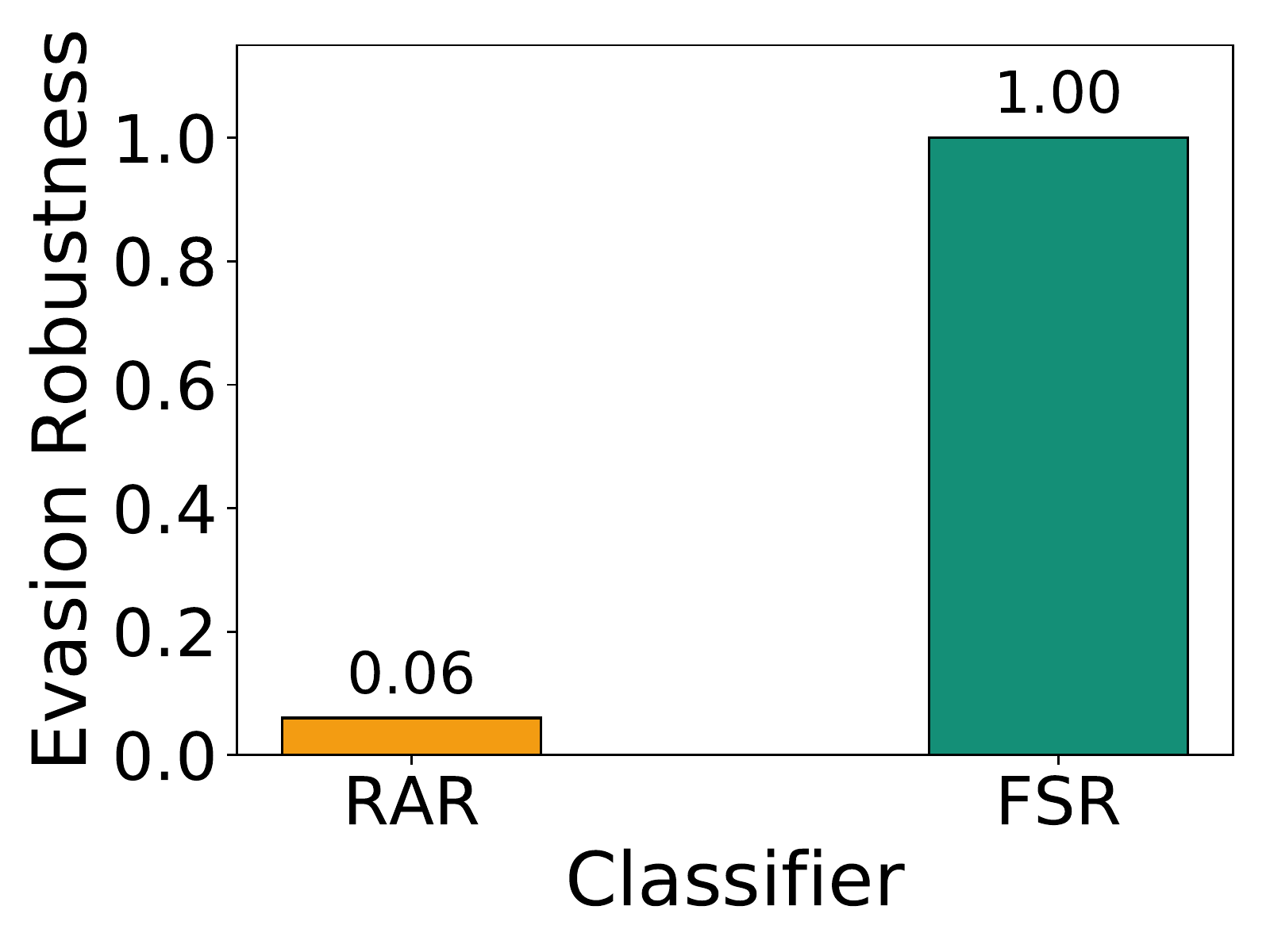} &
 \includegraphics[width=0.22\textwidth]{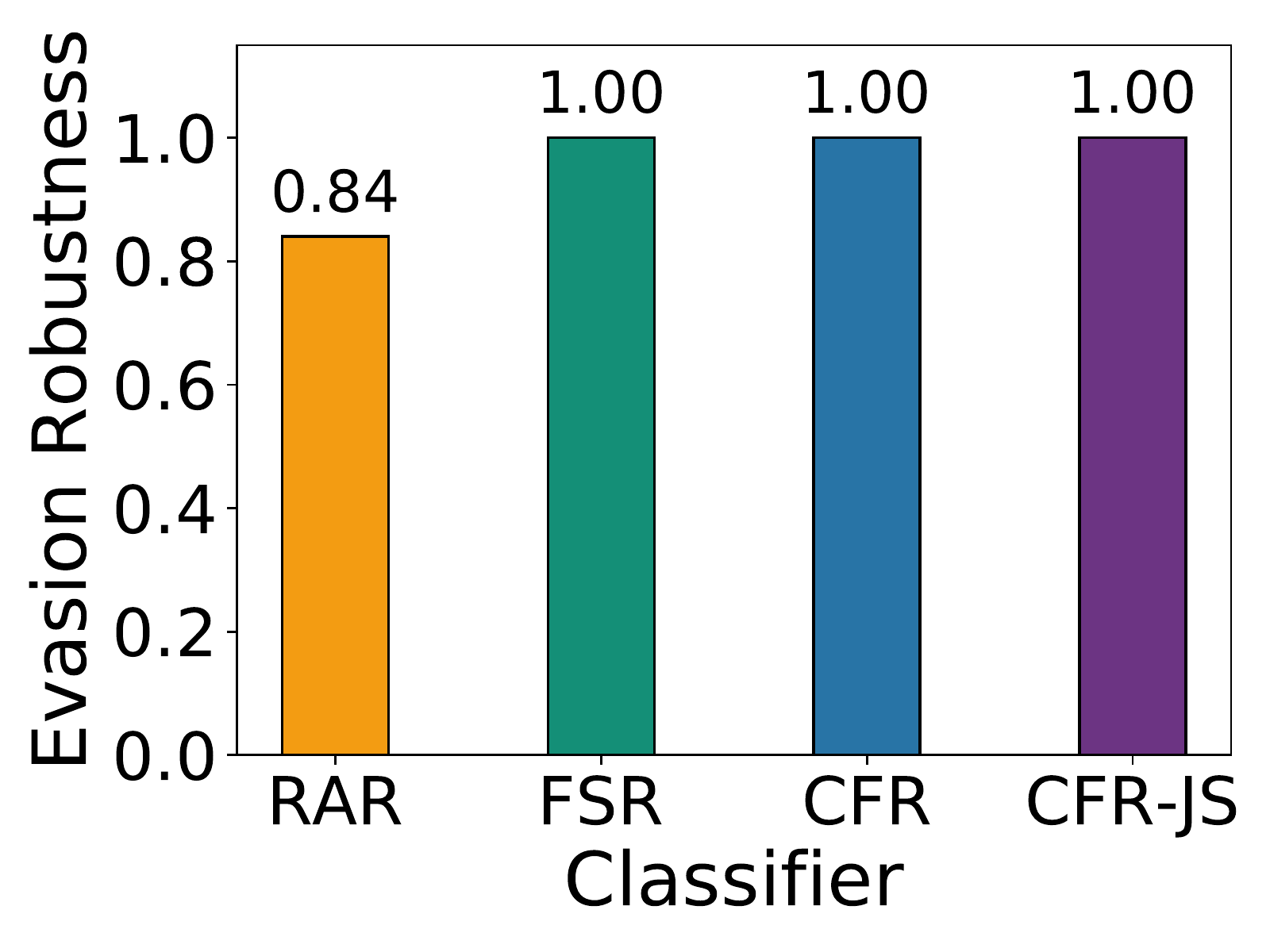}\\
\end{tabular}
	\caption{Robustness to Mimicry+ attack.  Left: PDFRate-R (note
          that our notion of CFR is not applicable here).  Right: PDFRate-B.}
	\label{F:mimicry-plus}
\end{figure}
The results are shown in Figures~\ref{F:mimicry}
and~\ref{F:mimicry-plus}, and offer two noteworthy findings.
First, as can be seen in Figure \ref{F:mimicry-plus}, RAR classifiers (hardened specifically against EvadeML, recall that the original PDFRate-B classifier is equivalent to RAR) can be quite vulnerable to the Mimicry+ attack, whereas both FSR and CFR classifiers remain robust.
Second, Mimicry+ is indeed a much stronger attack than Mimicry:
the original Mimicry fails to significantly degrade RAR performance,
whereas Mimicry+ largely evades the RAR variant of PDFRate-R, and is
somewhat more potent against PDFRate-B than Mimicry.
This demonstrates that besides its mathematical elegance, the abstract
feature-space evasion models, once appropriately anchored to the domain, are rather generally robust to evasion
attacks.
%, as they are not hardened against a specific evasion
%architecture.
%are promising: it appears that classifiers---either RAR or FSR, with and
%without conserved features---are robust to mimicry (recall that the original PDFRate-B classifier is equivalent to RAR).
%Nevertheless, FSR classifiers do appear to be slightly more robust,
%but overall it appears that EvadeML is at least as strong an attack
%as mimicry.
%It is perhaps somewhat notable that FSR/CFR are slightly more robust to this attack for both real-valued and binarized PDFRate, but the differences are relatively small.

%\textcolor{red}{
%Next, we evaluate the robustness against the Mimicry+ attack of the PDFRate variants.
%The results are shown in Figure~\ref{F:mimicry-plus}.
%We can observe that a significant degradation on evasion robustness for the classifiers hardened with RAR,  from 94\% to 6\% for PDFRate-R and from 91\% to 84\% for PDFRate-B.
%In contrast, the classifiers hardened with feature space models---with and without conserved features---are robust to the Mimicy+ attack.
%}

\subsection{MalGAN Attack}

Next, we consider the MalGAN attack on the three 
classifiers over binary feature space we have previously studied:
SL2013, Hidost, and PDFRate-B, with RAR and FSR/CFR versions that have
been shown robust to EvadeML.

\begin{figure}[t]
\centering
\begin{tabular}{cc}
  \includegraphics[width=0.22\textwidth]{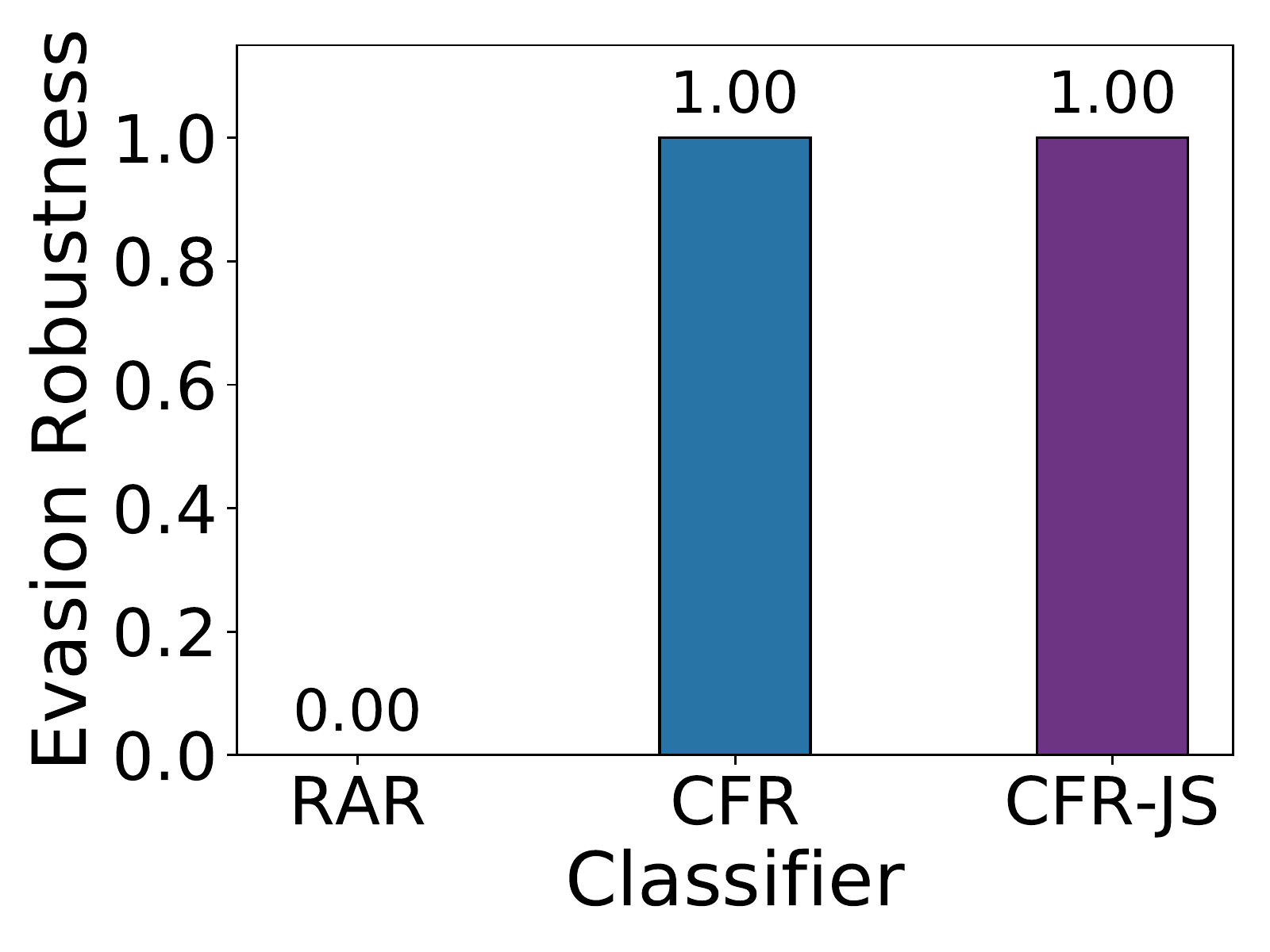} &
 \includegraphics[width=0.22\textwidth]{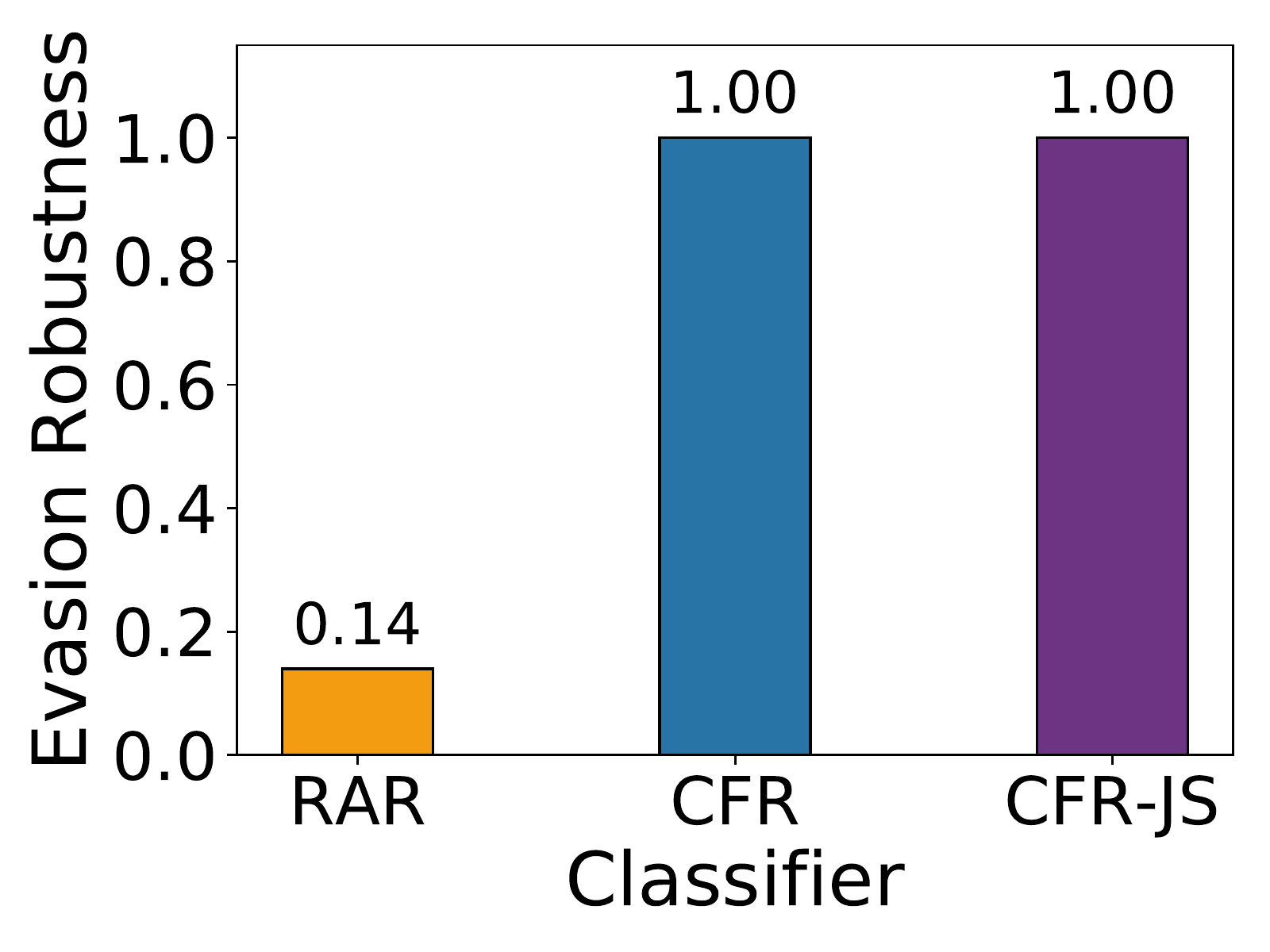}\\
\multicolumn{2}{c}{\includegraphics[width=0.22\textwidth]{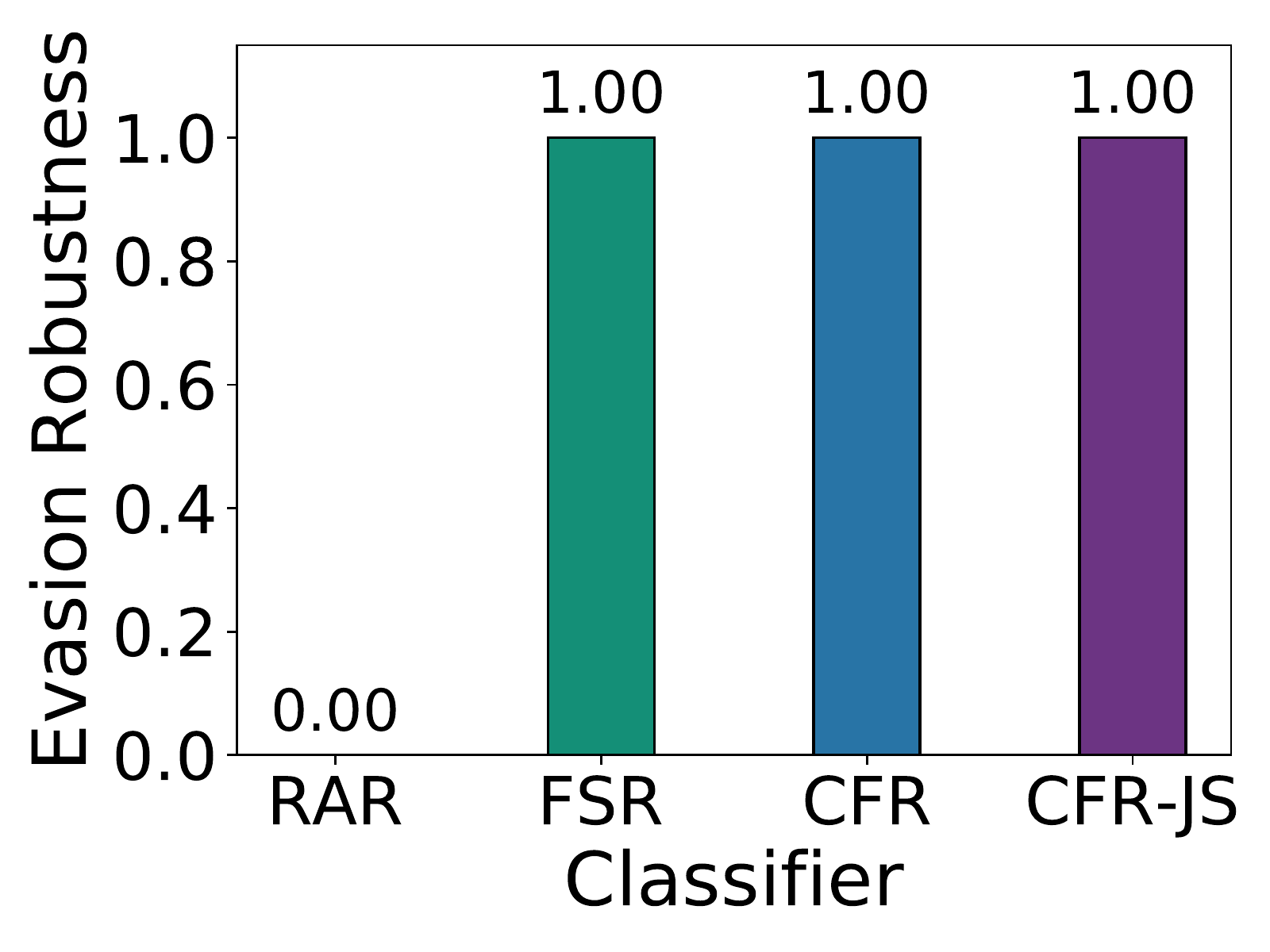}}
\end{tabular}
	\caption{Robustness to MalGAN attack. SL2013 (top left),
          Hidost (top right), PDFRate-B (bottom).}
	\label{F:malgan}
\end{figure}
The results, shown in Figure~\ref{F:malgan}, demonstrate that
%, are quite surprising.
despite EvadeML being a powerful attack, the RAR approaches which use
it for hardening (with resulting classifiers no longer very vulnerable
to EvadeML) are \emph{highly} vulnerable to MalGAN, with evasion
robustness of 0\% in most cases.
In contrast, CFR models which use conserved features remain highly
robust (100\% in all cases), just as we had observed earlier.

\subsection{Reverse Mimicry Attack}

Next, we employ the \emph{Reverse Mimicry} attack on the
EvadeML-robust variants of all the classifier types (SL2013, Hidost,
PDFRate-R, and PDFRate-B).

\begin{figure}[t]
\centering
\begin{tabular}{cc}
  \includegraphics[width=0.22\textwidth]{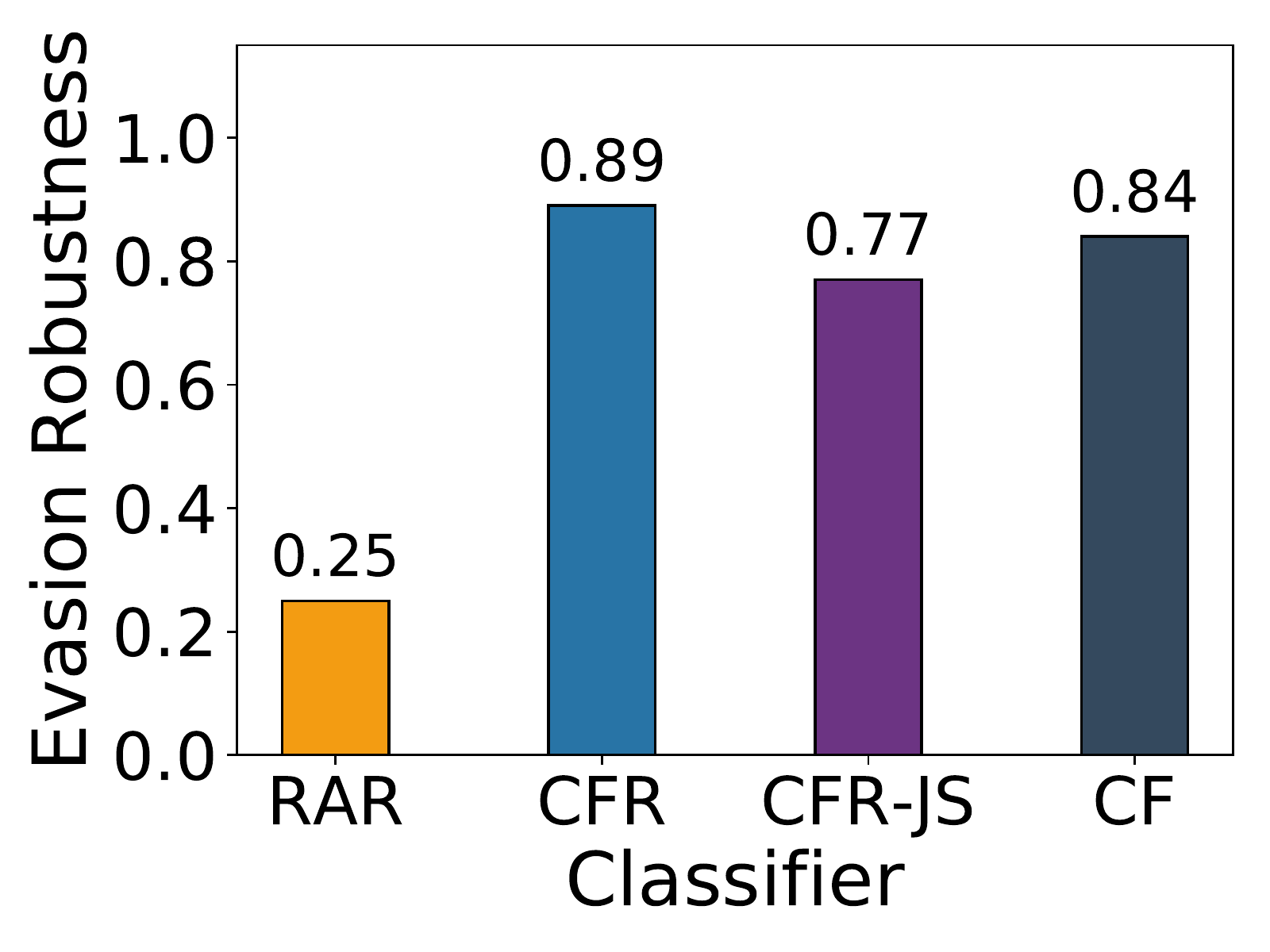} &
 \includegraphics[width=0.22\textwidth]{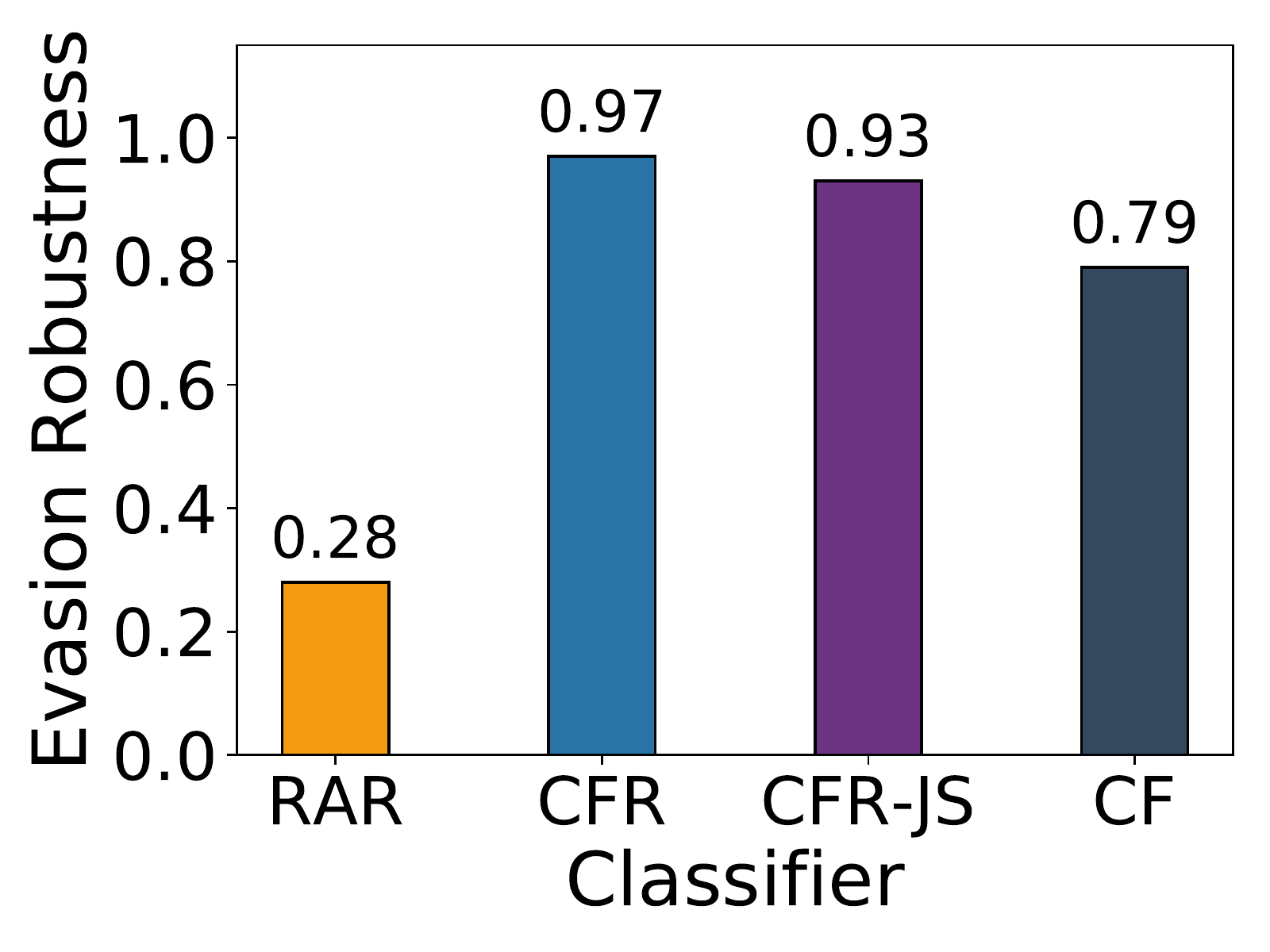}\\
   \includegraphics[width=0.22\textwidth]{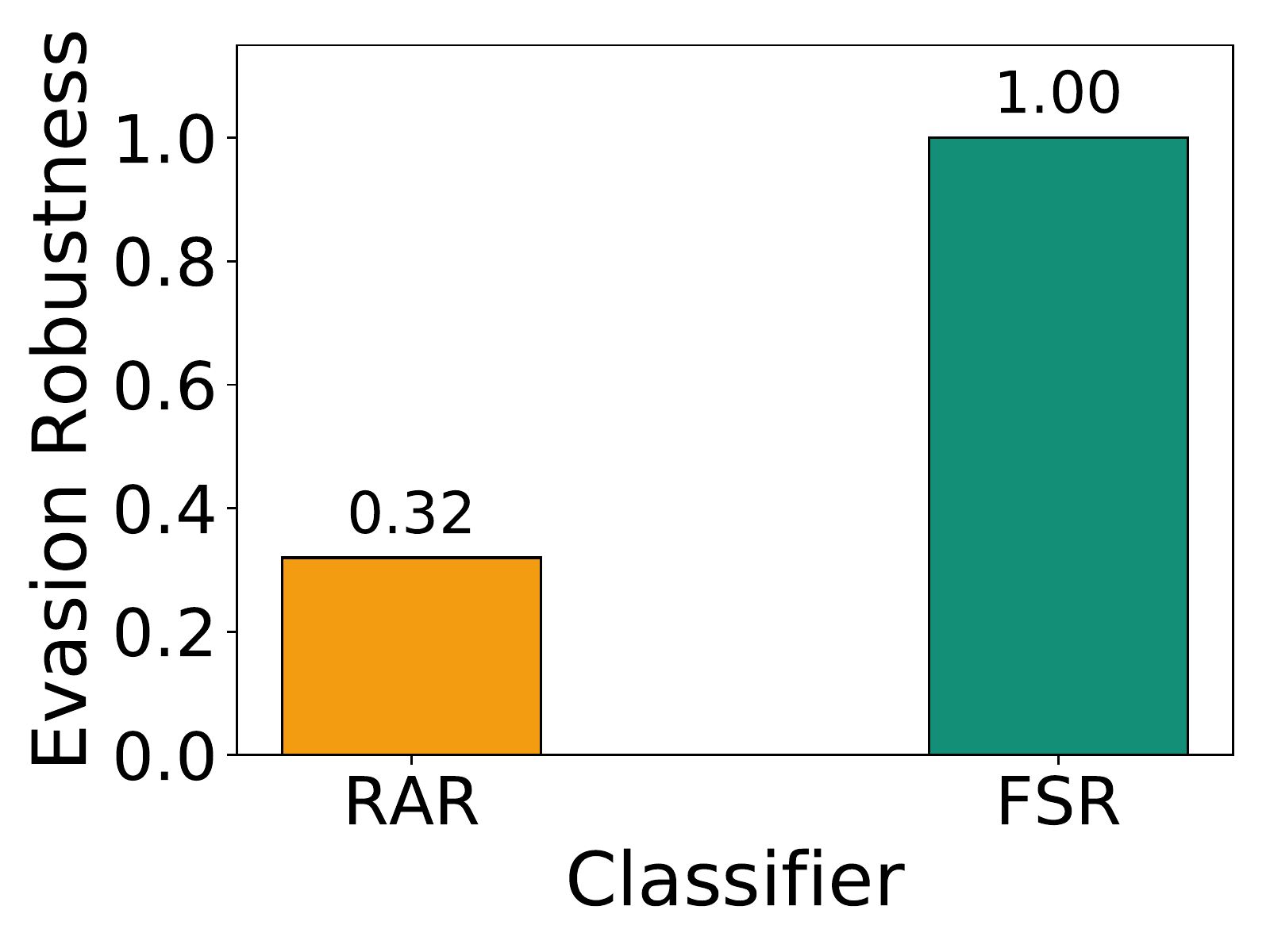} &
 \includegraphics[width=0.22\textwidth]{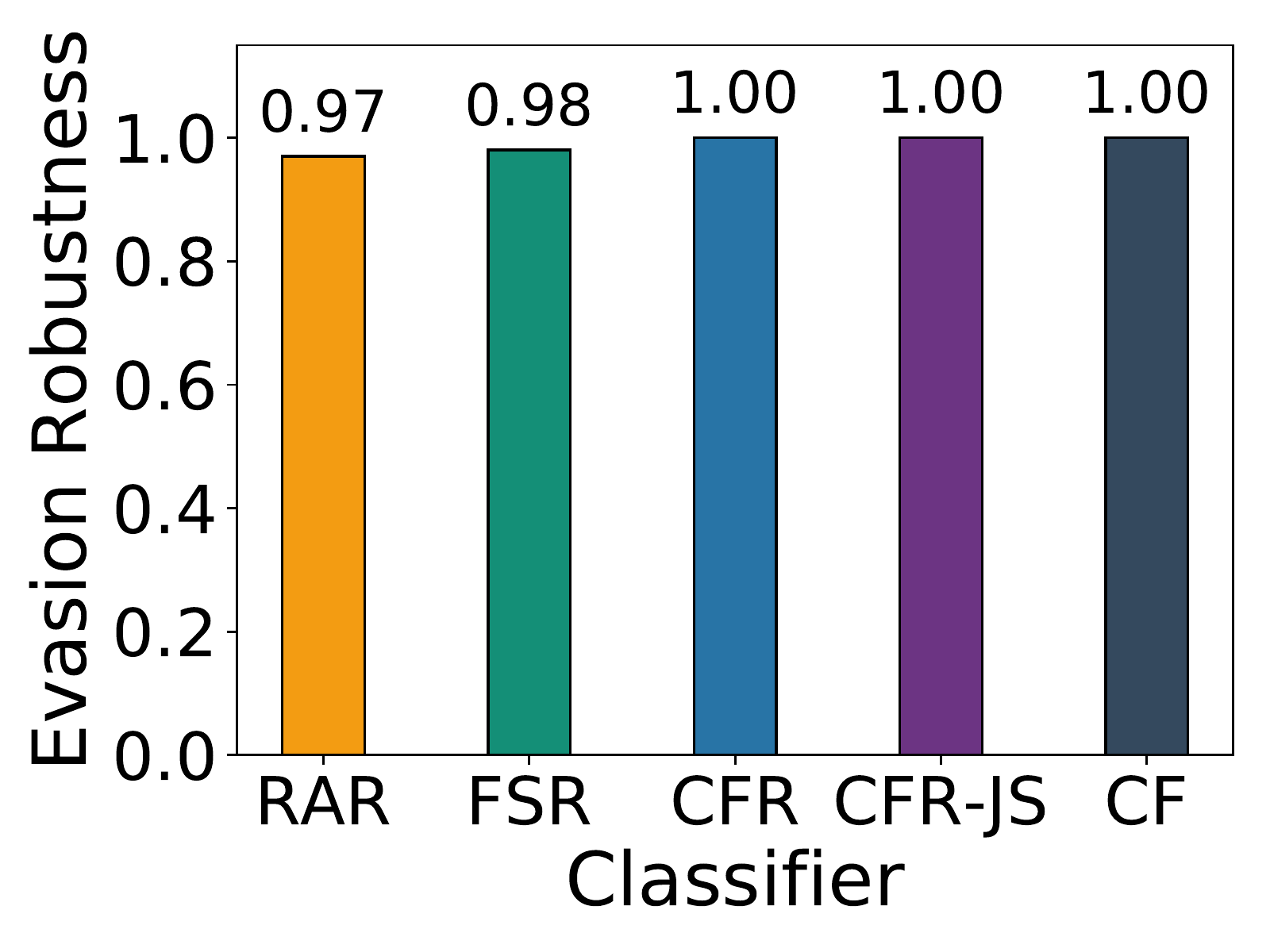} \\
\end{tabular}
	\caption{Robustness to Reverse Mimicry attack. SL2013 (top left),
          Hidost (top right), PDFRate-R (bottom left), PDFRate-B (bottom right).
          Note that our notions of CFR and CF for PDFRate-R is not applicable here.}
	\label{F:reverse-mimicry}
\end{figure}

%\begin{table}[h]
%\centering
%\scalebox{0.8}{
%
%\begin{tabular}{|c|c|c|c|c|c|}
%\hline
%\multirow{2}{*}{Classifier type} & \multicolumn{5}{c|}{Variant}       \\ \cline{2-6} 
%                            & RAR  & FSR  & CFR  & CFR-JS & CF   \\ \hline\hline
%SL2013                      & 0.25 & 0.69 & \textbf{0.89} & 0.77   & 0.84 \\ \hline
%Hidost                      & 0.28 & 0.47 & \textbf{0.97} & 0.93   & 0.79 \\ \hline
%PDFRate-R                   & 0.32 & 1.0  & -    & -      & -    \\ \hline
%PDFRate-B                   & 0.97 & 0.98  & 1.0  & 1.0    & 1.0  \\ \hline
%\end{tabular}
%}
%\caption{Evasion robustness to the Reverse Mimicry attack.}
%\label{table:reverse-mimicry}
%\end{table}
%
%\vspace{-0.1in}

Figure~\ref{F:reverse-mimicry} presents the results, which are revealing in several ways.
First, we again observe that RAR (hardened specifically against EvadeML) is roundly defeated in most cases.
%Second, consider the last column of the table, which provides robustness results for the classifier using only the conserved features (CF).
Second, consider the robustness results for the classifier using only the conserved features (CF), we can see that reverse mimicry succeeds in defeating conserved features for a non-trivial proportion of instances.
It does so by including Javascript tags in structural paths that are not used as features by SL2013/Hidost (since these classifiers only consider commonly occurring sets of structural paths).
%(which only consider a commonly
%occurring subset of
%possible structural paths).
Thus, this attack reveals an important vulnerability in the feature extraction approach employed by these classifiers; indeed, it suggests that structure-based classifiers may be \emph{inherently} difficult to harden.
Remarkably, CFR remains more robust than CF despite these vulnerabilities.
The case of Hidost is particularly stark: CFR is nearly 20\% more robust than CF!

%\begin{table}[h]
%\centering
%\begin{tabular}{|c|c|c|c|c|c|}
%\hline
%\multirow{2}{*}{Classifier type} & \multicolumn{5}{c|}{Variant}       \\ \cline{2-6} 
%                            & RAR  & FSR  & CFR  & CFR-JS & CF   \\ \hline\hline
%SL2013                      & 0.25 & 0.69 & \textbf{0.89} & 0.77   & 0.84 \\ \hline
%Hidost                      & 0.28 & 0.47 & \textbf{0.97} & 0.93   & 0.79 \\ \hline
%PDFRate-R                   & 0.32 & 1.0  & -    & -      & -    \\ \hline
%PDFRate-B                   & 0.97 & 1.0  & 1.0  & 1.0    & 1.0  \\ \hline
%\end{tabular}
%\caption{Evasion robustness to the Reverse Mimicry attack.}
%\label{table:reverse-mimicry}
%\end{table}

%\textcolor{red}{
%that boosts robustness to EvadeML.
%The results shown in Table~\ref{table:reverse-mimicry} indicate that most variants of the structure-based classifiers (SL2013 and Hidost), regardless of their robustness to EvadeML, are highly vulnerable to the Reverse Mimicry attack.
%In contrast, the CFR models achieve the highest robustness ($>89\%$) among all the variants.
%In other words, CFR enables generalized robustness to its counterparts (including those CF classifiers which only use conserved features).
%For content-based classifiers (PDFRate-R and PDFRate-B), we observe a boost of robustness from RAR to FSR/CFR, which is consistent with the results for Mimicry and MalGAN attacks. 
%}

\subsection{The Custom Attack}

Our final attack specifically targets a feature extraction bug in the
Mimicus implementation of PDFRate in order to defeat the corresponding
CF classifier.

\begin{figure}[t]
\centering
\begin{tabular}{cc}
  \includegraphics[width=0.22\textwidth]{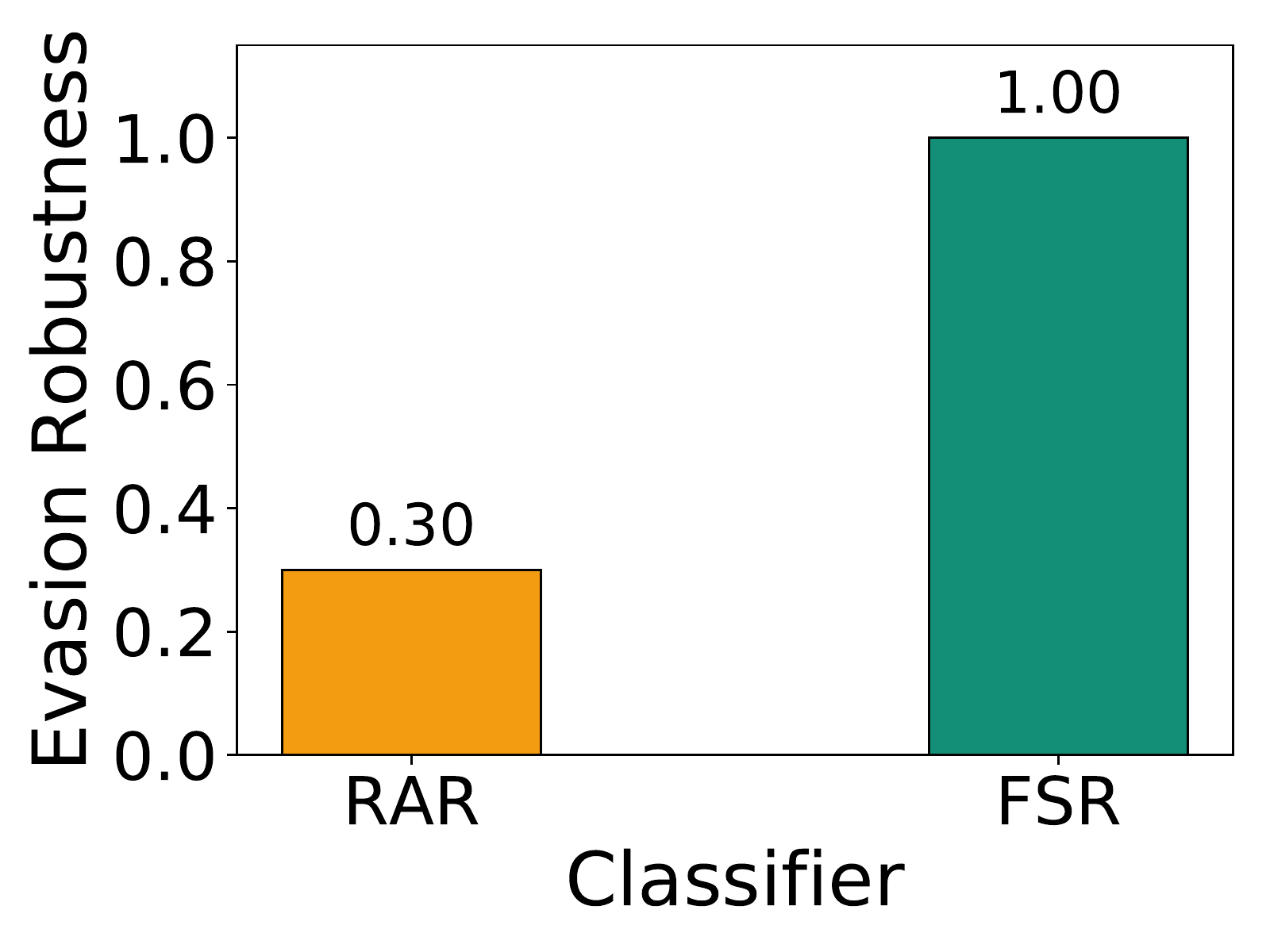} &
 \includegraphics[width=0.22\textwidth]{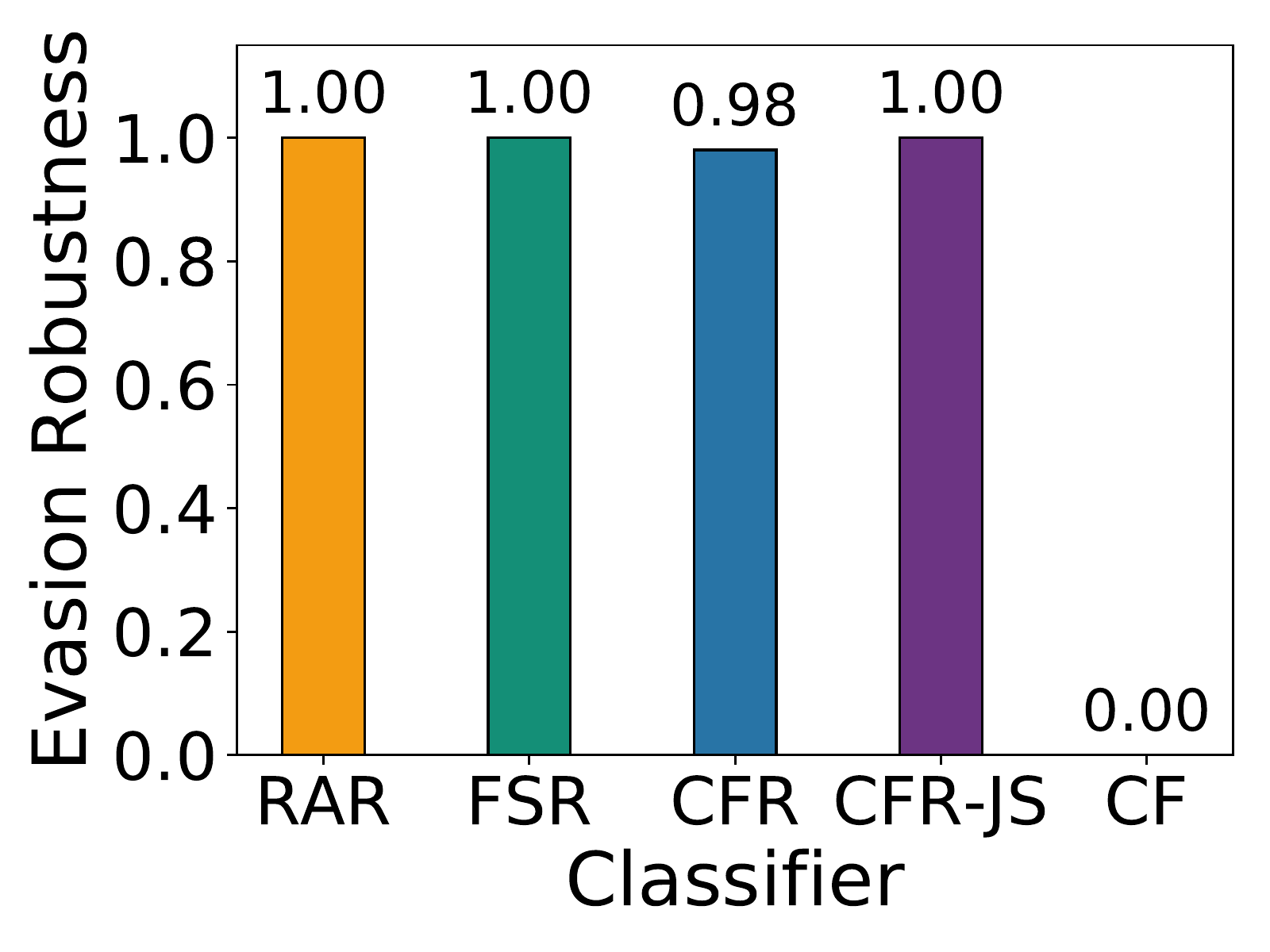}\\
\end{tabular}
	\caption{Robustness to the custom attack.  Left: PDFRate-R (note
          that our notions of CFR and CF are not applicable here).  Right: PDFRate-B.}
	\label{F:custom}
\end{figure}

The results are shown in Figure~\ref{F:custom}.
We find that after this attack, CF robustness is 0.
We also observe that the robustness of RAR classifier for PDFRate-R
also drops, although to 0.3 rather than 0.
Significantly, the FSR classifiers for both PDFRate-R and PDFRate-B
remain 100\% robust, and the CFR variant of PDFRate-B has nearly
perfect robustness (0.98) against this attack.
Our latter observation is particularly remarkable: although the
conserved features are roundly defeated by this attack, the use of
these as a part of a holistic retraining approach yields a classifier
that remains robust.
Thus, not only is it possible to construct a robust
malware classifier without unduly relying on conserved features, but
we can accomplish this through iterative retraining in feature space.
\section{Related Work}
\label{related-work}

%\subsection{Realizable Methods for Classifier Evasion}

%\noindent{\bf Evasion Attacks: }

Below we briefly describe some of the related literature on adversarial evasion or adversarial example attacks and defenses; we refer readers to Vorobeychik and Kantarcioglu~\cite{Vorobeychik18book} for a broader and more in-depth treatment of the subject of ML attacks and defenses.

%\vspace{-0.1in}
\noindent{\bf Evasion and Adversarial Example Attacks:} 
An early realizable evasion attack on machine learning was devised by Fogla et al.~\cite{Fogla06,Fogla06b}, who developed an attack on anomaly-based intrusion detection systems.
%Fogla et al. subsequently generalize and systematize the polymorphic blending attack~\cite{Fogla06b}.
{\v{S}}rndic and Laskov~\cite{oakland2014} present a case study of an evasion attack on a state-of-the-art PDF malware classifier, PDFRate.
%Their mimicry attack, which primarily adds content to a PDF to make it appear as benign as possible, actually leverages an initial feature-space evasion, which is subsequently modified to effect the actual PDF source.
Xu et al.~\cite{ndss2016} propose EvadeML, a fully realizable attack on PDF malware classifiers which generates evasion instances by using genetic programming to modify PDF source directly, using a sandbox to ensure that malicious functionality is preserved.
Grosse et al.~\cite{Grosse17} develop a method for generating evasion attacks against a deep learning-based Android malware classifier, using a gradient-based approach which is also a form of iterative improvement heuristics, but chooses the best coordinate to improve in each iteration as evaluated by the gradient, rather than random coordinate as in our case.
Their approach likely requires fewer steps than coordinate greedy, but since we run coordinate greedy until convergence, this difference isn't important in our study.
Moreover, we also optimize among several local optima through random restarts, which is likely to obtain better evasion solutions (Grosse et al.~\cite{Grosse17} stop as soon as an evasion is found, rather than trying to identify the most benign looking malware).
This particular attack can be viewed as realizable, even though it wasn't implemented and evaluated in actual malware, since the attack space is significantly restricted to only add features that do not interfere with others already present. 
Similarly, MalGAN, an evasion attack based on generative adversarial networks developed by Hu and Tan~\cite{Hu2017}, only adds features from benign to malicious malware, and we treat it as a realizable attack (since it's not difficult to implement).
%The fully automated nature of EvadeML makes it a natural candidate for our in-depth exploration of the relationship between realizable and feature space evasion attacks.
%These are in a sense both problem and feature space, as actual pixels of the images are modified in the so-called \emph{adversarial examples} which are crafted to effect errors in deep classifiers.
%Several efforts explored the impact of indirect modification when an adversarial image must be printed, showing that effectiveness of such techniques can nevertheless be preserved~\cite{CCS2016,iclr2017}.
%Recently, an approach for printing specifically designed glass frames had been shown to mislead vision-based biometric systems to either mistakenly grant authorization, or enable evasion of face recognition techniques~\cite{CCS2016}.
%In addition to classifier evasion methods which change the actual malicious instances, a number of techniques have sprouted for evasion models acting directly on features~\cite{kdd2004,kdd2005,asiaccs2006,icml2016}.
%\noindent{\bf Feature Space Attacks: }

In addition to classifier evasion methods which change the actual malicious instances (or are relatively direct to implement as such), 
a number of techniques have sprouted for modeling adversarial examples in feature space~\cite{Athalye18,Carlini17,iclr2015,pkdd2013,kdd2004,kdd2005,asiaccs2006,icml2016,Biggio13,Biggio15,icml2016,nips2014,li2016,Nelson12,Vorobeychik14}.
Moreover, a series of efforts explore evasion in the context of image classification by deep neural networks~\cite{iclr2015,iclr2016,papernot2016,CCS2016}, although Gilmer et al.~\cite{Gilmer2018} question the common threat models used in these works.
Several recent approaches attempt to generate adversarial examples against computer vision systems in physical space, such as adding stickers to a stop sign to cause misclassification, or wearing printed glass frames to fool face recognition, and are therefore somewhat analogous to our notion of realizable attacks~\cite{CCS2016,Evtimov18}.

%\vspace{-0.1in}
\noindent{\bf Evasion-Robust Classification:}
Dalvi et al.~\cite{kdd2004} presented the first approach for evasion-robust classification.
%, making use of a model in which the attacker aims to transform feature vectors into benign instances in response to the original (non-robust) classifier, and then subsequently devising an approach which is robust to the former attack.
A series of approaches formulate robust classification as minimizing maximum loss (i.e., following a robust optimization paradigm), where maximization is attributed to the evading attacker aiming to maximize the learner's loss through small feature-space transformations~\cite{Teo07,kdd2012,Madry18,Raghunathan18,Wong18}.
%All of these effectively assume that the interaction between the learner and attacker is zero sum.
A number of alternative methods for designing classifiers 
%robust to evasion relax the somewhat unreasonable assumption that the adversary aims to maximize the defender's loss.
%Instead, these 
consider the interaction as a non-zero-sum game~\cite{Bruckner12,kdd2011,nips2014, li2016,icml2016}.
%either played simultaneously between the learner and the attacker, or a Stackelberg game, in which the learner is the leader, while the attacker the follower~\cite{kdd2011,nips2014,icml2016,aisec2016}.
Finally, a series of iterative retraining procedures have been proposed, both for general adversarial evasion~\cite{li2016,icml2016}, and specifically for deep learning methods for vision~\cite{iclr2015,iclr2016,Madry18} (note that Madry et al.~\cite{Madry18} fall into both robust optimization and retraining buckets, since their approach is equivalent to retraining if stochastic gradient descent simply continues by processing adversarial examples as they are added).
%In the deep learning literature involving adversarial manipulations of images, several procedures for retraining the classifier to boost its robustness have been proposed~\cite{iclr2015,iclr2016}, and these have been adapted to other classification models, such as decision tree classifiers~\cite{icml2016}.
%Recently, a systematic iterative retraining procedure had been proposed which leverages general-purpose adversarial evasion models, and offers a theoretical connection to an underlying Stackelberg game played between the learner and the evading adversary~\cite{li2016}.
These diverse efforts share one common property: attack models that they leverage use feature-space manipulations, which are only a proxy for realizable attacks on ML.

\section{Discussion and Conclusion}
\label{conclusion}

We undertook an extensive exploration of the extent to which robust ML that uses the conventional feature-space models of evasion attacks remains robust to ``real'' attacks that can be implemented in actual malware and preserve malicious functionality (what we call realizable attacks).
%Having first developed a general framework for systematically exploring such questions, we proceed to make a series of findings.
%, many quite surprising.
Our first intriguing observation is that defense based on feature-space models can fail to achieve satisfactory robustness.
This in itself raises some doubts about the nearly universal focus on such models as a means for ML defense, and suggests that practical usefulness of such approaches cannot be taken for granted.
However, we also show that changing the nature of the feature space can make a difference: robust ML with feature-space models is quite robust in content-based detection (which uses content, rather than structural paths, as features).
Additionally, we presented a refined version of the feature-space model that makes use of conserved features (which we can identify automatically, as shown in the Appendix), and showed that where feature-space defense previously failed, it now succeeds.
Our final finding may well be the most intriguing: feature-space approaches exhibit generalized robustness, in that the resulting robust ML (after appropriate refinement using conserved features) exhibits robustness to multiple realizable attacks.
This contrasts with defense that is hardened using a \emph{specific} realizable attack---even one quite powerful on the surface (EvadeML)---which can fail dramatically when faced with a different attack.
These findings demonstrate the power of effective mathematical abstractions in security.

It is natural to wonder how our approach and results are applied to other domains.  In computer vision, the analog of realizable malware attacks are physical attacks, whereby the physical environment is modified, rather than the digital object, such as an image.
Here, the corresponding foundational question is whether common robust ML methods based on small-$l_p$ attacks successfully protect against physical attacks.
The notion of conserved features can also be seen as more generally applicable.
For example, in a bag-of-words representation for spam filtering, these could correspond to the existence of URL or file attachments, and in SQL injection attacks, these may refer to the existence of specific SQL commands, such as \verb|Select|.

The main limitation of our study is in the specific choices we had to make to ensure that it is tractable.
We chose a particular defensive paradigm---iterative retraining.
As we have argued, it is the only paradigm that can fit every case that we investigate; for example, there is no other general approach for learning a robust SVM with non-linear kernels.
However, it is possible that approaches based on robust optimization, if they were developed, can improve performance by taking advantage of the special structure of this problem.
We implemented a particular class of feature-space attacks, using $l_2$ norm to measure the attacker's cost of feature manipulations, and stochastic local search to compute evasions.
It is possible that better attack algorithms for generating attacks over binary domains will be developed, and, indeed, some alternatives exist.
However, prior work suggests that this approach yields attacks that are close to optimal~\cite{li2016}, with the use of random restarts playing a crucial role.
Finally, our study was specific to PDF malware detection.
However, our framework is quite general, and could be used in the future to consider other similar questions, such as the effectiveness of robust deep learning against physical attacks. 
Several additional limitations offer further opportunities for future work.
One example is the fact that we only define conserved features when these are binary; it may be that finding meaningful conserved features in continuous feature spaces is inherently more difficult.
Another issue is the surprising finding that sufficient anchoring of feature-space defense in the domain using conserved features allows us to achieve robustness, \emph{even when conserved features can be circumvented}.
It may be that conserved features are ultimately only a part of the solution, and only help if they adequately capture the attack surface in the abstract feature space.
The extent to which small variations in the set of identified conserved features matters is also an open question: our evidence is mixed, with ``expert''-defined features usually, but not always, sufficient for robustness.

\section*{Acknowledgments}
This work was partially supported by the Army Research Office (W911NF1610069) and NSF CAREER award (IIS-1649972).

\appendix
\section*{Appendix}

\section{Identifying Conserved Features}
\label{sec:CF}

%Having demonstrated the effectiveness of conserved features in bridging the gap between realizable and feature space evasion models, 
We now describe a systematic automated procedure for identifying these. 
%\textcolor{red}{
We first introduce how to identify conserved features of SL2013, and then describe how to generalize the approach to extract conserved features of Hidost.
%present how these features are further employed to identify conserved features of other classifiers.  
%}

The key to identifying the conserved features of a malicious PDF is to discriminate them from non-conserved ones.
Since merely applying statistical approaches on training data is insufficient to discriminate between these two classes of features, as demonstrated above, we need a qualitatively different approach which relies on the nature of evasions (as implemented in EvadeML) and the sandbox (which determines whether malicious functionality is preserved) to identify features that are conserved.
%Instead, we use a systematic approach to identify conserved features. 

We use a modified version of pdfrw \cite{pdfrw}\footnote{The modified version is available at \url{https://github.com/mzweilin/pdfrw}.} to parse the objects of PDF file 
%with a logic structure as shown on the right-hand side of Figure \ref{pdf-structure} 
and repack them to produce a new PDF file. We use Cuckoo \cite{Guarnieri} as the sandbox to evaluate malicious functionality. 
In the discussion below, we define $x_i$ to be the malicious file, $\mathsf{S}_i$ the conserved feature set of $x_i$, and $\mathsf{O}_i$ the set of its non-conserved features. 
Initially, $\mathsf{S}_i = \mathsf{O}_i = \emptyset$.

At the high level, our first step is to sequentially delete each object of a malicious file and eliminate non-conserved features by evaluating the existence of a malware signature in a sandbox for each resulting PDF, which provides a preliminary set of conserved features. Then, we replace the object of each corresponding structural path in the resulting preliminary set with an external benign object and assess the corresponding functionality, which allows us to further prune non-conserved features. 
Next, we describe these procedures in detail.

\subsection{Structural Path Deletion}

In the first step, we filter out non-conserved features by deleting each object and its corresponding structural path, and then checking whether this eliminates malicious functionality (and should therefore be conserved). First, we obtain all the structural paths (objects) by parsing a PDF file. These objects are organized as a tree-topology and are sequentially deleted. Each time an object is removed, we produce a resulting PDF file by repacking the remaining objects. Then, we employ the sandbox to detect malicious functionality of the PDF after the object deletion. If any malware signature is captured, the corresponding structural path of the object is deleted as a non-conserved feature, and added to $\mathsf{O}_i$. On the other hand, if no malware signature is detected, the corresponding feature is added in $\mathsf{S}_i$ as a \emph{possibly} conserved feature.

One important challenge in this process is that features are not necessarily independent.
Thus, in addition to identifying $\mathsf{S}_i$ and $\mathsf{O}_i$, we explore \emph{interdependence} between features by deleting objects. 
As the logic structure of a PDF file is with a tree-topology, the presence of some structural path depends on the presence of other structural paths whose object refers to the object of the prior one. 
%For example in some PDFs, the presence of \verb|/OpenAction/JS| and \verb|/OpenAction/S| depend on \verb|/OpenAction|. If the object \verb|OpenAction| is deleted, so as its structural path \verb|/OpenAction|, then both \verb|/OpenAction/JS| and \verb|/OpenAction/S| are deleted as well. We call \verb|/OpenAction/JS| and \verb|/OpenAction/S| as the dependents of \verb|/OpenAction|. 
We define that a structural path is a dependent of another if unilateral deleting the object associated with the latter causes a flip from 1 to 0 on the feature value of the former.
For any feature $j$ of $x_i$, we denote the set of features that depend on $j$ by $\mathsf{D}_i^{j}$.
Note that for a given structural path (feature), there could be multiple corresponding PDF objects. In such a case, these objects are deleted simultaneously, so as the corresponding feature value is shifted from 1 to 0. 

\subsection{Structural Path Replacement}

In the second step, we subtract the remaining non-conserved features in the preliminary $\mathsf{S}_i$ and move them to $\mathsf{O}_i$. Similar to the prior step, we first obtain all the structural paths and objects of the malicious PDF file. Then for each object of the PDF that is in $\mathsf{S}_i$, we replace it with an external object from a benign PDF file and produce the resulting PDF, which is further evaluated in the sandbox. If the sandbox detects any malware signature, then the corresponding structural path of the object replaced is moved from $\mathsf{S}_i$ to $\mathsf{O}_i$. Otherwise, the structural path is a conserved feature since both deletion and replacement of the corresponding object removes the malicious functionality of the PDF file. Note that in the case of multiple corresponding and identical objects of a structural path, all of these objects are replaced simultaneously.

After structural path deletion and replacement, for each malicious PDF file $x_i$, we can get its conserved feature set $\mathsf{S}_i$, non-conserved feature set $\mathsf{O}_i$, and dependent feature set $\mathsf{D}_j$ for any feature $j \in \mathsf{S}_i  \cup \mathsf{O}_i$, which could be further leveraged to design evasion-robust classifiers.  

\subsection{Obtaining a Uniform Conserved Feature Set}

The systematic approach discussed above provides a conserved feature set for each malicious seed to retrain a classifier. 
Our goal, however, is to identify a single set of conserved features which is \emph{independent} of the specific malicious PDF seed file.
We now develop an approach for transforming a collection of $\mathsf{S}_i$, $\mathsf{O}_i$, and $\mathsf{D}_i^{j}$ for a set of malicious seeds $i$ into a \emph{uniform} set of conserved features.

\renewcommand{\algorithmicrequire}{ \textbf{Input:}} %Use Input in the format of Algorithm  
\renewcommand{\algorithmicensure}{ \textbf{Output:}} %UseOutput in the format of Algorithm 

Obtaining a uniform set of conserved features faces two challenges: 1) minimizing conflicts among different conserved features, as a conserved feature for one malicious instance could be a non-conserved feature for another, and 2) abiding by feature interdependence if a conserved feature should be further eliminated. 

\begin{algorithm}     
\begin{algorithmic}[1]  
\REQUIRE ~~\\
The set of conserved features for $x_i(i \in [1,n])$, $\mathsf{S}_i$;\\  
The set of non-conserved features for $x_i(i \in [1,n])$, $\mathsf{O}_i$;\\
The set of dependent features for $j \in \mathsf{S}_i \cup \mathsf{O}_i$ , $\mathsf{D}_i^j$; 
\ENSURE ~~\\   
The uniform conserved feature set for $\{x_1,x_2,...,x_n\}$, $\mathsf{S}$;
\STATE $\mathsf{S} \leftarrow \bigcup_{i=1}^n \mathsf{S}_i$;
\STATE $\mathsf{S}^{'} \leftarrow \mathsf{S}$;
\STATE $\mathsf{Q} \leftarrow \emptyset$;
\STATE $\mathsf{D}^j = \bigcup_{i=1}^n \mathsf{D}_i^j$;
\FOR{each $j \in \mathsf{S}^{'}$}
\IF{$j \notin \mathsf{Q}$}
\IF{$\sum_{i=1}^n \mathbbm{1}_{j \in \mathsf{O}_i} \geq \beta\cdot\sum_{i=1}^n \mathbbm{1}_{j \in \mathsf{S}_i}$}
\STATE $\mathsf{S} \leftarrow \mathsf{S} \setminus (\{j\} \cup \mathsf{D}^j)$;
\STATE $\mathsf{Q} \leftarrow \mathsf{Q} \cup (\{j\} \cup \mathsf{D}^{j})$;
\ENDIF
\ENDIF  
\ENDFOR  
\RETURN $\mathsf{S}$;  
\end{algorithmic}
\caption{ Forward Elimination for uniform conserved feature set.}
%\vspace{-0.15in}   
\label{alg:uniform} 
\end{algorithm}

To address these challenges,  we propose a \emph{Forward Elimination} algorithm to compute the uniform conserved feature set for a set of malicious seeds $\{x_1,x_2,...,x_n\}$, given the conserved feature sets, non-conserved feature sets and dependent sets for each seed. 
As Algorithm \ref{alg:uniform} shows, we first obtain a union of the conserved feature sets. Then, we explore the contradiction of each feature in the union with the others, by comparing the total number of the feature being selected as a non-conserved feature and conserved feature. If the former one is greater than $\beta$ times the latter one, then this feature, together with its dependents, are eliminated from the union. Otherwise, the feature is added to the uniform feature set. We use $\beta$ as a parameter to adjust the balance between conserved and non-conserved features. Typically, $\beta > 1$  as we are inclined to preserve malicious functionality associated with a conserved feature, even it could be a non-conserved feature of another PDF file. We set $\beta = 3$ in our experiments.

\subsection{Identifying Conserved Features for Other Classifiers}

Once we obtain conserved features of SL2013 for each malicious seeds, we can employ these features to identify conserved features for other classifiers using binary features. 
As our approach relies on the existence of malicious functionality and corresponding features, such a relation is not obvious for real-valued features; we therefore leave the question of how to define and identify conserved features in real space for future work.

\noindent{\bf Hidost} Hidost and SL2013 are similar in nature in such a way that they employ structural paths as features. 
The only difference is that Hidost consolidates features of SL2013 as described in Section~\ref{S:exp}.
Therefore, once the conserved features of SL2013 are identified, we can simply apply the \emph{PDF structural path consolidation rules} described in Srndic and Laskov~\cite{srndic2016} to transform these features to the corresponding conserved features for Hidost.

\noindent{\bf Binarized PDFRate} We identify the conserved features for PDFRate-B by using the conserved feature set $\mathsf{S}_i$ of each seed $x_i$. 
For each $x_i$, we generate $|\mathsf{S}_i|$ PDF files, each of which corresponds to the PDF file when an element (structural path) in $\mathsf{S}_i$ is deleted.
We then compare PDFRate-B features of these PDFs to the original $x_i$. 
If any feature value of $x_i$ is flipped from 1 to 0, then this feature will be added in the conserved feature set of $x_i$ for PDFRate-B.
Afterward, we use Algorithm \ref{alg:uniform} to obtain the uniform conserved feature set.
This approach can in fact be used for arbitrary PDF malware detectors over binary features (leveraging conserved structural paths identified using SL2013).

\subsection{Conserved Features}
\label{appendix:cf}
%In our experiments, we identified a subset of the features as the conserved feature set for the classifiers in our case study; see 
Table~\ref{table:cf} presents the full list of conserved features we identified for each classifier.
%\begin{itemize}%[noitemsep]
%
%\item \textbf{SL2013}:
%	\begin{flushleft}
%	\verb|/Names| \\
%	\verb|/Names/JavaScript| \\
%	\verb|/Names/JavaScript/Names| \\
%	\verb|/Names/JavaScript/JS| \\
%	\verb|/OpenAction| \\
%	\verb|/OpenAction/JS| \\
%	\verb|/OpenAction/S| \\
%	\verb|/Pages| \\
%	\end{flushleft}
%
%\item \textbf{Hidost}:
%	\begin{flushleft}
%	\verb|/Names| \\
%	\verb|/Names/JavaScript| \\
%	\verb|/Names/JavaScript/Names| \\
%	\verb|/Names/JavaScript/JS| \\
%	\verb|/OpenAction/JS| \\
%	\verb|/OpenAction/S| \\
%	\verb|/Pages| \\
%	\end{flushleft}
%
%\item \textbf{PDFRate-B}:
%	\begin{flushleft}
%	\verb|count_box_other| \\
%	\verb|count_javascript| \\
%	\verb|count_js| \\
%	\verb|count_page| \\
%	\end{flushleft}
%\end{itemize}

\begin{table}[t]
\centering
\scalebox{0.8}{

\begin{tabular}{|c|c|c|c|c|}
\hline
\textbf{Classifier}                 & \textbf{Conserved features}         & \multicolumn{3}{c|}{Involve JS?} \\ \hline\hline
\multirow{8}{*}{SL2013}    & /Names                     & \multicolumn{3}{c|}{No}                     \\ \cline{2-5} 
                           & /Names/JavaScript          & \multicolumn{3}{c|}{Yes}                    \\ \cline{2-5} 
                           & /Names/JavaScript/Names    & \multicolumn{3}{c|}{Yes}                    \\ \cline{2-5} 
                           & /Names/JavaScript/Names/JS & \multicolumn{3}{c|}{Yes}                    \\ \cline{2-5} 
                           & /OpenAction                & \multicolumn{3}{c|}{No}                     \\ \cline{2-5} 
                           & /OpenAction/JS             & \multicolumn{3}{c|}{Yes}                    \\ \cline{2-5} 
                           & /OpenAction/S              & \multicolumn{3}{c|}{No}                     \\ \cline{2-5} 
                           & /Pages                     & \multicolumn{3}{c|}{No}                     \\ \hline \hline
\multirow{7}{*}{Hidost}    & /Names                     & \multicolumn{3}{c|}{No}                     \\ \cline{2-5} 
                           & /Names/JavaScript          & \multicolumn{3}{c|}{Yes}                    \\ \cline{2-5} 
                           & /Names/JavaScript/Names    & \multicolumn{3}{c|}{Yes}                    \\ \cline{2-5} 
                           & /Names/JavaScript/Names/JS & \multicolumn{3}{c|}{Yes}                    \\ \cline{2-5} 
                           & /OpenAction                & \multicolumn{3}{c|}{No}                     \\ \cline{2-5} 
                           & /OpenAction/JS             & \multicolumn{3}{c|}{Yes}                    \\ \cline{2-5} 
                           & /Pages                     & \multicolumn{3}{c|}{No}                     \\ \hline \hline
\multirow{4}{*}{PDFRate-B} & count\_box\_other          & \multicolumn{3}{c|}{No}                     \\ \cline{2-5} 
                           & count\_javascript          & \multicolumn{3}{c|}{Yes}                    \\ \cline{2-5} 
                           & count\_js                  & \multicolumn{3}{c|}{Yes}                    \\ \cline{2-5} 
                           & count\_page                & \multicolumn{3}{c|}{No}                     \\ \hline 
\end{tabular}

}

\caption{Conserved features and  their relevance to JavaScript.}
\label{table:cf}
\end{table}

\subsection{Conserved vs. Regularized Features}
\label{appendix:regularized-features}
In our experiments, we empirically adjust the SVM parameter $C$ to study the overlap between \emph{conserved features} and those selected by $l_1$ regularization. 
We first adjust $C$ to perform feature reduction until the number of features is identical to the number of conserved features.
In this case, sparse versions of both SL2013 and Hidost include only 3 of the conserved features, while sparse PDFRate-B includes only 1.
In another experiment, we adjusted $C$ until all conserved features were selected.
In this case, SL2013 requires 510 features, Hidost needs 154, and PDFRate-B needs 83.
%We then adjust $C$ until the features selected by regularization contain all the conserved features.
%The results are summarized in Table~\ref{table:cf-regu}.

%\begin{table}[t]
%\centering
%\begin{tabular}{|c|c|c|}
%\hline
%Classifiers                & \# Features & \# Overlaps with CF \\ \hline \hline
%\multirow{2}{*}{SL2013}    & 8           & 3                                   \\ \cline{2-3} 
%                           & 510         & 8                                   \\ \hline
%\multirow{2}{*}{Hidost}    & 7           & 3                                   \\ \cline{2-3} 
%                           & 154         & 7                                   \\ \hline
%\multirow{2}{*}{PDFRate-B} & 4           & 1                                   \\ \cline{2-3} 
 %                          & 83          & 4                                   \\ \hline
%\end{tabular}
%\caption{Conserved vs. regularized features}
%\label{table:cf-regu}
%\end{table}

{\normalsize \bibliographystyle{acm}
\bibliography{mlvalidate}}

%\theendnotes

\end{document}